# Shifting the shear paradigm to take into account the atoms in the crystallographic theories of deformation twinning and martensitic transformations of metals and alloys

Cyril Cayron

*LMTM, EPFL (Ecole Polytechnique Fédérale de Lausanne), Neuchâtel, Switzerland*

Phone:+41.21.695.44.56; email: cyril.cayron@epfl.ch, cyril.cayron@gmail.com

ORCID number: orcid.org/0000-0002-3190-5158

Deformation twinning and martensitic transformations are displacive transitions; they are defined by high speed collective displacements of the atoms, the existence of a parent/daughter orientation relationship, and plate or lath morphologies. The current crystallographic models of deformation twinning in metals are based on the 150 year-old concept of shear. Simple shear is a deformation mode at constant volume relevant for deformation twinning. For martensitic transformations, a generalized version of simple shear called invariant plane strain takes into account the volume change; it is associated with one or two simple shears in the phenomenological theory of martensitic crystallography built more than 60 years ago. As simple shears would involve unrealistic stresses, dislocation/disconnection-mediated versions of the usual models have been developed over the last decades. However, fundamental questions remain unsolved. How do the atoms move? How could dislocations be created and propagate in a coordinated way at the speed of sound? In order to solve these issues an approach that is not based on simple shears nor on dislocation/disconnection has been applied to different displacive transformations over the last years. It assumes that the atoms are hard-spheres, which permits, for any specific orientation relationship, to determine the atomics trajectories, the lattice distortion and the shuffling (if required) as analytical functions of a unique angular parameter. The habit planes are calculated with the simple "untilted plane" criterion. A new way to understand and predict non-Schmid behaviour is proposed. The paper gives a brief historical review of the models based on the shear concept and of their dislocation-mediated versions, and it introduces the new paradigm of angular distortion. Examples in steels and magnesium alloys are taken in recent publications. The possibilities offers by this approach in mechanics and thermodynamics are briefly discussed.

# 1 The origin of the concept of simple shear

Mechanical twinning and martensitic transformations are known to form very rapidly, sometimes at the velocities close to the speed of sound; the atoms move collectively; the product phase (martensite or twins) appear as plates, laths or lenticles. These transformations are called "displacive" in metallurgy[1]. In his extensive review [1] Cahn wrote: *"Cooperative atom movements in phase transformations may extend over quite large volumes of crystals. When this happens, the transformation is termed martensitic (from Martens, who discovered the first transformation of this type in carbon steel). If the atom movements in a crystal results in a new crystal of different orientation, but identical structure, the process is termed mechanical twinning."* Christian and Mahajan [2] also insisted on the great similarities between deformation twinning and martensite transformations: *"all deformation twinning should strictly be regarded as a special case type of stress-induced martensitic transformation…"* Simple shears are the corner stone of the theories of deformation twinning and martensitic transformation developed during the last century [1]-[3], and the notion of simple shear can even be traced back to older times, as recalled by Hardouin Duparc [4]. One hundred and fifty years ago, in 1867, William Thomson and Peter Guthrie Tait's Treatise on Natural Philosophy [5] (from Ref. [4]) defined that "*'a simple shear' is the property that two kinds of planes (two different sets of parallel planes) remain unaltered, each in itself*". The mathematical formalism, with the nomenclature ($K_1$, $\eta_1$, $K_2$, $\eta_2$) was introduced by Otto Mügge in 1889 [6] (from Ref. [4]). At that time, the crystallographers and mineralogists already knew the periodic structure of minerals thanks to Haüy's works at the beginning of the 19th century [7]. The crystal periodicity was expressed by the lattice, but the concept of unit cell was yet not fully clarified; it was just a "*molecule intégrante*", like a "brick". Therefore, it was natural at that time to consider that the "bricks" could glide on themselves in order to rearrange them and form a new twinned lattice, as illustrated in Figure 1.

---

[1] Please note that the meaning of the term "displacive" is (unfortunately) different from the one used in physics where "displacive" implies only small displacements of atoms without breaking the atomic bonds.

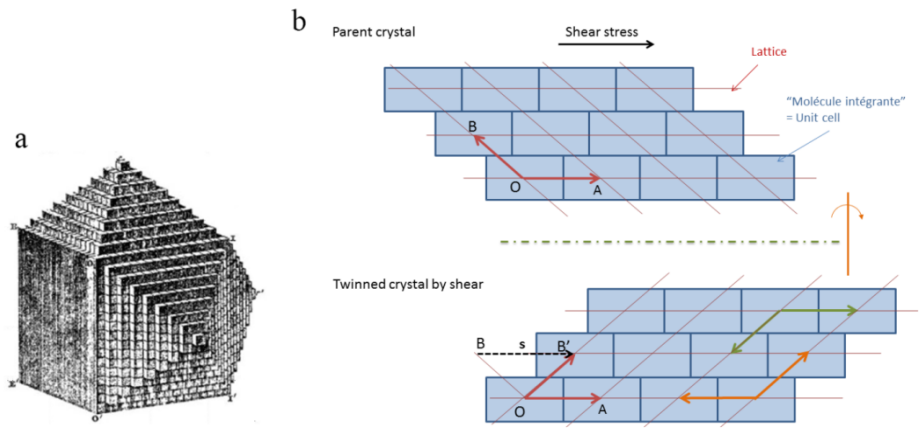

Figure 1. Crystal and deformation twinning. (a) Crystal imagined as a stacking of bricks called "*molecules intégrantes*" (today "unit cell" or "motif"), as published in 1801 by Haüy [7]. The way to envision deformation twinning at the end of the 19th century was probably as follows: b) twinning can be obtained by gliding the bricks on themselves in such a way that the lattice is restored after the glide. The basis vectors are represented by the red arrows, they are modified by the shear, but the newly formed red twinned lattice is equivalent to the initial one by reflection or by 180° rotation.

At the early beginning of the 20th century these notions originated from mineralogists were acquired by the metallurgists, without any possibility at that time to clarify further the nature of the "bricks". Remember that the atoms were still a speculative idea [8]. Democrite and ancient Greek philosophers supposed their existence, but the first evidence was given in the early 1800s by the chemist John Dalton who noticed that the chemical reactions always imply ratios of elements in small integers. Dalton also imagined the atoms as solid spheres. The idea of discontinuous matter was however not unanimously admitted. In 1827, botanist Robert Brown observed the erratic movements of dust grains floating at the water surface; and the Bownian motion was explained in 1905 by Albert Einstein using statistical physics. The mass and dimensions of the atoms were soon experimentally determined by physicist Jean Perrin. A new area then opened to specify the properties of the atoms. Very quickly, it appeared with the discoveries of the electron by Thomson, the protons by Rutherford, and the quantum physics born with Bohr, that the atoms were far more complex than Dalton's solid spheres. However, in many metals, the hard-sphere assumption holds satisfactorily. William Barlow in 1883 [9] noticed that the symmetries of some metals and alloys can be represented by packings of hard-spheres, with among them the body centred cubic (bcc), the face centred cubic (fcc) and the hexagonal close-packed (hcp) structures. The precise determination of the lattice parameters thanks to X-ray or electron diffraction revealed that in many bimetallic solid solutions, the lattice parameter of the mixture of two elements is the average of the lattice parameter of each pure element, in agreement with a hard-sphere packing model. The c/a ratios of magnesium, cobalt, zirconium, titanium, rhenium and scandium differ from the ideal packing ratio $\sqrt{\frac{8}{3}}$ by less than 3%. During fcc-bcc martensitic phase transitions, the difference of lattice parameters expected by a hard-sphere model is only 4%. The critical shear required to activate dislocations was studied in 1947 by Bragg and Nye [10] with sub-millimeter bubble rafts, and the hard sphere model is still used in classrooms to explain to our students the ABCABC and ABABAB packing of the fcc and hcp phases respectively. Hard-sphere is also an important approximation made in molecular dynamics. However, the atoms, even with their simple hard-sphere image, are not explicitly present in the crystallographic theories of twinning and martensitic transformations, which are all based on the lattices and their transformations by shears (with or without dislocations), as in the initial old Mügge's model. The atoms are the "big losers" of these theories.

## 2   Shears used in the theories of deformation twinning

*Cahn wrote about deformation twinning: "That part of the parent crystal which is thus transformed undergoes a macroscopic change of shape which can be described exactly as a simple shear"* [1]. The shear is defined by its shear plane $K_1$, its shear direction $\eta_1$, and its amplitude $s_1$. The shear leaves a second plane $K_2$ undistorted (but rotated). The direction $\eta_2$ belongs to $K_2$ and is perpendicular to the intersection between $K_1$ and $K_2$. The plane $K_2$ is rotated by an angle $\phi$ and the shear amplitude is given by $s_1 = 2\tan(\phi/2)$. It is classical to define two types of twins. Type I are twins where the plane $K_1$ and the direction $\eta_2$ are rational. Type II are twins where the plane $K_2$ and the direction $\eta_1$ are rational. For type I twins, $K_1$ is a mirror plane common to both parent and twin crystals. For type II twins $\eta_1$ is the two-fold axis common to both parent and twin crystals. For any mode I twin, defined by $K_1$ along $\eta_1$, one can associate a conjugate mode II twin defined by substituting $K_2$ in place of $K_1$ and $\eta_2$ in place of $\eta_2$ [1][6]. The four components $K_1$, $K_2$, $\eta_1$, $\eta_2$ and the classical geometric representation of these elements with the shearing ellipsoid, as represented by Hall in 1954 [11] (Figure 2a), date from Thomson and Guthrie Tait who clearly noted *"the planes of no distortion in a simple shear are clearly the [two] circular sections of the strain ellipsoid"* [5] (from Ref. [4]). From the 1950s, the twinning theory was mathematically developed and refined, mainly by Kihô in 1954 [12], Jaswon and Dove in 1956 [13], and later by Bilby, Crocker and Bevis [14]-[16], following some of the notions and crystallographic tools used by Bowles and Makenzie in the Phenomenological Theory of Martensite Transformation (PTMC) detailed in a next section. The master equation of the crystallographic theory of

twinning can be summarised by $\mathbf{C} = \mathbf{R}.\mathbf{S}$, where $\mathbf{C}$ is the correspondence matrix, $\mathbf{R}$ is a rotation, and $\mathbf{S}$ is a simple shear. These three matrices have a determinant equal to ± 1. The correspondence matrix gives the coordinates in the twin basis of the parent basis vectors once distorted by the shear matrix. As these vectors should be vectors of the twin lattice, the coordinate should be integers, or rational if the cell of the Bravais lattice contains more than one atom, as it is the case for hcp metals, or if a supercell is chosen for the calculations. For these cases, the atoms inside the cell that are not positioned at the nodes of the cell don't follow the same shear trajectory as those positioned at the nodes; one says that they "shuffle" (we will come later on this term). An additional mathematical restriction comes from the fact that the rotation matrix should check $\mathbf{R}.\mathbf{R}^T = \mathbf{I}$, with $\mathbf{I}$ identity matrix, which permits to establish a list of correspondence matrices. Among the numerous (but finite) possible correspondence and shear matrices given by the theory, only those with the lowest shear amplitude and minimum shuffles are considered as realistic. For example, fifteen deformation twinning modes could be listed in titanium [16], some of them are reproduced in Figure 2b. The theory built by Bilby, Crocker and Bevis is still used today (see for example [17]). Despite its mathematical sophistication and rigour, one must admit that its fundaments do not differ from the initial 150-years old view. Despite the progress, the theory does not reply to the simple question: how do the atoms move during twinning? In the case of a cell made of one atom, it is assumed that the atoms follow the same shear as the nodes of the lattice, but for multiple-atom (super)cells, the shuffling of the atoms is not satisfactorily treated. Some possible shuffling modes were proposed by Bilby and Crocker [14]; they often consist in finding the different ways of translating the positions obtained by the shear in order to reach those obtained by the reflections or by the 180° rotation symmetries. Shuffles in complex twinning of titanium are shown by the arrows of Figure 2b.

Shuffling was not examined anymore in the last version of the theory proposed by Bevis and Crocker [15].

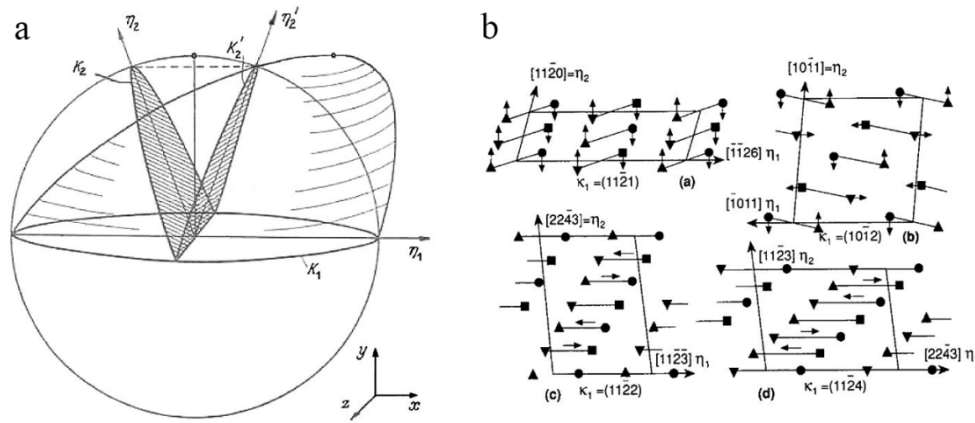

Figure 2. Deformation twinning. (a) Lattice deformation and associated ($K_1,\eta_1,K_2,\eta_2$) crystallographic parameters, represented by Hall in 1954, from Ref. [11]. There is no evolution of the fundamental concept in comparison with Mügge's work published in 1889. (b) Complex deformation twinning in titanium involving supercells and shuffles, from Crocker and Bevis in 1970 [16].

Actually, even if not clearly admitted, it seems unrealistic that the atoms can simply move along straight lines such as drawn by the arrows of Figure 2b; even in the case of one atom per cell. For steric reasons the atoms can't follow simple translations dictated by a simple shear. The way the atoms move is of prime importance, but for a long time, the theories were focused on the way the lattices are deformed, omitting to confront to the question of the atomic trajectories. Therefore, the crystallographic theory of twinning is also a *phenomenological* theory, exactly as the PTMC that will be detailed in the next section. Besides, assuming that twinning results from a simple shear implies that the Schmid's law [18] should be verified. Deformation twinning should occur when the shear stress resolved along the slip direction on the slip plane reaches a critical value, called the critical resolved shear stress. However, this law is difficult to confirm experimentally (see for example the section 5.1 "*Orientation dependence: is there a*

*CRSS for twinning?"* of Ref. [2]). In addition, it was experimentally noticed that some twinning modes in magnesium have "abnormal", i.e. "non-Schmid" behaviour [19]. The reason was attributed to the fact that local stresses differ from global ones [20], but to our knowledge, there are no convincing quantitative experimental results that could confirm this explanation. Non-Schmid behaviour was also observed at least for slipping deformation in bcc metals. In 1928, Taylor wrote "*in β-brass resistance to slipping in one direction on a given plane of slip is not the same as resistance offered to slipping in the opposite direction*" [21]. This effect, called twinning-antitwinning effect, was attributed to the core structure of the dislocations in the bcc metals [22][23], but the fundamental reason of the dissymmetry is that Schmid's law is a continuum mechanics law that doesn't take into account the atomic structure of the metals, or more precisely the configuration of the atoms in the stacking of the glide planes or twinning planes. The normal to the {112} bcc twinning plane is not a two-fold rotation axis, which means that a shear along a <111> direction is not equal to a shear in its opposite direction. The same situation exists for deformation twinning in fcc metals: as the direction normal to the {111} plane is a three-fold axis but not a two-fold axis, the shears in the <110> directions are not equal to their opposites. The use of simple or pure shear (both are deformation at constant volume) was proved to be highly efficient in continuum mechanics to describe the plastic behaviour of materials, but its direct application to describe lattice transformation of crystals (packings of discontinuous atoms) raise some important questions that should not be discarded too quickly.

## 3 Shears used in the theories of martensitic transformations

### 3.1 Early models of martensitic transformations

For martensitic transformations, the notion of shear was also so pregnant that some great metallurgists did not hesitate to write that "*shear transformations are synonymous with martensitic transformations*" [24]. However, the first crystallographic model of fcc-bcc transformation proposed by Bain in 1924 [25], with it is well-known Bain (stretch) distortion does not imply shear. The orientation relationship (OR) expected from the Bain's model is not observed. Indeed, few years later, Young in 1926 [26], Kurdjumov and Sachs in 1930 [27], and Wassermann in 1933 [28] and Nishiyama in 1934 [29], could determine by X-ray diffraction the orientation relationship between the parent fcc and the bcc martensite. Young worked on Fe-Ni meteorites, Kurdjumov and Sachs on Fe-1.4C steel, and Nishiyama on a Fe-30Ni alloy. The OR discovered by Young was very close to that discovered by Kurdjumov and Sachs, but history of metallurgy only kept the names of the two latter researchers. The now called KS OR (for Kurdjumov-Sachs) and NW OR (for Nishiyama-Wassermann) are at 5° from each other, and at 10° away from the OR expected from a direct Bain distortion. The difference is non negligible; that's why Kurdjumov, Sachs and Nishiyama proposed in their respective papers a model of fcc-bcc transformation. Kurdjumov and Sachs imagined a transformation made by two consecutive shears: $(111)_\gamma [\bar{1}\bar{1}2]_\gamma$ followed by $(1\bar{1}2)_\alpha [\bar{1}11]_\alpha$. Nishiyama proposed a slightly different sequence in which the first shear $(111)_\gamma [\bar{1}\bar{1}2]_\gamma$ is followed by a stretch. These models are quite close and summarized by Nishiyama in his 1934 paper [29] with a summary figure that we report in Figure 3. Therefore, in those models, the fcc-bcc transformation is not a Bain distortion, but it is neither a simple shear; it is a combination of shears or shears and stretch.

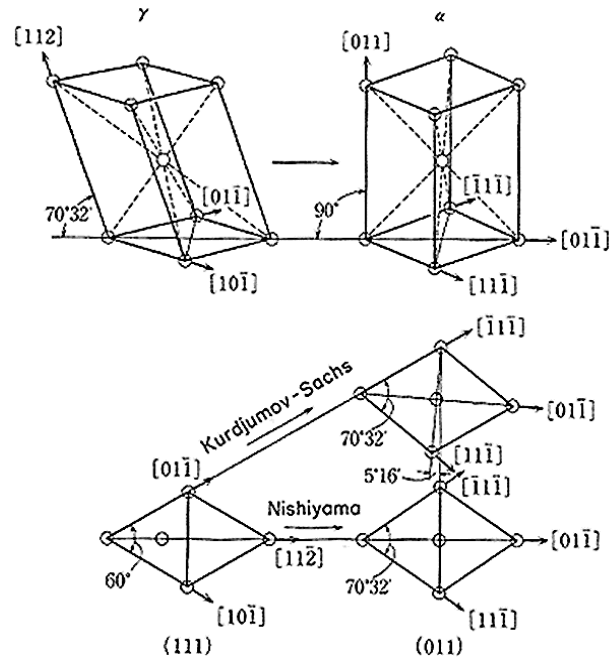

Figure 3.  KSN model of fcc-bcc phase transformation proposed by Kurdjumov and Sachs (1930) and Nishiyama (1934), from Ref.[29].

The important point of the KSN model is the distortion of the dense plane $\{111\}_\gamma$ into a dense plane $\{110\}_\alpha$. Such planar distortion was also noticed and largely discussed by Young who wrote *"On account of the marked resemblance of the (111) plane of the solid solution [taenite, fcc] and the (110) plane in kamacite [bcc], it is possible to form the solid solution by simply shearing rows of atoms and rearranging the atoms in adjacent planes, as already described, a crystal of kamacite which is only a few planes in thickness but of considerable area."* [26].

In 1934 Burgers determined by X-ray diffraction the OR between the high temperature parent bcc phase and the low-temperature martensite hcp phase in zirconium [30]; and, as Young, Kurdjumov, Sachs and Nishimaya did for fcc-bcc transformations, he proposed a crystallographic model of the bcc-hcp transformation. Actually, he proposed three models, with one of them implying an intermediate fcc phase, but only the first one is now widely accepted. This model is reproduced in Figure 4. Burgers explains it as the combination of *"a shear parallel to a {112} plane in the [111] direction lying in this plane, followed by a definite displacement of alternate atomic layers [shuffle] and a homogeneous contraction (eventual dilatation) parallel to definite crystallographic directions."* Burgers also calculated the associated shear value and found *s* = 0.22. His description is not exactly that of his figure, as two shears are actually applied on two different {112} planes. In order to overcome this problem, it is now usual to replace the shears by a diagonal distortion of the orthorhombic cell marked by the bold lines in part B of Figure 4, as described for example in Kelly and Groves' book [31]. This distortion is analogous to the Bain distortion in the fcc-bcc transformation, but the difference is that a shuffle is required to put half of the atoms in their good positions and obtain the hcp phase, as shown in the part noted by the letter D in Figure 4.

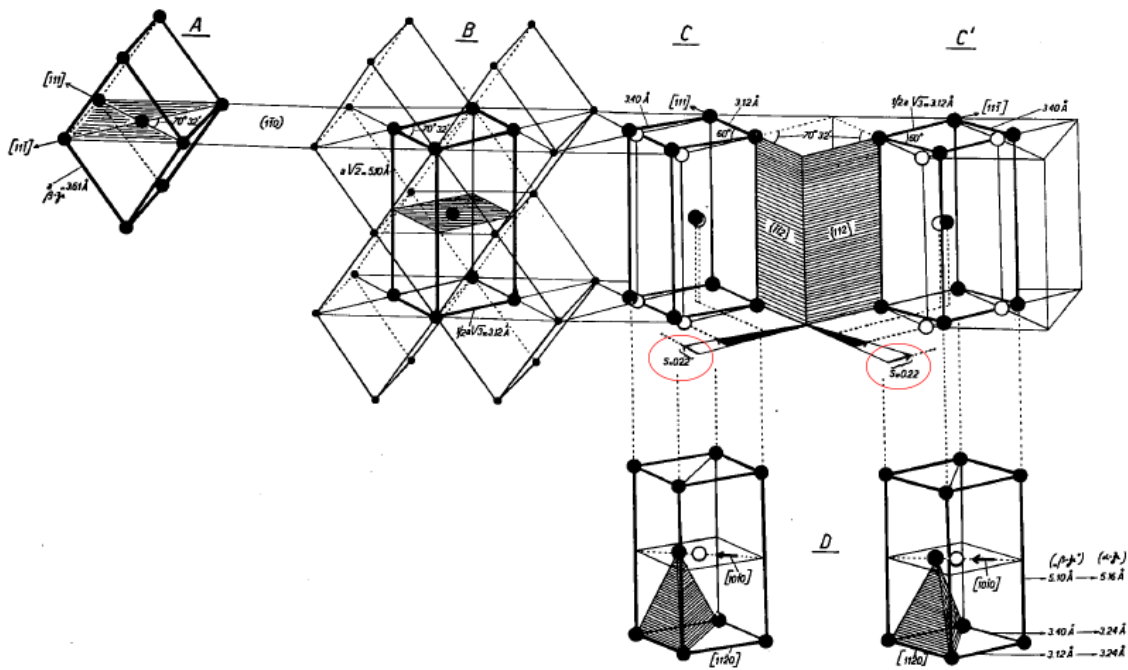

Figure 4.  Burgers' model of bcc-hcp transformation, from Ref. [30]. The two simultaneous shears are marked by the red ellipses.

## 3.2 The phenomenological theory of martensite crystallography (PTMC)

The KSN model of martensitic fcc-bcc transformation was criticized by Greninger and Troaino in their 1949 paper [32] because the KSN model could not explain the observed habit planes in Fe-22Ni-0.8C and because of the *"relatively large movements and readjustments"* needed to obtain exactly the fcc structure. They proposed that *"The martensite crystal is formed from austenite crystal almost entirely by means of two homogeneous shears. The function of the first shear is to create a lattice containing a unique set of parallel atomic planes whose interplanar spacing and atom positions are the same as those of a set of planes in the martensite lattice; a second shear on this unique plane will then generate the martensite lattice"*. Thus, the main idea is to combine two shears, one is supposed to explain the shape, i.e., the habit plane of martensite; it is the invariant plane strain (IPS); while the other one deforms the lattice without changing the shape because of an "invisible" compensating lattice invariant shear (LIS). The idea of composing two homogeneous shears to obtain an inhomogeneous structure with an invariant plane was then soon associated with the notion of correspondence matrix previously introduced by Jaswon and Wheeler in 1948 [33]. This matrix establishes a correspondence between the directions of the fcc and bcc lattices after a Bain distortion. The result forms a mathematical and unified theory that "predicts" both the habit planes and the orientation relationships. This theory was proposed by Weschler, Liebermann and Read [34], and by Bowles and Mackenzie [35][36] in 1953-54. This theory, now called Phenomenological Theory of Martensite Crystallography (PTMC), has been adopted by most of the metallurgists, thanks to exhaustive review papers and books, as those written by Christian [37] and Nishiyama [38], and thanks to the very didactic books written by Bhadeshia [39][40]. The theory of deformation twinning (see previous section) reintroduces most of the mathematical tools used in the PTMC.

For the classical fcc-bcc martensitic transformations in steels, the master equation of PTMC is $\mathbf{RB} = \mathbf{P}_1\mathbf{P}_2$, where $\mathbf{RB}$ is the product of the lattice deformation constituted of the symmetric Bain stretch matrix $\mathbf{B}$ by an additional rotation matrix $\mathbf{R}$, and where $\mathbf{P}_1$ is an IPS and $\mathbf{P}_2$ is a simple shear. PTMC assumes that the simple shear $\mathbf{P}_2$ is exactly compensated by a LIS $(\mathbf{P}_2)^{-1}$ produced by twinning or dislocation gliding such that the martensite shape is only given by $\mathbf{P}_1$. The IPS $\mathbf{P}_1$ appears as a generalized notion of simple shear that takes into account the volume change of the phase transformation, i.e. the dilatation or contraction component in the direction normal to the shear plane. It gives the shape of the martensite product (lath, plate or lenticle); the shear plane of $\mathbf{P}_1$ is the habit (interface) plane. Both parts of the equations, $\mathbf{RB}$ and $\mathbf{P}_1\mathbf{P}_2$ are invariant line strains (ILS), and the rotation $\mathbf{R}$ is the rotation added to render unrotated the line undistorted by the Bain stretch $\mathbf{B}$. This is usually geometrically illustrated in classical textbooks by showing how a sphere is deformed into an ellipsoid, and by finding the intersection points between the sphere and the ellipsoid. Didactic schemes of the IPS and strain matrices used in the PTMC are given in Figure 5 according to Ref. [40].

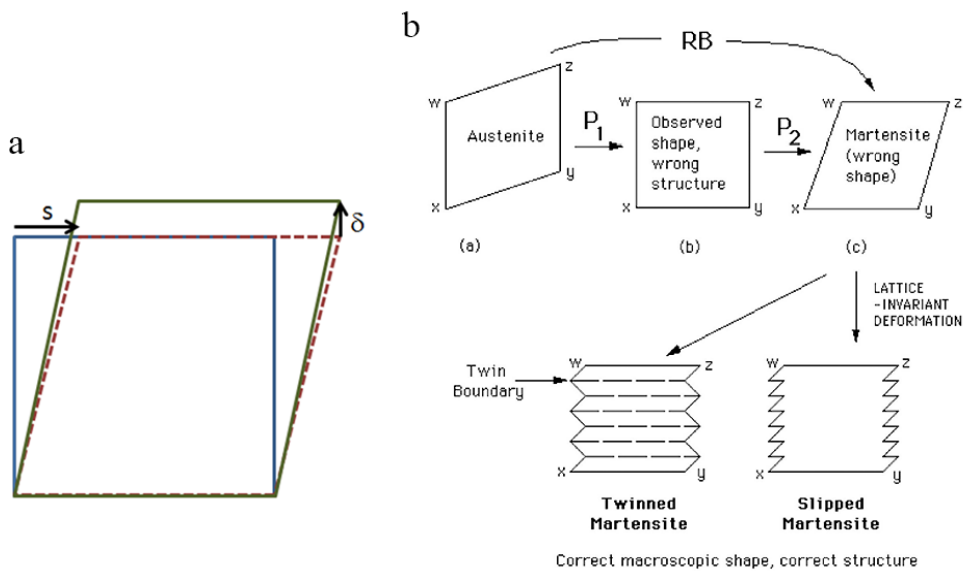

Figure 5. The use of shears in PTMC. (a) Schematic view of an invariant plane strain (IPS) with the shear value s and the dilatation part δ perpendicularly to the shear plane. (b) Combination of the IPS $P_1$ (also called shape strain) with a simple shear $P_2$ that generates the correct and well oriented lattice RB. The simple shear $P_2$ *is* compensated by a lattice invariant shear to get the correct shape $P_1$, from Ref. [40].

The PTMC is very subtle and their main inventors introduced important modern concepts in crystallography, such as the coordinate transformation matrix, the transformations from direct to reciprocal spaces, clear references to the bases used for the calculations, correspondence matrices, equations that link the shear matrices with the correspondence and coordinate transformation matrices etc. That is true that PTMC predicted the existence of twins inside the bcc or bct martensite formed in steels and that these twins were not visible by optical microscopy. Their discovery by Nishiyama and Shimizu at the early era of transmission electron microscopy (TEM) in 1956 [41] probably made Nishiyama finishing to give up his two-step shear-stretch model (Figure 3) to fully adopt the PTMC. That was not a complete change of mind according to Shimizu [42]: "*he [Nishiyama] foresightedly expected the existence of lattice invariant deformation in martensite a few years before the phenomenological crystallographic theory of martensitic transformation was proposed*", but it is plausible that Nishiyama's rallying to PTMC had a huge impact on the rest of the scientific community and finished to convince the most recalcitrant metallurgists of that time about the importance of this theory. However, the predictive character of PTMC is often "oversold". Some researchers claim that PTMC "predicts ALL" the martensite features, but that affirmation should be considered thoughtfully. For example, one can be impressed by the apparent prediction of the orientation relationships, but it should be reminded that PTMC starts from the Bain OR, which is at 10° from the experimentally observed KS or NW ORs, and then PTMC imposes the existence of an invariant line, which makes it closer to KS. If one considers the strong internal misorientations (up to 10°) inside the martensitic products, "finding" an OR close to one belonging to the continuous list of experimentally observed ORs should not be considered as a "prediction". Concerning the habit planes, one should be clear: PTMC did not *predict* these planes because these planes were already observed before the establishment of the theory; PTMC "only" tried to *explain* them. PTMC did not make predictions but postdictions; which is completely different when the reliability of a theory is considered. The {259} habit planes were explained in 1951 by Machlin and Cohen [43] following the Greninger and Troiano's initial idea of composing shears. The {225} habit planes were explained by Bowles and Mackenzie [35] with a $(225)_\gamma[\bar{1}\bar{1}2]_\gamma$ IPS followed by $(112)_\alpha[11\bar{1}]_\alpha$ shear, but with the help of a dilation parameter that was later subject of controversies. Numerous papers were published on {225} martensite by increasing the level complexity with extra shears (see for examples [44][45]). This "saga" of the {225} habit planes was summarized in 1990 by Wayman [46], and more recently in 2009 by Dunne and by Zhang and Kelly in Ref. [47][48]. Surprisingly, the fact that the {225} habit planes could be explained by using different methods and different parameters did not raise interrogations on the real relevance of the PTMC approaches. It is also important to note here that all the PTMC studies never cite the Jaswon and Wheeler's study [33] for the explanation of the {225} habit planes (sometimes they refer to this paper but only in the introduction for the use of Bain correspondence). It is worth recalling that Jaswon and Wheeler's model was discarded by Bowles and Barrett in 1952 [49] because: *"Jaswon and Wheeler's picture of the transformation as a simple homogeneous distortion of the lattice is not consistent with the observed relief effects"*. We will come back later on this "discarding" that we considered as an unfortunate missed opportunity. Apart from the {225}, the {557} habit planes have also been the objects of many researches for more than 60 years without consensus; most of them include additional shears, see for example Ref. [50][51]. The development of the PTMC mainly consisted in adding or varying the shears and their combinations, i.e. by adding complexity, without being more effective than their initial BM and WLR models. The "predictions" are sometimes written with 6 or 8 digit numbers and compared with experimental results where accuracy rarely exceeds one digit.

PTMC was applied to other transformations, such as the bcc-orthorhombic and bcc-hcp transformations [52], and here again the same criticisms can be raised. The theory does not predict the orientation relationships because it starts from Bain-type distortions which already agree with experimental orientations. In the case of the bcc-hcp transformation for example, PTMC starts with the orthohexagonal cell that was already defined by the Burgers OR and shown in Figure 4. The habit planes are not predicted, but "explained" by choosing the LIS twinning systems among those reported by experimental observations and by adjusting the dilatation parameter and the twinning amplitude in order to fit the calculated habit planes with those experimentally reported, as in Ref. [53]. The success of PTMC for shape memory alloys in which the transformation strains are lower than in steels is probably more convincing. Importantly, as admitted by its creators and by most of its promoters, PTMC is and remains *phenomenological*. For example, Bhadeshia [39] clearly wrote that *"the theory is phenomenological and is concerned only with the initial and final states. It follows that nothing can be deduced about the actual paths taken by the atoms during transformation: only a description of the correspondence in position between the atoms in the two structures can be obtained."* Therefore, it is the same problem as for deformation twinning; the theory does not answer the simple question: how do the atoms move during the transformation?

Bogers and Burgers developed in 1964 a hard sphere model in order to respond to this essential question for the fcc-bcc martensitic transformations [54], as illustrated in Figure 6. They noticed that if a shear on a $(111)_\gamma$ plane is applied to a fcc crystal and stopped at the midway between the initial fcc structure and its twin, the operation distorts the two adjacent $\{111\}_\gamma$ planes into $\{110\}_\alpha$ planes, as it is the case for Bain distortion. This idea was actually not new, as Cottrell in his book [55] reported that Zener [56] thought that *"The face-centred cubic metals, for example, pass through twinning and body-centered cubic configurations when sheared on their slip planes"*. However, Bogers and Burgers noticed that actually the midway structure is not exactly bcc, and another shear on another $(111)_\gamma$ plane is required to obtain the final and correct bcc structure.

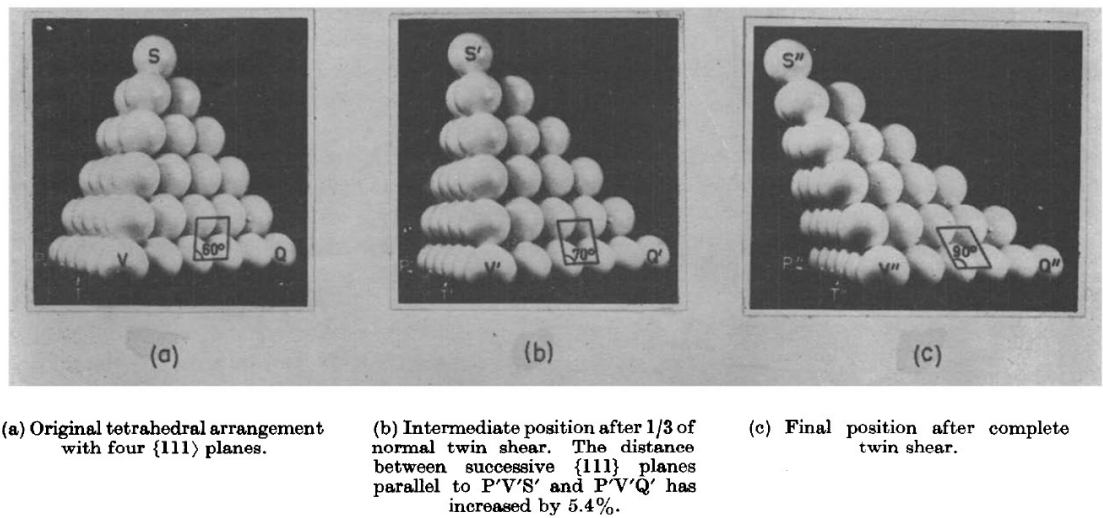

Figure 6. Hard-sphere model proposed by Bogers and Burgers for the fcc-bcc transformation and fcc-fcc deformation twinning. The initial fcc structure (a) becomes after distortion a new twinned fcc structure (c). The bcc structure is at the midway between the two structures (b), but additional shears are required to place the atoms at their correct positions. From Ref. [54].

Their work was later corrected/refined by Olson and Cohen in 1972 and 1976 by the introduction of two shears [57][58]. It can be summarized in Figure 7 as follows: the first shear on a $\{111\}_\gamma$ plane is achieved by $1/6 <112>_\gamma$ partial dislocations averaging one over every second $(111)_\gamma$ slip plane, and the second shear on another $\{111\}_\gamma$ plane along is achieved by $1/6 <112>_\gamma$ partial dislocations averaging one over every third $(111)_\gamma$ slip plane. The former is noted T/2 and the latter T/3. This approach is in qualitative agreement with the observations of the martensite formation at the intersection of hcp plates or stacking faulted bands on two $(111)_\gamma$ planes. The model has an interesting physical basis but its intrinsic asymmetry between the $\{111\}_\gamma$ planes with T/2 and T/3 seems to be too strict to be obtained in a real material. The model is quite complex and does not answer the question about the atomic trajectories. It also raises questions about the origin of the partial dislocations, but we will come back on this point in the next section.

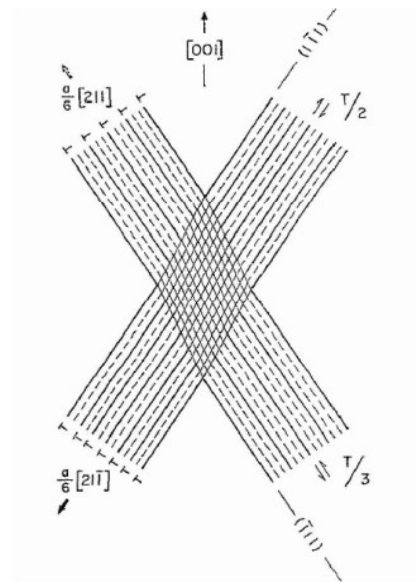

Figure 7.  Olson-Cohen's model. The α martensite is formed at the intersection of the two planar faults T/2 and T/3. From Ref.[57].

Another hard-sphere model of fcc-bcc transformation was proposed by Le Lann and Dubertret [59]. It contains some essential ingredients, such as the fact that one of dense directions remains invariant $<110>_\gamma = <111>_\alpha$, and the authors qualitatively envisioned the transformation as a wave propagating perpendicularly to this direction. In addition, the authors proposed atomistic structures of the well-known $\{225\}_\gamma$ and $\{3,10,15\}_\gamma$ habit planes. However, the model is quite difficult to understand because it implies the distortion of a regular octahedron made of 19 atoms; and it is mainly geometric, it does not explain how to calculate the atom trajectories or the distortion matrix.

More than 150 years after Mügge's model of deformation twinning, and more than 60 years after the birth of PTMC, the exact continuous paths related to the collective displacements of the atoms during the lattice deformation remain beyond the possibilities of the classical crystallographic theories of displacive transformations in metals. We think that the absence of progress is due to the main paradigm of these theories, i.e. the assumption that the lattice distortion should be a shear or a composition of shears.

## 4 Simple shear saved by the dislocations/disconnections?

Simple shears or its derivative lattice invariant strain is perfectly adapted to model the collective displacements of the atoms that move all together at the same time. However, simple shear has raised important issues for a long time. The atoms are not bricks that can glide on themselves as represented in Figure 1. A collective shear displacement of the atoms would imply a shear stress of the same order of magnitude as the Young modulus. It is worth recalling the usual "demonstration" that was given by Frenkel [60] in 1926 and reported in different books [61][62]. For example in Ref. [61], it is explained that *"The shearing force required to move a plane of atoms over the plane below will be periodic, since for displacements x<b/2, where b is the spacing of atoms in the shear direction, the lattice resists the applied stress but for x>b/2 the lattice forces assist the applied stress. The simplest function these properties is a sinusoidal relation of the form $\tau = \tau_m \sin(\frac{2\pi x}{b}) \approx \tau_m \frac{2\pi x}{b}$, where $\tau_m$ is the maximum shear stress at the displacement = b/4. For small displacements the elastic shear strain given by x/a is equal to $\tau/\mu$ from the Hooke's law, where µ is the shear modulus, so that $\tau_m = \frac{\mu}{2\pi}\frac{b}{a}$ and since b ≈ a, the theoretical strength of a perfect crystal is of the order of µ/10."* The sinusoidal form was also used by Peierls in his famous paper introducing the friction force on a dislocation [63]. However, Frenkel's demonstration is actually misleading. Indeed, there is no reason to believe that the periodicity along the x-axis could explain the stress value at $x = 0$. What could justify that the value $\tau(x = 0)$ is correlated to the maximum value $\tau(x = \frac{b}{4})$? Actually, the "trick" of the demonstration is hidden in the sinusoidal function. This function seems to be harmless, but it already contains the result because it is such that its derivative is proportional to its maximum. Other periodic functions lead to completely different results. Let us consider for examples the soft and rigid functions defined as follows:

(Soft) $\tau = \tau_m \sin^3(\frac{2\pi x}{b})$ (1)

(Rigid) $\tau = \begin{cases} \tau_m\sqrt{y} & \text{if } y = \sin\left(\frac{2\pi x}{b}\right) > 0 \\ -\tau_m\sqrt{-y} & \text{if } y = \sin\left(\frac{2\pi x}{b}\right) < 0 \end{cases}$

The sinusoidal function, and the soft and rigid functions are shown in Figure 8.

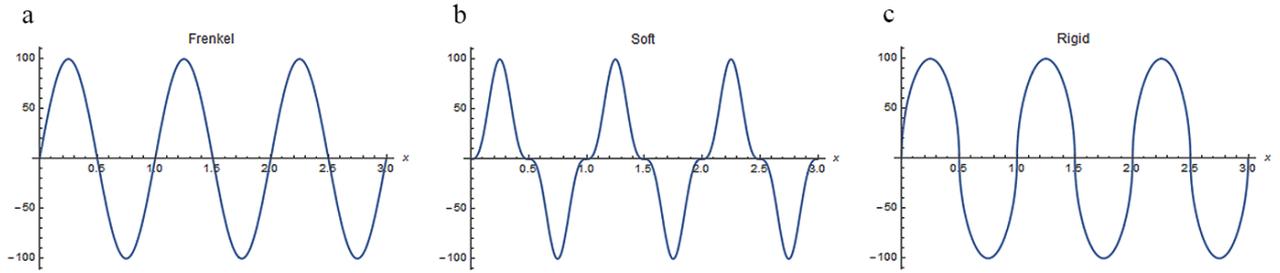

Figure 8. Periodic functions τ with different slopes μ at the points τ = 0. (a) Sinusoidal function, μ = 1, (b) soft function, μ = 0, (c) infinitely rigid function, μ = ∞.

The soft case gives $\mu = \frac{d\tau}{dx}_{x=0} = 0$, and the rigid case gives $\mu = \frac{d\tau}{dx}_{x=0} = \infty$. In these two extreme cases, there is no correlation between the shear modulus μ and the maximum stress $\tau_m$. One could also find different functions for which the shear modulus is constant but the maximum shear stress are very different. Thus, the result is given in the premise; the Frenkel's demonstration is biased; any argument based on the periodicity of the structure can't be correct. Cottrell mentions in the first chapter of his excellent book [55] that Orowan told him in a private communication that *"the critical shear strain should be less than this when a realistic law is taken for the force between the atoms"*. Cottrell then made a rough estimation using a central law force and came to estimate this strain is about μ/30. He also gave the advice that "the sinusoidal function [..] should be replaced by a [central force] relation". However, he did not specify that the periodicity argument should be definitely discarded. Despite Frenkel's error, and as shown by Cottrell, it is correct to assume that the maximum shear stress associated with a simple shear is of order of a tenth of magnitude of the shear modulus. Nabarro [64] also took a more appropriate function, as suggested by Cottrell, and showed that the order of magnitude of the friction force estimated by Peierls is correct. The fundamental reason is not the periodic nature of the lattice but the fact that the interactions between atoms located in neighboured unit cells create a lattice friction. This can be understood by considering the case of a one-atom unit cell of Figure 9a, for which the atom interaction is assumed to be fully elastic.

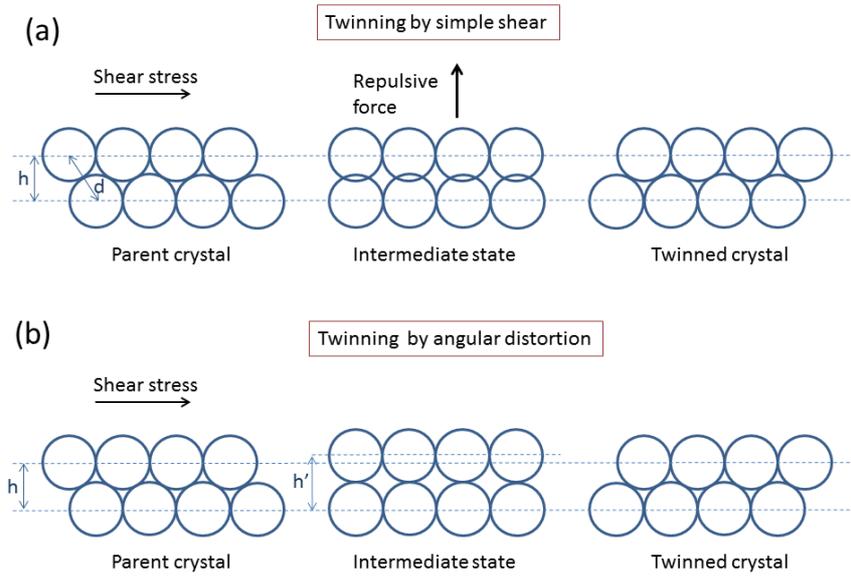

Figure 9.   Schematic 2D representation of a collective displacement of the atoms during deformation twinning according to (a) simple shear, (b) angular distortion.

If the trajectory of the atoms could continuously follow a simple shear strain, they would interpenetrate so much that the repulsive force of the surface would reach very high values. In the 2D example of Figure 9a, by noting $d$ the diameter of the atoms, the rate of interpenetration would be $1 - \frac{h}{d} = 1 - \sqrt{3}/2 \approx 13.4\%$, i.e. far above the usual elastic limit (around 0.1 to 1%). In the 3D case of a ABCABC stacking, the layers are separated by a distance $h = \frac{1}{2}\sqrt{\frac{3}{8}}d$, and a simple shear displacement of the atoms of the upper layer in-between the two atoms of the lower layer would reduce the distance between the atoms down to $d' = \frac{1}{2}\sqrt{h^2 + \frac{d^2}{4}}$, which induces an interpenetration of $1 - \frac{d'}{d} = 1 - \sqrt{\frac{11}{12}} = 8.3\%$. Here again, this value is too high to be compatible with the elastic limits of the metals. Thus, even if incorrectly demonstrated by Frenkel (at least the sinusoidal function should be replaced by a more realistic function as suggested by Cottrell), it is true that it is not possible with reasonable stress to induce a collective movement of the atoms on a crystallographic plane. This result was also confirmed with the experiments made by Bragg and Lomer with sub-millimeter soap bubble rafts [65]. The conclusion seems unescapable: dislocations are required for gliding and twinning. We will see that the conclusion actually relies on another assumption: it is supposed that all the atoms move simultaneously, i.e. *all at the same time*; but we will come back later on this point.

The impossibility of simple shear with reasonable stress value immediately made the scientists reach the turning point. Dislocation theory was emerging at that time, and soon went successful in the 1940-1950 to explain plasticity of metals, recrystallization and many other phenomena in materials science. Brilliant confirmations of the existence of dislocations were also obtained by TEM in the early 1950' [66]. Consequently, most of the researchers came to assume that dislocations were the cause and the fundamental part of deformation twinning. According to Hardoin Duparc [4], the term "twin dislocation" was coined by Seitz and Read in 1941 [67], and this concept has been pursued till now by Vladimirskii [68], Frank and van der Merwe [69], Sleeswyk [70], Christian and Mahajan [2][71], and many other great scientists. Let us cite Cottrell [55] when he explains the arguments for this turning point and the reasoning at the origin of the concept of twinning dislocation. "*It has often been suggested that mechanical twinning takes place by the continuous growth on an atomic scale of twinned materials, and the general arguments for a dislocation mechanism are the same as in the case of slip; first, it is scarcely believable that the atom should all move simultaneously and, second, twinning occurs at stresses comparable with those for slip, i.e. far below the*

*theoretical strength of the perfect lattice.*" Cottrell continues directly by proposing a first version of what will become the "pole mechanism": "*Since a new configuration is produced by twinning, the dislocations that cause it must be imperfect. While it is usually not difficult to discover a suitable imperfect dislocation to cause the required shear of neighbouring planes as it passes between them, the problem is to explain how twinning develops homogeneously through successive planes. The homogeneous shear required a twinning dislocation on every plane without exception, which seems unlikely, or the motion of a single dislocation from plane to plane in a regular manner. Cottrell and Bilby have recently suggested a mechanism, based on that of Frank and Read, whereby the latter process can occur in certain crystals containing dislocations. Fig. 51 [reported in Figure 10b] illustrated the mechanism. Here OA, OB, and OC, represent three dislocation lines and CDE is a slip or twinning plane. The dislocation OC ("the sweeping" dislocation) and its Burgers vector both lie in this plane (the "sweeping" plane) and the dislocation can rotate in the plane about the point O. If it is a unit dislocation and remains in the plane as it rotates a slip band is formed. The requirements for twinning (or a shear transformation) to occur are as follows: 1. The sweeping dislocation must be imperfect and produce the correct shear displacement on the sweeping plane. 2. Successive sweeping planes must be joined to form a helical surface.*" The Frank and Read model mentioned by Cottrell was explained a little earlier in his book; it is a former model of what is now known as "Frank-Read" source of dislocations. In this model, Frank and Read imagined a dislocation rotating around a point producing a new slip for each revolution, as explained by Cottrell in the figure 49 of his book [55] and reported in Figure 10ab. Cottrell made a parallel with the spirals formed at the surface of a growing crystal, even if it was clearly stated that this spiral dislocation results from growth and not from deformation. The Frank-Read model, applied by Cottrell to explain twinning was the ancestor of the "pole mechanism" model. It was followed by Cottrell and Bilby's model [72], refined by Sleeswyk [70][73], and later by Venables [74] (Figure 10c). Sleeswyk [70] also imagined how the twinning dislocations at the interface could dissociate and move, and he introduced the concept of "emissary dislocations" at the interface. General reviews on twinning mechanisms and twinning dislocations in metals were given in Ref. [2][3]. An updated review was recently published by Mahato *et al.* in section 4 of Ref. [75].

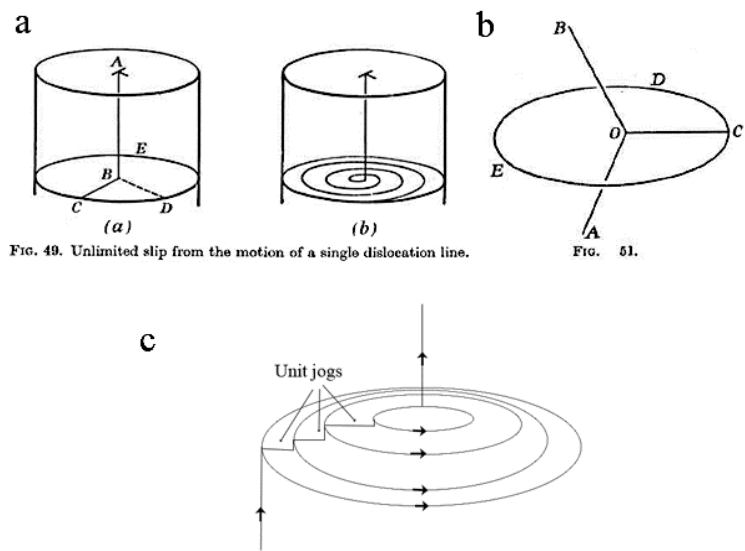

Figure 10. Schematic representation of twinning dislocations and pole mechanism. (a) A early model made by Frank and Read explains the dislocation "multiplication" in a crystal, from Cottrell's book [55]. (b) This model was used by Cottrell to explain how dislocations can build twins. (c) Scheme of the pole mechanism proposed by Venables, adapted from Ref. [74]

TEM observations fully confirmed the existence of dislocations, dislocation pile-ups, dislocation dissociations, climb etc., but to our best knowledge, the spiralling dislocations that could be expected from a pole mechanism were never put in evidence in metals. Spiralling/helical dislocations were observed by TEM in cast Al-Cu alloys [76] but they come from a vacancy collapse and not from stresses. What are often presented in TEM as "twinning dislocations" are dislocation pile-ups in front of microtwins [75][77]-[80], such as those shown in Figure 11. The origin of the dislocations is generally interpreted in term of dissociations of full dislocations associated with complex pole mechanisms, but, to the best of our knowledge, the initial spiralling dislocation source has never been shown.

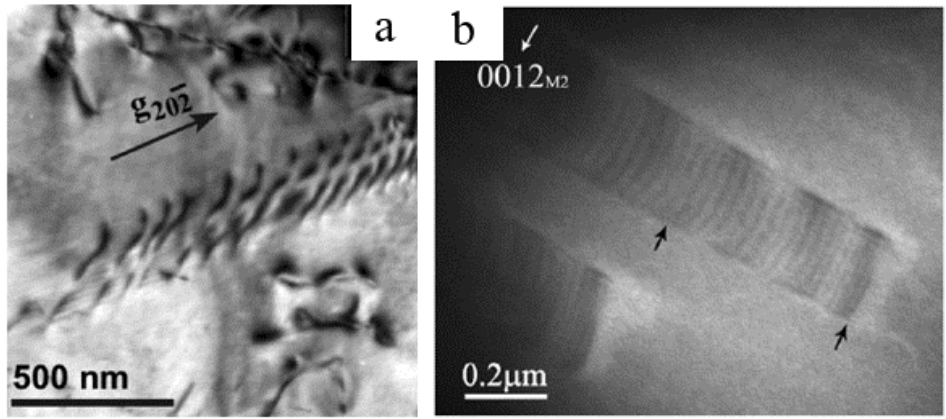

Figure 11.  TEM image of dislocation pileups in front of microtwins presented as "twinning dislocations". (a) in a Fe-Mn-Si-Al TRIP steel, from Ref. [75], (b) in a tetragonal Ni-Mn-Ga alloy, from Ref. [80].

The notion of "twinning dislocations" has been progressively enlarged to displacive phase transformations. For fcc-hcp and hcp-fcc transformations, the classical models imply the coordinate displacement of partial Shockley dislocations that are supposed to be created by a pole mechanism or similar complex mechanism [81]-[84]. For fcc-bcc transformations, we have seen in Figure 7 that Olson and Cohen [57][58] used partial dislocations in order to correct the initial hard-sphere Bogers and Burgers' model [54]. The use of dislocations as a fundamental part of the transformation models was generalized with the introduction of the concept of "disconnection", which became the core of the "topological model" (TM) developed by Hirth, Pond and co-workers [85]-[89]. A disconnection is a kind a dislocation located at the interface of two misoriented crystals that can be of same phase (for twinning) or different phases (for phase transformation); it is the assembly of a classical dislocation and a "step", both required to assure the compatibly at the interface. The dislocations/disconnections are assumed to be glissile because twinning and martensitic transformations are envisioned as the consequence of the propagation of the interface, and researchers came to introduce the concept of "glissile interfaces" [71] by extrapolating the concept of glissile dislocations. Despite its efficiency to describe locally at nanometer scale the structure of the interface between the parent matrix and the daughter or twin crystal, the TM cannot answer the most important questions: where the dislocations come from? How are they created? How can they move collectively and in coordinate way to build a new phase? How can they move at speeds close to the sound velocity? Why is twinning activated when the temperature decreases? Why hcp metals exhibit so many twinning modes with so few dislocation modes? Atomistic simulations [90][91] show that some dislocation loops can be created "ex-nihilo" (helped by the external stress field and dislocation pile-ups) and that these loops can be the very first nucleation step of twinning. These studies propose partial responses to the first two questions, but the others remain unsolved. Let us now explain now the way we envision deformation twinning and martensitic transformations.

## 5   Angular distortive transformations

### 5.1   *Twinning dislocations replaced by transformation waves and dislocations induced by twinning*

We think that the TEM images of the dislocation pile-ups in front the microtwins such as those of Figure 11 do not prove that the dislocations are the *cause* of twinning. They can be as well a *consequence* of twinning. Despite the fact that for the last 60 years nearly all the theoretical models and experimental observations are interpreted in term of twinning dislocations or disconnections, it is difficult to believe in the models of the pole mechanism and that dislocations could be created in a periodic sequential way, propagate at the speed of sound and produce a new crystallographic structure (twin or martensite). The high speed of twin/martensite propagation and the fact they are favoured by low temperature and high deformation rates should constitute important arguments against theories based on "twinning dislocations". Cottrell is right when he says that "*it is scarcely believable that the atom should all move simultaneously*" because it would imply that all the bonds break instantaneously, which would require very high energy for an instantaneous moment. Nevertheless, an important point should be considered: in non-quantum physics, the word "instantaneously" has no meaning. Any information cannot travel faster than a limited speed, i.e. the speed of light for electromagnetism or speed of sound for waves in materials (even if supersonic dislocations are reported for extreme conditions). Let us explain this general idea with the classical Ising model. Reversing a spin has an influence on the neighbouring spins, and when the temperature is close to the transition (Curie) temperature, i.e. when the local spin-spin interaction energy is not masked by the high temperature Brownian motion nor counterbalanced by the low temperature global magnetic field, this influence becomes predominant on the system, the magnetic susceptibility and the correlation length become infinite [92]. The effect of flipping a spin on the other spins is not instantaneous but travels at the light velocity; the spin-up and spin-down domains are thus not formed instantaneously; they appear under the effect of a solitary phase transition wave, i.e. a soliton. It can be imagined as a domino-cascade or as a seismic wave. This idea is not new, but appears only sporadically in literature, with long periods of oversight. Machlin and Cohen [43] envisioned martensitic transformation in Fe-30Ni alloys as a strain wave; Nishiyama wrote in his book that "*martensite transformation is like Shôgidaoshi (a Japansese word meaning "falling one after another in succession)*

*rather than like military notion*". Meyers [93] proposed an equation of martensite growth that takes into account the velocities of the longitudinal and shear elastic waves. Le Lann and Dubertret [59] developed their crystallographic model of fcc-bcc transformation by maintaining the direction $<110>_\gamma = <111>_\alpha$ invariant because they considered it as the wave vector of the transformation. With this way to envision displacive phase transformations, there is no need of dislocations, at least in a free single crystal. Barsch and Krumansl introduced in 1984 the concept of soliton creating boundaries without interface dislocation for ferroelastic transitions [94][95]. Solitary waves were also introduced by Flack the same year in a one-dimension shear model [96]. Unfortunately, this physical idea is often forgotten or ignored in metallurgy. A Russian team lead by Kashchenko and Chashchina [97] are developing the concept of dynamics and wave propagation of martensite in steels, but the physical details (l-waves, s-waves) are not easy to understand and would need experimental confirmations. We consider here that during deformation twinning or martensitic transformation in a free crystal, the atoms move following a solitary "phase transformation wave", i.e. a phase soliton, as geometrically shown in Figure 12.

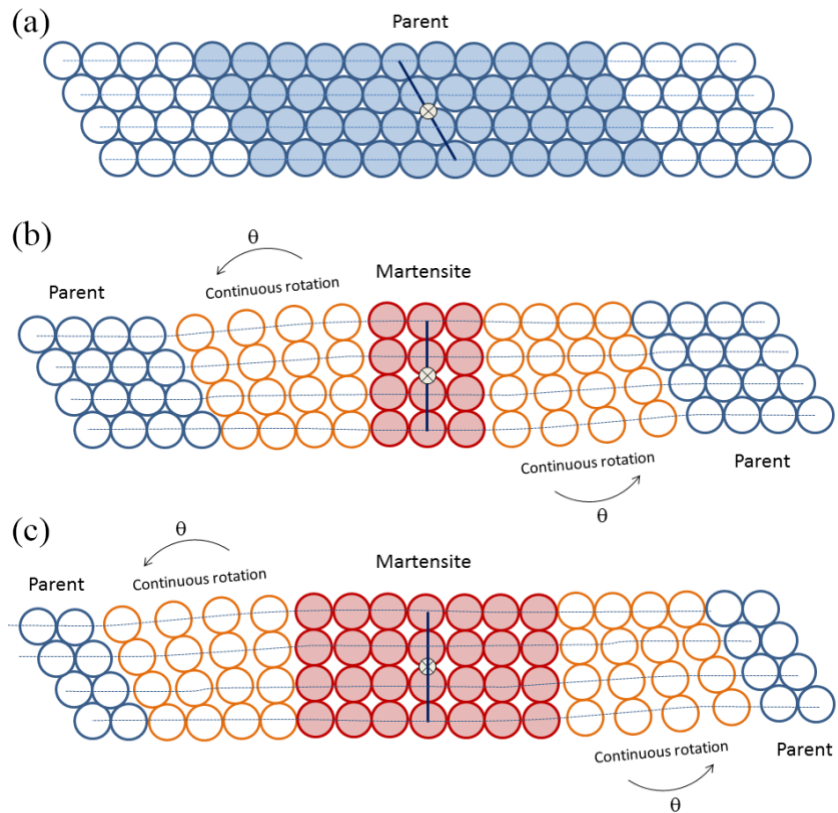

Figure 12. Schematic representation of a 2D hexagonal → square displacive transformation viewed as a "transformation wave". The square phase is stabilized by chemical reduction of energy or by mechanical work. (a) Initial hexagonal phase, (b) the atoms in a small area (in blue) collectively move to form the square martensite phase (in red). An accommodation zone (in orange) is created between the parent hexagonal and the martensite square phases. The accommodation takes here the form of a progressive 60°→ 90° rotation. (c) The transformation wave propagates at high velocity and continues forming the martensite phase; the previous accommodation zone becomes martensite, and a new accommodation zone is formed in front of the wave.

The displacement of each atom has an influence of the neighbouring atoms, such as a spin flip has an influence of the neighbouring spins in an Ising model. If the parent single crystal is free, the lattice distortion becomes a macroscopic shape distortion. This idea is quite new to us and should be developed; however, it seems possible that the parent/martensite accommodation area, in orange in Figure 12, is spread over a large area such that the distances between the neighboured atoms remain lower than the limit imposed by the critical yield strain, and the interface is accommodated purely elastically. There is no reason to believe that the dislocation-free model proposed by Barsch and Krumansl is limited to ferroelastics. We will come back on this point in the section 0.

For polycrystalline materials or non-free crystals, one must distinguish two cases depending on the martensite size. When the size of the martensite domain and its elastic accommodation zone are significantly lower than the grain size, the distortion can still be elastically accommodated inside the grain, at least during high speed transitory stages of the transformation process. The elastic zone is then quickly relaxed by formation of dislocation arrays in the surrounding matrix and by interfacial dislocations (disconnections). When the martensite grows and the size of elastic accommodation zone becomes comparable with the grain size, the incompatibilities must then be plastically accommodated, as schematized in Figure 13. The volume change associated with the phase transformation is averaged in all the grains and distributed in whole sample (it can be measured by dilatometry), but the deviatoric parts of the lattice distortion are "blocked" by the grain boundaries. The accommodation zone cannot be spread anymore and dislocations become necessary.

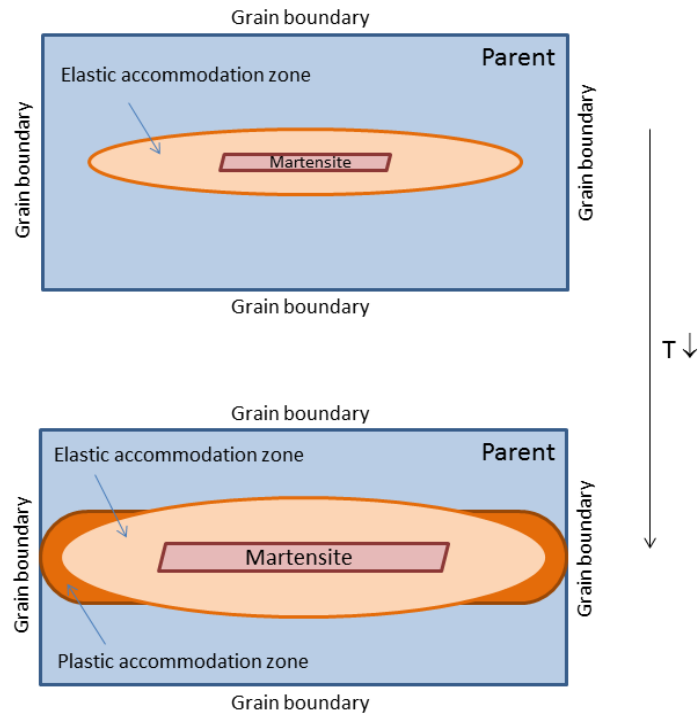

Figure 13. Elastic and plastic accommodation zone around the martensite product in a parent grain of a polycrystalline material. (a) During the transitory stages of the martensite propagation, the parent/martensite incompatibilities can be spread to be elastically accommodated, martensite can be formed as an acoustic wave, (b) when the martensite size becomes comparable to the grain size, the incompatibilities must be plastically accommodated.

## 5.2 *The concept of angular distortion*

According to this point of view, freed of the limitations imposed by the shear paradigm and its associated twinning dislocations, one can now imagine how the atoms move during displacive transformations. Of course, atoms are not hard-spheres, but in some metals (fcc, bcc, hcp with ideal packing ratio), the hard-sphere assumption is a good starting point [9]. Simple (or pure) shear are deformation modes at constant volume, which is well adapted for plasticity by dislocation, but not suited for displacive transformations in these metals. Indeed, it is known from Kepler in 1611 (the Kepler's conjecture became a theorem in 1998 thanks to Hales' demonstration) that the two dense-packed structures are only fcc and hcp, which means that any intermediate state between two fcc crystals, between two hcp crystals, or between a fcc and a hcp crystals should have a higher volume than that of fcc or hcp. Simple shear is not compatible with the hard-sphere assumption is one is looking for continuous deformation models. Let us consider again Figure 9a. When a simple shear is applied continuously to a crystal, the distance $h$ between the atomic layers is supposed to be constant, as if the crystal were constrained between two infinitely-rigid horizontal walls. Now, with a hard-sphere model, we consider the case in which the crystal is free to move in the direction normal to the atomic layers, as illustrated in Figure 9b. Of course, the atoms can slightly interpenetrate each other according to a reasonable elastic limit (that should be defined), but in first approximation it is assumed that the sphere size is only slightly elastically reduced before becoming "hard" spheres (infinitely rigid), such that the trajectories of the atoms are the same as those of hard spheres with a slightly lower diameter than the initial one, as shown in Figure 14.

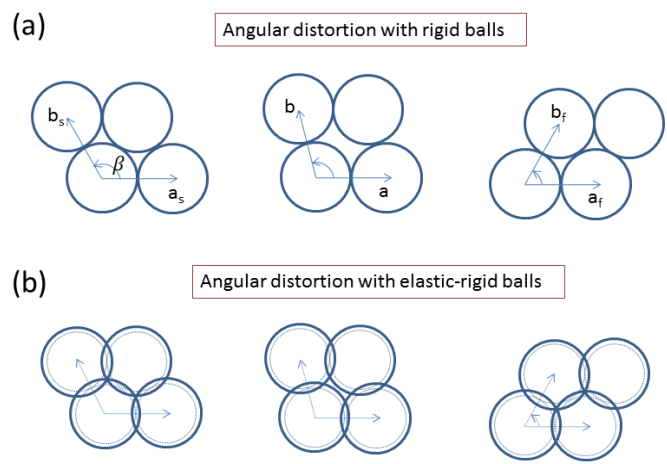

Figure 14.   Angular distortive model, (a) with hard-spheres, (b) with elastic-hard spheres.

Thus, in first approximation, the atomic displacements are not essentially modified by the elasticity of the spheres; the trajectories remain close to that of "rolling balls". Now, if the lattice noted by the vectors **a**, **b** in Figure 14 is considered, the distortion is not anymore a simple shear along **a**; it is a distortion in which the length of the vector **b** remains constant; it is and remains the atom diameter and the angle between the vectors **a** and **b**, noted here β, continuously changes. This angle becomes the unique parameter of the lattice distortion; in physics we would say that it is the *order parameter* of the transition. The trajectories of the atoms become naturally arcs of circle, and the lattice distortion is not anymore a shear strain but an angular distortion. With such a simple approach, the shape of the crystal is given by the same distortion as that of the atomic displacements. Of course in polycrystalline materials, a grain is not free to be arbitrarily deformed due to the surrounding grains; and complex accommodation modes by variant coupling and by dislocations are required.

*5.3  Accommodation dislocations and disclinations*

The accommodation dislocations are not randomly distributed; they are such that their strain field compensates the deviatoric parts of the angular distortion; they constitute the plastic trace of the mechanism. To our opinion, these dislocations are those shown in TEM images of Figure 11. The elastic field associated with the dislocation arrays should be equal to the back-strains between the martensite product and the parent crystal. These dislocation arrays are very probably at the origin of the continuous features observed in the EBSD pole figures of martensitic steels (see Figure 19 in the next section). It can be expected that for soft parent / hard daughter phases, the distortion is accommodated in the parent phase, and for hard parent / soft daughter phases, the distortion is accommodated in the daughter phase, and one can expect an equi-proportion for deformation twinning. According to this model, the mesoscopic strain field should be dictated by the phase transformation distortion, and the exact and detailed knowledge of the structure of the dislocations adopted by the material to produce this strain field is not required to explain it. As the lattice distortion is now modelled by an angular distortion, the dislocation arrays should be formed in order to compensate the angular deficit (or excess) implied by the distortion. The most appropriate plastic mode for such types of accommodation implies the concept of "disclination". It was first introduced in 1907 by Volterra [98] who considered two types of defects in a periodic solid: the rotational dislocations (disclinations) and the translational dislocations (simply referred as dislocations nowadays). The disclination strength is given by an axial vector *w*, called Frank vector, encoding the rotation needed to close the system, in a similar way that the dislocation strength is given by its Burgers vector **b** encoding the translation needed to close the Burgers circuit. If dislocations constitute a fundamental part of metallurgy, disclinations are less known. The theory has been developed by Romanov [99], and by Kleman and Friedel [100], but the applications are mainly limited to highly deformed metals [101]. To our knowledge, only Müllner and co-workers use disclinations to calculate the strain field and strain energy of hierarchically twinned structures formed by deformation twinning [102] and by martensitic transformation in Ni-Mn-Ga Heusler alloys [103]. As their models are based on the classical concepts of shears and twinning dislocations, that could of interest to see whether they could be substituted by the concept of angular distortion. Indeed, disclinations are accommodation modes of positive or negative angular misfits, as shown in Figure 15; they are the most appropriate tools to complement the angular distortion related to phase transitions in metals.

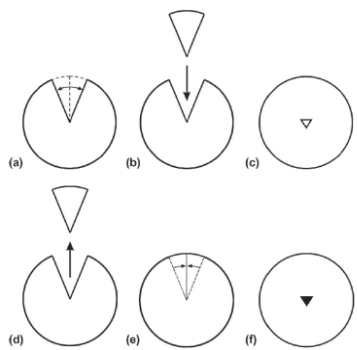

Figure 15. Volterra cut for a negative wedge disclination where material is inserted (a–c), and alternative Volterra cut for a positive wedge disclination where material is removed (d–f). Figure and caption from Müllner and King [103].

In this approach, martensitic transformations and deformation twinning are not the result of a *cause* that would be the sweep of "twinning dislocations" or of the propagation of a "glissile interface", but they directly result from a lattice distortion driven: (a) for martensite, by the difference of chemical energy between a parent phase and its daughter phase, and (b) for twinning, by the difference of mechanical energy given by the work of the applied stress along the distortion path. The interface and the accommodating dislocations in its surrounding then come as the *consequences* of the distortion. Large volumes of parent phase are suddenly transformed into daughter phase (at the acoustic wave velocity), and the parent-daughter interface *created by* the distortion is made of glissile and sessile dislocations/disconnections. At mesoscale these dislocations induce rotation gradients (disclinations), such as those schematically represented in orange in the hexagon-square transformation of Figure 12. This way to envision deformation twinning and martensite transformation completely reverses the usual paradigm that gives a central driving role to the dislocations in the transformation mechanism. There is no need here to imagine a hypothetical complex pole mechanism because now the dislocations are created by the twinning or by the martensitic transformation; and there no need to impose glissile properties to the interface. That is true that plasticity is required to accommodate the lattice distortion in polycrystalline metals; but in no way it means that dislocations/disconnections are the cause or can explain the distortion.

# 6   Angular distortion vs PTMC illustrated with a simple example

The concept of angular distortion is so simple and natural that it can be directly applied to martensitic transformations, without requiring the four (or more) matrices of the PTMC. Let us use again the example of a 2D hexagonal-square phase transformation of Figure 12. The classical PTMC decomposes this transformation into two distinct paths. The first path is the matrix product **R**.**U**, which says that first a (Bain) stretch **U** is applied, and then a rotation **R** must adjust the stretch in order to maintain an invariant line (here the **x**-axis). The notation **U** is now preferred to **B** (Bain) to avoid confusion with our notation of the bases. The second path is $\mathbf{P}_1.\mathbf{P}_2$ where $\mathbf{P}_1$ is an IPS and $\mathbf{P}_2$ a simple shear. Here, $\mathbf{P}_2$ is reduced to identity because only $\mathbf{P}_1$ is required to get the final square lattice. However, even by keeping only $\mathbf{P}_1$, one must admit that the case is not direct because there are two parameters in $\mathbf{P}_1$, i.e. the simple shear $\mathbf{P}_s$ and the dilatation $\mathbf{P}_\delta$ (see Figure 5a). The directions of the vectors **s** and **δ** are known, but nothing is said on how the norms of these two vectors evolve during the transformation. The PTMC treats the two paths **R**.**U** and $\mathbf{P}_s\mathbf{P}_\delta$ as two independent steps, as shown in Figure 16. One can understand with this simple example why PTMC is still phenomenological and continues to be mute on the atomic displacements during the transformation, even seventy years after its birth. Do the atoms follow the trajectory **R**.**U** or the trajectory $\mathbf{P}_s\mathbf{P}_\delta$? Only an atom with a gift of ubiquity or behaving as a quantum particle could follow both paths at the same time! The only way to tackle this issue is to impose rules between the parameters involved in **R**, **U**, $\mathbf{P}_s$, and $\mathbf{P}_\delta$, such that an equality of the two paths is obtained *during* the transformation for all the intermediate states. These calculations will be given a little later in the text.

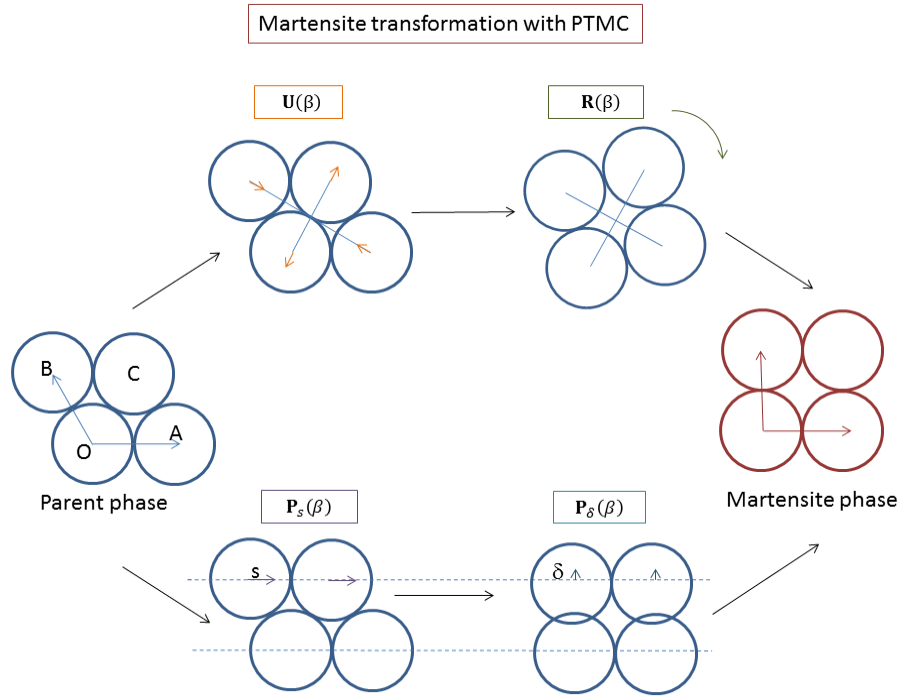

Figure 16. Hexagonal-square transformation with PTMC.

Let us first consider the natural alternative to PTMC. The trajectories of the atoms can be determined easily by using an angular distortion matrix. The only thing to do is to write the way that the lattice is distorted by the atomic displacements. In that aim, we use the crystallographic basis $\mathbf{B}_c = (\mathbf{a}, \mathbf{b})$ and the orthonormal basis $\mathbf{B}_0 = (\mathbf{x}, \mathbf{y})$. During the distortion the vector $\mathbf{a}$ remains invariant, and the vector $\mathbf{b}$ is rotated. Its image is noted $\mathbf{b}$'. The angle between $\mathbf{a}$ and $\mathbf{b}$' is β. The angle β decreases during the transformation; in the starting state it is $\beta_s = \frac{2\pi}{3}$ and in the final state $\beta_f = \frac{\pi}{2}$, as shown in Figure 17.

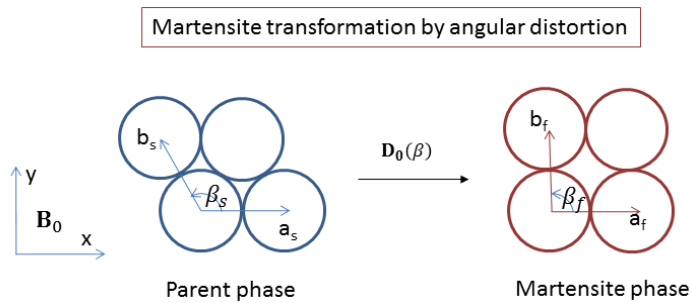

Figure 17. Hexagonal-square transformation with angular distortion

The vectors $(\mathbf{a}, \mathbf{b}')$ define a crystallographic basis $\mathbf{B}_c(\beta)$ expressed in $\mathbf{B}_0$ by

$$\mathbf{B}_c(\beta) = [\mathbf{B}_0 \to \mathbf{B}_c(\beta)] = \begin{pmatrix} 1 & Cos(\beta) \\ 0 & Sin(\beta) \end{pmatrix} \qquad (2)$$

The matrix of lattice distortion $\mathbf{D}_c(\beta)$ between the starting basis $\mathbf{B}_c = \mathbf{B}_c(\beta_s)$ and any distorted basis $\mathbf{B}_c(\beta)$, expressed in the crystallographic basis $\mathbf{B}_c$, is simply $\mathbf{D}_c(\beta) = [\mathbf{B}_c \to \mathbf{B}_0][\mathbf{B}_0 \to \mathbf{B}_c(\beta)] = \mathbf{B}_c^{-1}\mathbf{B}_c(\beta)$. This matrix can now be expressed in the orthonormal basis $\mathbf{B}_0$ by using the usual coordinate change equation; i.e. $\mathbf{D}_0(\beta) = [\mathbf{B}_0 \to \mathbf{B}_c].\mathbf{D}_c(\beta).[\mathbf{B}_c \to \mathbf{B}_0]$. It becomes

$$\mathbf{D}_0(\beta) = \mathbf{B}_c(\beta)\, \mathbf{B}_c^{-1} \qquad (3)$$

Equation (3) is very general and can be applied to any displacive transformation. In the hexagon-square example, combined with equation (2), it directly gives

$$\mathbf{D}_0(\beta) = \begin{pmatrix} 1 & \dfrac{1+2Cos(\beta)}{\sqrt{3}} \\ 0 & \dfrac{2Sin(\beta)}{\sqrt{3}} \end{pmatrix} \qquad (4)$$

The matrix of complete distortion $\mathbf{D}_0$ is obtained for $\beta = \beta_f$

$$\mathbf{D}_0 = \mathbf{D}_0(\beta_f) = \begin{pmatrix} 1 & \dfrac{1}{\sqrt{3}} \\ 0 & \dfrac{2}{\sqrt{3}} \end{pmatrix} \qquad (5)$$

The volume (here surface) change during the distortion is given by the determinant of the distortion matrix:

$$\frac{\mathcal{V}'}{\mathcal{V}}(\beta_f) = Det\, \mathbf{D}_0(\beta) = \frac{2Sin(\beta)}{\sqrt{3}} \qquad (6)$$

We take the opportunity given by formula (6) to mention an important point. It was assumed by Weschsler, Liebermann and Read in their seminal paper [34] that gave birth to PTMC, that the average distortion matrix $\mathbf{D}$ resulting from two distortion matrices $\mathbf{D}_1$ and $\mathbf{D}_2$ of variants in proportions $\lambda$ and $1 - \lambda$, with $0 \leq \lambda \leq 1$ is

$$\mathbf{D} = \lambda\mathbf{D}_1 + (1-\lambda)\mathbf{D}_2 \tag{7}$$

This formula works only in the special condition called "kinematical compatibility"; in the general case, the volume is not conserved, i.e. $\det(\mathbf{D}) \neq \lambda.\det(\mathbf{D}_1) + (1-\lambda).\det(\mathbf{D}_2)$. We note that, to our best knowledge, Bowles and Mackenzie never used this formula and they always composed matrices by multiplication; that is why we are not convinced when it is said that WLR and BM forms of the PTMC are the same; they differ at least by the way the distortion matrices are averaged. This is probably one of the reasons that both PTMC versions have continued their own developments without intermixing and nearly ignoring each other.

The approach based on hard-spheres is very effective in comparison with the PTMC because it gives in one simple step the continuous analytical expression of the lattice distortion that is compatible with the atom size; i.e. realistic from an energetic point of view. The PTMC can determine the distortion matrix only once the transformation is finished, i.e. equation (5). Of course, one could try developing a complex continuous infinitesimal version of the PTMC as in Ref. [104], but the method is artificial and needlessly heavy. For example, let us consider what it would be with the simple hexagon-square example. We note O, A, B, C, the nodes of the lattice with the vectors $\mathbf{OA} = [1,0]$, $\mathbf{OB}(\beta) = [Cos(\beta), Sin(\beta)]$, $\mathbf{OC}(\beta) = \mathbf{OA} + \mathbf{OB}(\beta)$ as shown in Figure 16. The stretch component $\mathbf{U}_L$ in a local orthonormal basis positioned along the diagonal $\mathbf{OC}$ and $\mathbf{AB}$ is

$$\mathbf{U}_L(\beta) = \begin{pmatrix} \frac{\|\mathbf{OC}(\beta)\|}{\|\mathbf{OC}(\beta_s)\|} & 0 \\ 0 & \frac{\|\mathbf{AB}(\beta)\|}{\|\mathbf{AB}(\beta_s)\|} \end{pmatrix} = \begin{pmatrix} 2Cos(\frac{\beta}{2}) & 0 \\ 0 & \frac{2}{\sqrt{3}}Sin(\frac{\beta}{2}) \end{pmatrix} \tag{8}$$

This local basis is obtained from the reference basis $\mathbf{B}_0$ by a rotation of angle $\frac{2\pi}{6}$

$$\mathbf{U}_0(\beta) = \mathbf{R}\left(\frac{2\pi}{6}\right) \cdot \mathbf{U}_L(\beta) \cdot \mathbf{R}\left(\frac{-2\pi}{6}\right) = \frac{1}{2}\begin{pmatrix} Cos(\frac{\beta}{2}) + \sqrt{3}Sin(\frac{\beta}{2}) & \sqrt{3}Cos(\frac{\beta}{2}) - Sin(\frac{\beta}{2}) \\ \sqrt{3}Cos(\frac{\beta}{2}) - Sin(\frac{\beta}{2}) & 3Cos(\frac{\beta}{2}) + \frac{1}{\sqrt{3}}Sin(\frac{\beta}{2}) \end{pmatrix} \quad (9)$$

The compensating rotation matrix $\mathbf{R}$ is a rotation of angle $\frac{1}{2}\left(\beta - \frac{2\pi}{3}\right)$

$$\mathbf{R}(\beta) = \frac{1}{2}\begin{pmatrix} Cos(\frac{\beta}{2}) + \sqrt{3}Sin(\frac{\beta}{2}) & \sqrt{3}Cos(\frac{\beta}{2}) - Sin(\frac{\beta}{2}) \\ -\sqrt{3}Cos\left(\frac{\beta}{2}\right) + Sin(\frac{\beta}{2}) & Cos(\frac{\beta}{2}) + \sqrt{3}Sin(\frac{\beta}{2}) \end{pmatrix} \quad (10)$$

The other path of the PTMC is made of the shear displacement along the **x**-axis $s(\beta) = Cos(\beta) - Cos(\frac{2\pi}{3})$ of the point B located at a distance $h$ from the shear axis. For a simple shear, $h$ is constant and equal to $\sqrt{3}/2$. Thus

$$\mathbf{P}_s(\beta) = \begin{pmatrix} 1 & \frac{s(\beta)}{h} \\ 0 & 1 \end{pmatrix} = \begin{pmatrix} 1 & \frac{1 + 2Cos(\beta)}{\sqrt{3}} \\ 0 & 1 \end{pmatrix} \quad (11)$$

The second part of the path is an extension along the **y**-axis that dilates the distance $h$ and makes it becomes $h(\beta) = h + Sin(\beta) - Sin(\frac{2\pi}{3})$. The dilatation matrix is thus

$$\mathbf{P}_\delta(\beta) = \begin{pmatrix} 1 & 0 \\ 0 & \frac{h(\beta)}{h} \end{pmatrix} = \begin{pmatrix} 1 & 0 \\ 0 & \frac{2Sin(\beta)}{\sqrt{3}} \end{pmatrix} \quad (12)$$

The first path is continuous and its analytical expression is given by the product $\mathbf{R}(\beta)\,\mathbf{U}_0(\beta)$ with the matrices given in equations (9) and (10). The second path is continuous and its analytical expression is given by the product $\mathbf{P}_\delta(\beta)\mathbf{P}_s(\beta)$ with the matrices given in equations (11) and (12). The atoms do not need ubiquity anymore because the two paths are actually the same: it is the path corresponding to the angular distortion given by equation (4). We let the reader make his own opinion on the physical relevance and simplicity of Figure 17 and equation (4), and that of Figure 16 and the set of four equations (9), (10), (11) and (12).

# 7 The angular distortive model for fcc-bcc martensitic transformation

## 7.1 The intriguing continuities in the pole figures

When fully transformed, many steels and titanium, zirconium alloys are only constituted of the daughter phase, without sufficient amount of retained parent phase to know the sizes and orientations of the prior parent grains that had existed at high temperatures before the transformation. This apparently lost information is however crucial to get a better understanding of the fatigue, impact and corrosion properties of steels because the prior parent grain boundaries are preferential location sites of impurity segregation. A method to reconstruct the prior parent beta grains in titanium alloys from EBSD maps was proposed by Gey and Humbert [105], but at that time it was based on misorientations between grains and was not applicable to steels due to the highest number of symmetries and variants. New constrains had to be found to reduce the influence of the tolerance angle. These constrains were found in the algebraic structure of the variants. The orientational variants and their misorientations form a groupoid structure [106], and the theoretical groupoid composition table can be used to automatically reconstruct the parent grains from EBSD data [107][108]. An example of reconstruction is given in Figure 18.

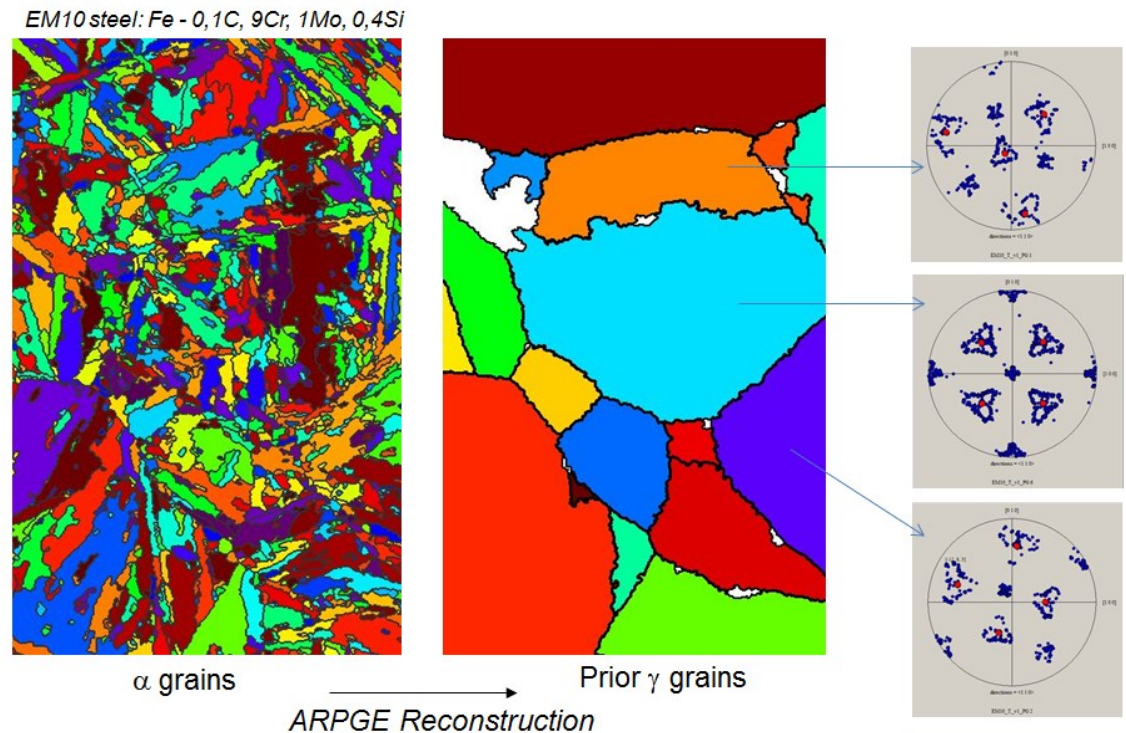

Figure 18. Reconstruction of the parent grains from EBSD maps. (a) Initial EBSD map of a fully martensitic steel. (b) Prior parent austenitic grains reconstructed with the software ARPGE. The orientations of the austenitic grains and those of the martensitic grains they contain are plotted in the pole figures, with the calculated $<111>_\gamma$ directions in red and the experimental $<110>_\alpha$ directions in blue.

The reconstruction method was applied to many alloys, and something was striking: for low-alloyed steels, continuous features were observed in the pole figures of the martensitic grains contained in the prior parent grains. That was surprising because only a discrete distribution of orientations was expected from the 24 KS variants. As these features were observed in many different steels, and were also reported by X-ray diffraction and EBSD in Fe-Ni meteorites [109][110], we made the hypothesis that they were the plastic trace of the distortion mechanism itself. The features could be simulated "phenomenologically", by applying two continuous rotations **A** and **B** to the 24 KS variants, one around the normal to the common dense plane, and one around the common dense direction, with rotation angles continuously varying in the range [0-10°].

## 7.2 A two-step model developed to explain the continuous features

What is the physical meaning of these rotations? PTMC tells nothing about them and they are not correlated to the usual plastic deformation modes of bcc or fcc structure. We found that the rotations **A** and **B** could be associated with the fcc-hcp and hcp-bcc transformations, respectively; thus, a two-step model was established implying the existence of an intermediate fleeting hcp phase, even in alloys without retained ε phase [111]. It was later realized that this two-step model has some similarities with the initial KSN model shown in Figure 3, the first shear in the KSN model being replaced by the fcc-hcp step made by sequential and coordinates movements of partial Shockley dislocations supposed at that time to be originated by a pole mechanism [112]. However, further studies could not confirm the two-step model. No trace of the intermediate hcp phase could be found in the microstructures of the martensitic steels; and ultrafast in-situ X-ray diffraction experiments in two synchrotrons (ESRF and Soleil) could not put in evidence the hypothetical fleeting hcp phase [113]. In addition, the pole mechanism appeared more and more doubtful; and many questions were remaining unanswered. All these "negative" results imposed to consider other models of fcc-bcc transformation.

## 7.3 One-step hard-sphere model with Pitsch OR

Would it be possible to establish a mechanistic model, as simple as possible, without combining shears and without dislocations in its core, a model in which the atoms would move collectively in one step? The important thing to start such a model is the orientation relationship between the fcc and bcc phases. It was noticed that the KS, NW and Pitsch ORs were located in the continuous features observed in the EBSD pole figures, and we made the assumption that all these ORs actually result from a unique OR, a "natural" spontaneous OR of the transformation for stress-free samples. What could be this "natural" OR? The Bain OR obtained by lattice stretching maintains the highest number of common symmetries between the parent fcc and the daughter bcc, and thus could a good candidate if one imagine that all the chemical bonds "break" instantaneously at the same time. However, to our best of knowledge, a criterion that predicts the OR only from symmetry considerations does not exist, and is probably not relevant. Indeed, the transformation does not occur instantaneously; it is envisioned here as a wave. The existence of common dense directions and/or planes obtained for special ORs, as it is the case for KS OR, is probably favourable to the wave propagation (see section 5.1 and 10.2), whereas stretch distortions in general does not maintain the parallelism of dense directions. Besides the theoretical considerations, the ORs experimentally determined are more than 10° far away from the Bain OR. Could the Pitsch OR be the good candidate? This OR is less known than KS or NW; it was observed by Pitsch in 1959 after cooling a TEM lamella of iron nitrogen alloy [114], which means that it was formed without the surrounding stresses that exist in bulk samples. Maybe Pitsch OR was missing link to build a theory. The Pitsch OR is at the midway between two low-misoriented KS variants, in the same way that the NW OR is at the midway between the two other low-misoriented KS variants. The simulations of the pole figures with 24 KS variants and the rotations **A** and **B** with angles limited to 0-5° are shown in Figure 19a. The distortion matrix associated with the Pitsch distortion was calculated, and the rotations **A** and **B** were qualitatively explained by the distortion, even if the simulations were based on KS OR and not with Pitsch OR. For the calculations, the lattice parameters of the fcc and bcc phases had to be chosen, and those of hard-sphere packing with a constant atomic radius for the iron atoms were taken as a first approximation. Even if a hard-sphere model is not perfect because the size of Fe atoms are not exactly the same in the fcc and bcc crystals (4% of difference), it is a good starting point to build an atomistic model. The Pitsch distortion matrix could be diagonalized and the strains were surprisingly lower than those of Bain, but the basis of diagonalization is not orthonormal [115]. One year later, a method that shows in the EBSD maps the regions oriented according to Pitsch, KS, NW ORs with a red, green, blue (RGB) colour coding [116] was developed and applied to various steels; it confirmed that each bcc martensitic grain exhibits internal gradients between these orientations, as shown in Figure 19b. However, the Pitsch distortion model would make expect gradients of type NW/KS/Pitsch/KS/NW, but only a Pitsch/KS/NW gradients could be observed in the RGB-coloured EBSD maps of martensitic steels.

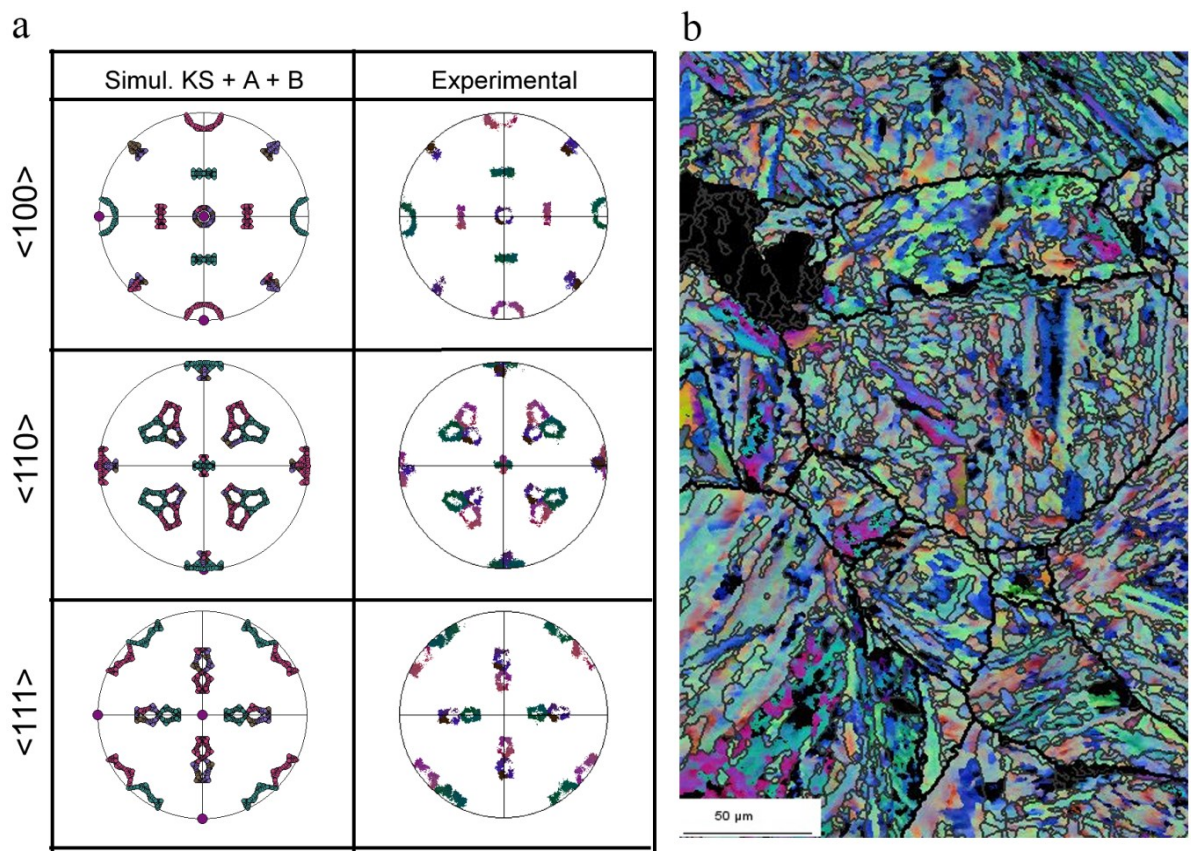

Figure 19. Continuous gradients of orientations inside bcc martensitic grains. (a) Simulations of the pole figures made with 24 KS variants and two rotations **A** and **B** with angular range 0-5° (left column), compared to the experiments (right column), from Ref. [115]. (b) RGB colouring of the Pitsch, KS and NW ORs inside the martensitic grains of the EMT10 steel shown in Figure 18. From Ref. [116].

Could KS be actually the "natural" OR? The KS OR was also reported in quenched TEM lamella; KS OR gives the best simulations of the continuous features in the pole figures; it is the only OR allowing at the same time the parallelism of the dense directions and the parallelism of the dense planes of the fcc and bcc phases. What if, contrarily what could be thought, nobody had tried to build a one-step model of transformation based on the KS OR, i.e. a continuous version of the KSN model shown in Figure 3? It was thus decided to develop such a model.

### 7.4 *One-step hard-sphere model with KS OR*

A one-step continuous model of fcc-bcc lattice distortion leading to a KS OR was built similarly as the one made with Pitsch OR in Ref. [115]. It also uses the hard-sphere assumption to *explicitly* calculate the trajectories of the atoms *during* the transformation, i.e. to obtain a *continuous* atomistic model of the fcc-bcc transformation [117]. The main idea can be briefly explained as follows. During the Bain distortion the fcc lattice is contracted along a $<100>_\gamma$ direction; the two other $<100>$ directions are expanded; when the transformation is complete half a bcc crystal is formed, as shown in Figure 20.

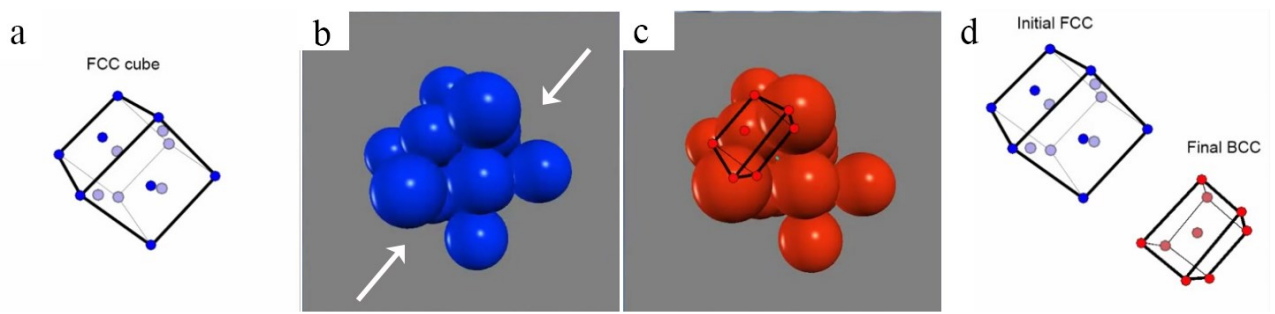

Figure 20. Bain distortion with hard-spheres. (a) Initial fcc lattice. (b) Same crystal with hard spheres. The direction of the contraction axis along a $<100>$ direction is marked by the white arrows. (c) Final distorted crystal, which is half a bcc crystal

cut on a (110) plane. (d) Initial fcc crystal and final bcc crystal after Bain distortion. Movie at http://lmtm.epfl.ch/research

For the distortion that leads to KS OR, simply called KS distortion, the trajectories of the atoms are slightly different from those obtained with Bain OR. Let us consider Figure 21. The KS OR implies the parallelisms of a dense plane (POK) = $\{111\}_\gamma$ // $\{110\}_\alpha$ and a dense direction **PO** = ½ $<110>_\gamma$ = ½ $<111>_\alpha$. The KS distortion is simply obtained as follows. The atom is P is fixed. The angle β made by the dense directions (**PO**, **PK**) opens from 60° to 70.5° while maintaining the direction **PO** invariant; the atoms O and K loose contact, and the atom M in the upper layer located above O and K move ("roll") on them to go closer to the atom P. When the transformation is complete half a bcc crystal is formed, as for Bain distortion, but now directly in KS OR.

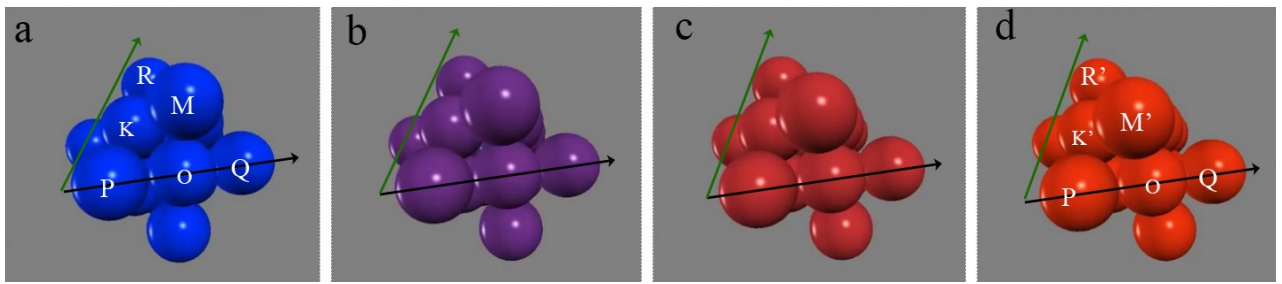

Figure 21. KS distortion explained with hard-spheres. The continuity is given by a unique parameter, the angle, β = (**PO**, **PK**) that changes from 60° (bcc) to 70.5° (fcc). Movie at http://lmtm.epfl.ch/research

For the KS OR expressed by $(\bar{1}11)_\gamma$ // $(\bar{1}10)_\alpha$ & $[110]_\gamma = [111]_\alpha$ the continuous form of the distortion matrix is

$$\mathbf{D}^{KS}(\beta) = \begin{pmatrix} \frac{1}{\sqrt{6}}\sqrt{\frac{1-X}{1+X}}\left(\sqrt{2}X+2\sqrt{X}\right)+X & 1-X-\frac{1}{\sqrt{6}}\sqrt{\frac{1-X}{1+X}}\left(\sqrt{2}X+2\sqrt{X}\right) & \sqrt{\frac{1-X^2}{3}}-\frac{1}{\sqrt{6}}\sqrt{\frac{1-X}{1+X}}\left(\sqrt{2}X+2\sqrt{X}\right) \\ -\frac{1}{\sqrt{6}}\sqrt{\frac{1-X}{1+X}}\left(\sqrt{2}X+2\sqrt{X}\right)+X & 1-X+\frac{1}{\sqrt{6}}\sqrt{\frac{1-X}{1+X}}\left(\sqrt{2}X+2\sqrt{X}\right) & -\sqrt{\frac{1-X^2}{3}}+\frac{1}{\sqrt{6}}\sqrt{\frac{1-X}{1+X}}\left(\sqrt{2}X+2\sqrt{X}\right) \\ \frac{2}{\sqrt{6}}\sqrt{\frac{1-X}{1+X}}\left(\sqrt{2}X-\sqrt{X}\right) & -\frac{2}{\sqrt{6}}\sqrt{\frac{1-X}{1+X}}\left(\sqrt{2}X-\sqrt{X}\right) & 2\sqrt{\frac{1-X^2}{3}}-\frac{2}{\sqrt{6}}\sqrt{\frac{1-X}{1+X}}\left(\sqrt{2}X-\sqrt{X}\right) \end{pmatrix} \quad (13)$$

where x = Cos($\beta$). It can be checked that this distortion matrix is the identity matrix for the starting state $\beta = 60°$ ($X = 1/2$). When the transformation is complete, $\beta = 70.5°$ ($X = 1/3$), it becomes:

$$\mathbf{D}^{KS} = \begin{pmatrix} \frac{2}{3}+\frac{\sqrt{6}}{18} & \frac{\sqrt{6}}{18}+\frac{1}{3} & \frac{\sqrt{6}}{6}-\frac{1}{3} \\ -\frac{\sqrt{6}}{18} & 1+\frac{\sqrt{6}}{18} & \frac{1}{3}-\frac{\sqrt{6}}{6} \\ \frac{\sqrt{6}}{9}-\frac{1}{3} & \frac{1}{3}-\frac{\sqrt{6}}{9} & \frac{1}{3}+\frac{\sqrt{6}}{3} \end{pmatrix} \quad (14)$$

If one prefers using the equivalent KS OR defined by $(111)_\gamma$ // $(110)_\alpha$ & $[\bar{1}10]_\gamma$ = $[\bar{1}11]_\alpha$, the distortion matrix becomes by a change of reference frame

$$\mathbf{D}^{KS} = \begin{pmatrix} \frac{2}{3}+\frac{\sqrt{6}}{18} & \frac{\sqrt{6}}{18}+\frac{1}{3} & \frac{\sqrt{6}}{6}-\frac{1}{3} \\ -\frac{\sqrt{6}}{18} & 1+\frac{\sqrt{6}}{18} & \frac{1}{3}-\frac{\sqrt{6}}{6} \\ \frac{\sqrt{6}}{9}-\frac{1}{3} & \frac{1}{3}-\frac{\sqrt{6}}{9} & \frac{1}{3}+\frac{\sqrt{6}}{3} \end{pmatrix} \quad (15)$$

Even if the trajectories of the atoms are different, the calculations prove that the KS distortion matrix $\mathbf{D}^{KS}$ is linked to the Bain distortion $\mathbf{B}$ by a polar decomposition $\mathbf{D}^{KS} = \mathbf{RB}$. All the calculations used in the simulations are analytical; they result from simple geometrical considerations, and not from molecular dynamics (MD). That is true that MD simulations help the understanding of displacive transformations, but it can doubted about the capacity of brute force MD to simulate the coordinated "wave-like" displacements of atoms, even if it was proved that the phonon dispersion curves can be obtained by MD simulations [118]. Simulating a "phase transition wave" seems more complex than an acoustic wave, and it is probable that a better understanding of the transformations is required to add new crystallographic constrains to MD. The first MD simulations proposed by different groups for the last decade are going in the good direction [119]-[122]. For example, in 2008 Sinclair and Hoagland could simulate by MD the formation of bcc martensite in Pitsch OR at the intersection of stacking faults [119]. Sandoval, Urbassek and Entel showed in 2009 [120] that a pure Bain path would necessitate compressive stresses five times higher than with a NW path (see KSN model described earlier). Wang *et al.* could reproduce in 2014 the effects of the carbon content and cooling rates on the Ms temperature [121]. It is sure that MD will be an indispensable tool in the future to compare the "realisms" of the different crystallographic models, so, efforts will be done in the future to combine the angular distortion with MD simulations.

PTMC has for output the habit planes of martensite, but as already discussed, PTMC made post-dictions, not pre-dictions. Can a simple criterion be introduced in order to explain the habit planes only from the unique value of the distortion matrix? As there is no common plane between the fcc and bcc structures (even if this has never been formally proved), the IPS condition cannot be used. PTMC made the choice to use it anyway by combining it with other shears. The other possibility consists in finding a condition that is less restrictive than IPS. A criterion in which the habit plane is only untilted is physically relevant because it avoids accumulating defects on long distances. The habit plane should be an eigenvector of the distortion matrix calculated in the reciprocal space, i.e. of the inverse of the transpose of $\mathbf{D}^{KS}$. The two only untilted planes determined with the matrix (15) are $(111)_\gamma$ and $(11\sqrt{6})_\gamma$; the latter plane is only at 0.5° away from the expected $(225)_\gamma$ habit plane. The calculation is direct, without any fitting parameter, without knowing the details of the accommodation mechanisms in this plane. While writing the paper [117] it was realized that Jaswon and Wheeler [33] already postdicted the $(225)_\gamma$ habit planes by calculating the distortion matrix related to KS OR and by using a "untilted plane" criterion; so that is true that part of the work [117] is actually an independent rediscovery of Jaswon and Wheeler's work. One should keep in mind however that Jaswon and Wheeler's study is indeed rarely cited and when it is, it is for the use of the Bain correspondence matrix and for introducing matrix algebra, and not for their calculation of the KS distortion nor for the "unrotated plane" criterion. As explained previously, Bowles and Barrett discarded Jaswon and Wheeler's model very early in 1951 [49], that's why it is rarely mentioned in PTMC books; for example it is not cited by Bhadeshia in his excellent and didactic book on martensitic transformations [39]. Bowles, Dunne and many other metallurgists did not try to come back to the Jaswon and Wheeler's model despite the huge difficulties they encountered to explain the $(225)_\gamma$ habit planes, as recalled by Dunne [47]: *"In contrast [to Mackenzie] John Bowles retained his intense concentration on martensite crystallography until he retired from the University of New South Wales in 1983. Bowles' focus, moreover, was predominantly on the seemingly intractable $\{225\}_F$ problem and six of his nine PhD students works on aspects of this problem: Peter McDougall, Allan Morton, Druce Dunne, Don Dautovich, Peter Krauklis and Barry Muddle"*. Recently, an anonymous reviewer claimed that the model of the KS distortion proposed in Ref. [117] was an *"outright copy of Jaswon and Wheeler's 1948 approach"*. This is unfair if one considers the way followed to come to the result and if one compares the papers. Experienced scientists like repeating Satayana's aphorism "*Those who cannot remember the past are condemned to repeat it* », but practically, how is it possible to remember or even know more than 6'000 papers devoted to martensitic transformations in steels (calculated from Scopus)? More importantly, rediscovering also allows exploring new directions. Only the matrix of *complete* transformation was given by

Jaswon and Wheeler, whereas the *continuous* KS distortion and analytical expression of the paths of the atoms were calculated in Ref. [117].

Now, if one compares the model of Ref. [117] with the PTMC, the former gives directly and continuously the correct bcc structure without adjusting parameters, without choosing addition shears, whereas the latter gives only the final structure and needs other parameters (such as the lattice parameters and a choice of LIS systems). All the steps of the transformation are now given analytically by a unique continuous parameter: the angle $\beta$ made by the dense directions (**PO**, **PK**). This angle is the natural *order parameter* that is usually so difficult to define for first-order transitions. The distortion matrix has singular properties. Contrarily to the Bain and Pitsch distortion matrices, it is not diagonalizable; it has only two eigenvalues 1 and 1.088 (which is the volume change due to the hard-sphere assumption); the eigenvector associated with 1 is the parallel dense direction **PO**, and the eigenvector associated with 1.088 is a direction that lies in the parallel dense plane (POK), which means that contrarily to an IPS where the volume changes occurs perpendicularly to the invariant plane, here the volume change occurs inside the untilted (but distorted) plane. We will come back later to this property. This result is new and implies that martensitic transformations should not be considered as a shear transformation. The notion of angular distortive transformation was thus introduced in order to replace the notion of shear.

With a large step back, the continuous hard-sphere displacive model of fcc-bcc transformation appears as a convergent model of nearly all the models reported in the past; it substitutes a one-step angular distortion for the two shears of the 1930-34 KSN model, or for the 2010 two-step fcc-hcp-bcc model [111]; it is a continuous version of the underestimated 1948 Jaswon and Wheeler's model of the KS distortion [33]; it replaces the four matrices of the 1952-53 PTMC models [34][35] by only one matrix. The model uses hard spheres as in the 1969 Bogers and Burgers' model [54], but now it opens the good angle on the good dense plane[2] such that there is no need of complementary shears as those proposed by Olson and Cohen in 1972-76 [57][58]. In addition, the important suspected crystallographic similarity with deformation twinning can be detailed and quantified. For example, fcc-bcc martensitic transformation and fcc-fcc deformation twinning can be geometrically represented in Figure 22, and quantitatively compared [123].

---

[2] In the Bogers & Burgers model shown in Figure 6, it is the angle between the untilted (111) dense plane noted VQP and another {111} plane, noted VSP, that is changed; the 60° angle between the dense directions in the plane VQP is not changed, i.e. the plane VQP is fully invariant before the second shear that allows forming the bcc structure. In the model shown in Figure 21, the untilted (111) plane is noted POK, and it is the angle between two of the three the dense directions of this plane (PO and PK) that is changed to directly obtain the bcc structure.

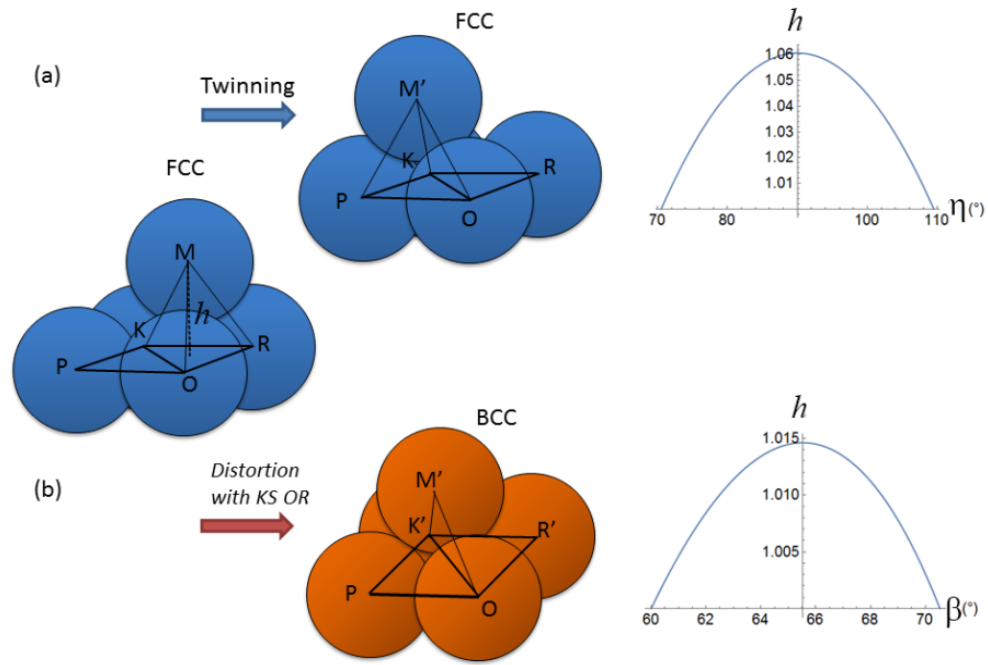

Figure 22. Hard-sphere model of (a) fcc-fcc deformation twinning and (b) fcc-bcc martensitic transformation. The distance $h$ is the distance between the planes initially POK = $(111)_\gamma$. This distance changes during the transformation process. Both fcc-fcc twinning and fcc-bcc martensitic transformations are modelled as functions of a unique angle of distortion, $\eta$ and $\beta$, respectively. In both transformations, the dense plane POK and the dense direction PO remain untilted. From Ref. [123].

They shows that twinning result from a simple atomic displacement on each $\{111\}_\gamma$ layer, as in the usual representation, except that the continuous atomic trajectory is not a simple shear strain because the atom M must go over the atoms O and K (and the angle $\beta$ remains locked at 60°). This trajectory is familiar to any student in metallurgy who already used balls stacked in ABCABC, ABCBA or ABAB sequences in practical classes. The fcc-bcc transformation results from the same atomic displacement, except that now the movement of M is combined with an increase of the distance between the atoms O and K (the angle $\beta$ is unlocked and changes from 60° to 70.5°); which explains why the "jump" *h* of the atom M (the distance between the atom M and its projection on the POK plane) is lower than for twinning. For fcc-fcc deformation twinning, as the surface of the plane POK is constant, the change of the distance *h* implies a volume change during the distortion, even if, of course, when twinning is complete the volume comes back to its initial value. This result is actually expected from a hard-sphere model because for fcc-fcc twinning the intermediate states are necessarily less dense than fcc by Kepler-Hales' theorem. The curve of Figure 22b shows that the distance *h* also changes during the fcc-bcc distortion. When the distortion is complete, and as for fcc-fcc twinning, this distance comes back to its initial value despite the fact the bcc phase has a volume higher than for the fcc phase. This can be understood by the fact that volume change between the initial and final states is completely due to the change of the surface of the plane POK→ POK' because of the angular distortion in which β = (**PO**, **PK**) increases for 60° to 70.5°, as shown in Figure 23.

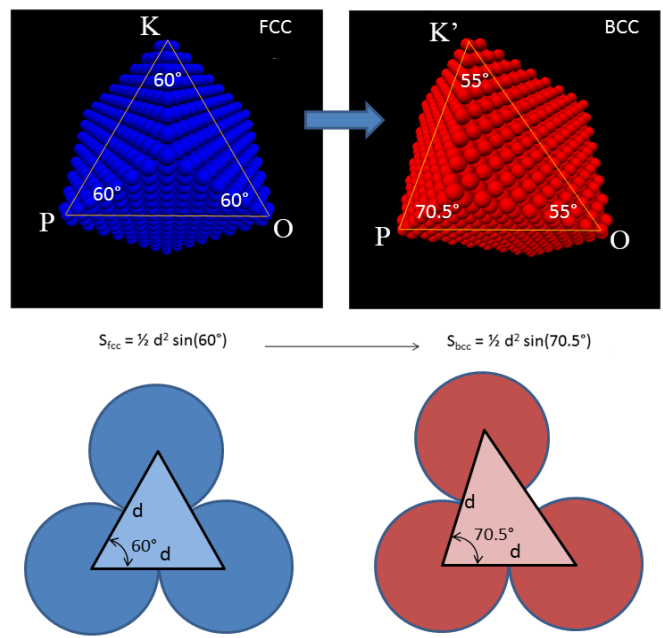

Figure 23. Representation of the fcc-bcc distortion with KS OR. The volume change is entirely due to the angular distortion in the plane POK→ POK', for which β = (**PO**, **PK**) increases for 60° to 70.5°.

In addition, it is qualitatively understood that the two components of the KS distortion, one given by the rotation of the direction **PK** around the normal to the dense plane POK, and the other by the rotation of the plane POM around the direction **OK**, give the continuous rotations **A** and **B** observed in the pole figures Figure 19a. In this approach, the special arrangements of dislocations are generated by the distortion: they form the disclinations at the origin of the rotations **A** and **B** retained in the material as plastic traces of the distortion mechanism. We are presently working to quantitatively simulate the continuous special features in the pole figures only from the analytical expressions of the distortion.

Even if in first approximation, the mechanism of fcc-bcc transformation can be modelled without knowing the exact nature of the accommodation processes, one can ask whether or not the "untilted plane" condition can be transformed into an IPS condition. The PTMC often transformed a lattice distortion **RB** into an IPS by using a LIS on $(112)_\alpha$ plane, explaining that it corresponds to a mechanical twinning of the bcc phase; however bcc metals generally twin only at very low temperatures (far below the usual Ms temperatures). The hypothesis made in Ref. [115] is that these apparent mechanical twins are in fact KS twin-related variants. This possibility was investigated in the case of the $(225)_\gamma$ martensite and we discovered that the "untilted" $(225)_\gamma$ plane could be indeed transformed into an IPS by combining two twin-related KS variants [124], as shown in Figure 24. These two KS are misoriented by 60° around the common dense direction **PO** = $<110>_\gamma$ // $<111>_\alpha$. A movie of the continuous formation of a (225) martensite composed of two KS variants was simulated.

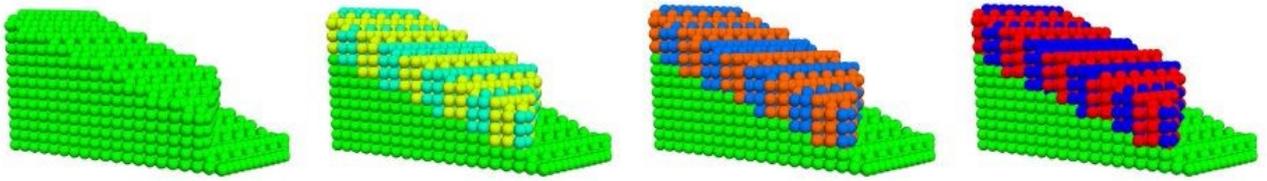

Figure 24.  Continuous distortion model of the fcc-bcc transformation on a (225) habit plane. This plane, which is only untilted during the formation of one variant, becomes an IPS when two twin-related KS variants are formed together. The atoms of the fcc crystal are in green, and those of the two final twin-related bcc variants are in red and blue. Movie available at https://www.nature.com/article-assets/npg/srep/2017/170120/srep40938/extref/srep40938-s1.avi. From Ref. [124].

These simulations are good agreement with the TEM observations of the midrib structure observed in the centre of lenticular martensite products [125][126]. However, one should keep in mind that far from the midrib, martensite is only constituted of one variant, which means that during martensite growth the transformation obeys an angular distortion accommodated by dislocations, and not an IPS anymore.

A model of the $\{557\}_\gamma$ martensite in low-carbon steels was also established [127]. In this model, the natural transformation is still the KS distortion, as for the $\{225\}_\gamma$ martensite in high carbon steels, but the average is now made between the two low-misorientated variants that share the same dense plane; these two variants form an assembly called "block". The average distortion is not an IPS as for the $\{225\}_\gamma$ martensite, but it is an "untilted" plane with strains lower than it would be without the variant coupling. It is assumed that this mode is complemented by additional accommodation dislocations. Nicely, the average distortion is exactly the distortion associated with the NW OR; and an eigenvector of the NW distortion matrix expressed in the reciprocal space is the $(11\sqrt{2})_\gamma$ plane, which is only 0.3° far away from the expected (557) plane. However, the model has been rejected twice and is still under review 18 months after its first submission because of a detail in its predictions: the calculated $\{557\}_\gamma$ habit plane is close to the common dense plane, in agreement with the experiments, but it does not contain the fcc-bcc parallel dense directions of any of the the two KS variants in the block, which seems to be in contradiction with some experimental papers. For the moment, we consider that the experimental results published in the past are not precise enough to reject our model of {557} martensite. It is indeed very difficult to determine precisely the habit plane of a martensite block by TEM, (a) because of the difficulty to align the interface along the electron beam due to of internal deformation inside the grains, (b) because the studies were done on fully transformed steels without retained austenite, which means that the habit plane was determined in the bcc reference frame and not in the fcc one, and (c) because in the habit planes was measured on the laths but not on the blocks. The $\{557\}_\gamma$ habit plane predicted by the model is the average habit plane, i.e. the habit plane of the blocks (pairs of low-misorientated variants sharing a common dense plane) and not the habit plane of the individual KS variants.

### 7.5 *Habit plane and stress relief*

Assuming that the habit plane is an invariant shear plane at mesoscopic scale (made of the martensite product associated with twins, or twin-related variants, or periodic arrangements of dislocations) is the classical paradigm of PTMC. We have seen that Bowles and Barrett in 1952 [49] discarded Jaswon and Wheeler's model of (225) martensite because of this paradigm. Two years later, Bowles and Mackenzie [35] explained the reasons for which, according them, the habit plane should be invariant. First they stated, as Jaswon and Wheeler had done, that, "*the amount of plastic deformation would be [unreasonable] extensive even for small rotations of the habit plane [...] It can therefore be concluded that the habit plane is not rotated*". Then, they stated that "*no line in the [habit] plane can be rotated [...] A direct test of the accuracy of this conclusion is furnished by the observation that a martensite plate and the neighbouring matrix can kept in focus under a microscope while traversing the whole length of plate.*" This last affirmation is essential because it implies the fact that the habit plane, in addition to be untilted is also undistorted, i.e. fully invariant. Unfortunately, this affirmation is not clear and does not seem to necessarily imply that the interface plane is undistorted. Peet and Bhadeshia [128] could observed by atomic force microscopy the surface relief produced by the formation of bainite at 200°C. The surface, shown in Figure 25a, is interpreted by these authors as the resultant of a homogeneous shear (Figure 25b), in agreement with Bowles and Mackenzie. However, such a relief can be explained as well by an angular distortion, as illustrated in Figure 25c,d.

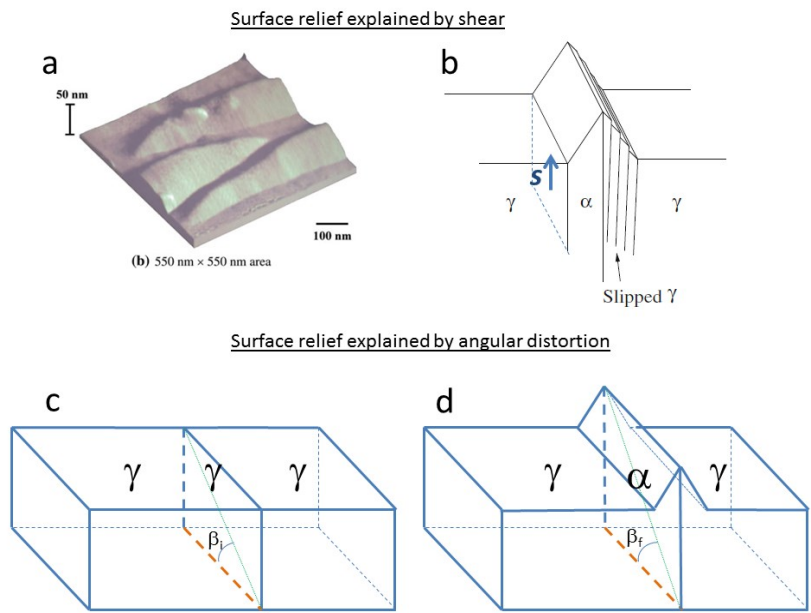

Figure 25. Interpretation of the surface relief formed by fcc-bcc transformation. (a) Image made by atomic force microscopy of bainite plates, and (b) their interpretation by a shear mechanism, from Ref. [128]. (c,d) Alternative explanation with an angular distortive mechanism. (c) Surface before transformation, and (d) surface after transformation: the tent-shape is compatible with an intra-planar distortion of the habit plane (without rotation).

It is probable that in many cases, mainly for the initial stages and for the midrib formation, the habit plane is rendered invariant by the co-formation of pair of variants (as in Figure 24), but that condition does not seem to be necessary for the growth of individual martensite; and cannot be used as a fundamental part of a mechanistic theory of martensitic transformation.

In brief, discarding the assumption of shear permits to establish over the last years simple and coherent models of the $\{225\}_\gamma$ and $\{557\}_\gamma$ martensite; which is encouraging if one considers the very long history and the high number of publications on this subject. Besides, it was stated from the first quantitative metallographies that the habit planes of martensite are irrational, even if noted with high index rational numbers. Irrational values can now be proposed: according to our model the $\{225\}_\gamma$ observed in high-carbon steels and the $\{557\}_\gamma$ habit planes in low-carbon steels are in fact $\{11\sqrt{6}\}_\gamma$ and $\{11\sqrt{2}\}_\gamma$ planes, respectively. The former are fully accommodated by coupling the twin-related variants that share the same dense direction, and the latter by coupling the low-misoriented variants that share the same dense plane and by additional dislocations. The difference between these two modes might come from the difference of Ms temperatures: in the former case, the low Ms temperatures do not allow dislocation plasticity but promote variant-pairing IPS accommodation, whereas in the latter case, dislocation plasticity allows for conditions less restrictive than pure IPS.

## 8 Generalization of the angular distortive model to fcc-bcc-hcp martensitic transformations and to fcc-fcc deformation twinning

The hard-sphere model allows describing with the same formalism the mechanisms of deformation twinning and martensite transformation. It permits to naturally introduce an *order parameter* in the transition, which is simply the angle of distortion (and not an ill-defined multidimensional strain state). The work done for the fcc-bcc transformation was thus quickly generalized to other displacive transformations between fcc, bcc and hcp phases observed in other alloys [123]. By noting fcc = γ, bcc = α and hcp = ε, as it is usually done for steels, the classical ORs between the three phases are written:

- KS:      $[110]_\gamma = [111]_\alpha$ and $(\bar{1}11)_\gamma // (\bar{1}10)_\alpha$  (16)
- Burgers: $[111]_\alpha = [100]_\varepsilon$ and $(\bar{1}10)_\alpha // (001)_\varepsilon$
- SN :     $[110]_\gamma = [100]_\varepsilon$ and $(\bar{1}11)_\gamma // (001)_\varepsilon$

These ORs are the KS OR for fcc→bcc, the Burgers OR [129] for bcc→hcp, and Shoji-Nishiyama (SN) [130] for fcc→hcp transformations, respectively. They respect the parallelism of the close-packed directions; they form a close set of ORs that is of interest to build a unifying theory of martensitic transformations, as already noticed by Burgers in 1934, and illustrated in Figure 26a. The three phases projected along the normal to their common dense plane is shown in Figure 26b. The use of angular distortion matrices offers the possibility to similarly treat these transformations in a unique framework, with the same paradigm. The fcc→hcp and bcc→hcp transformations involve shuffling because the parent and daughter phases do not have the same number of atoms in their Bravais cell (the bcc and fcc primitive cells contain one atom, whereas the hcp primitive cell contains two atoms). The distortion matrices and the shuffle trajectories (when required for the hcp phase) were analytically determined as function of an angular parameter that depends on the transformation. The choice of angular parameter and the details of the calculations are given in Ref. [123]. It is important to note that both lattice distortion and shuffling are the two sides of the same coin; there is no reason to disconnect them as it is done in some studies.

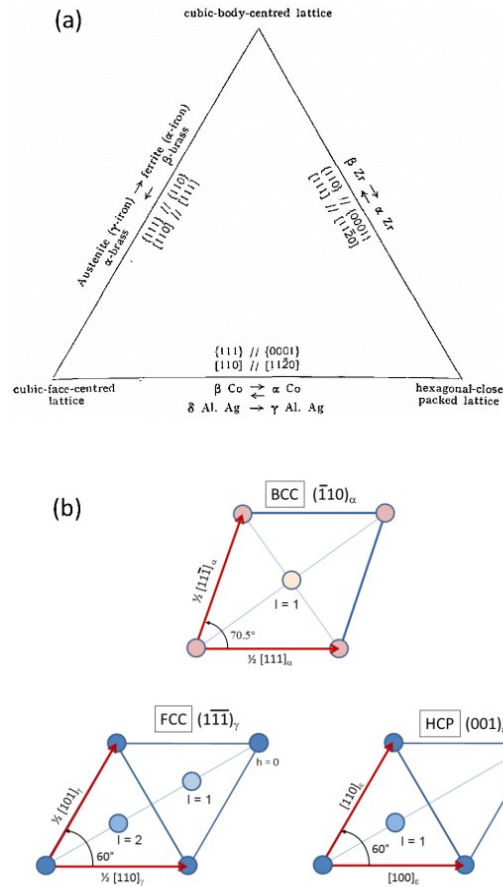

Figure 26. Phase transformations in the fcc-hcp-bcc system. (a) As represented by Burgers in 1934 [129]. (b) Planar representation of the fcc, bcc and hcp. The positions $l = 1$ and $l = 2$ represent the level of the atoms in their stacking perpendicularly the dense planes $(\bar{1}11)_\gamma // (\bar{1}10)_\alpha // (001)_\varepsilon$. From Ref. [123].

Some movies showing the atomic displacements were simulated, with some snapshots shown in Figure 27. They illustrate the atomic displacements during the transformation process by using an initial single crystal with a cubic shape; the exact morphology of the martensite product and the details of the accommodation mechanisms were not yet considered in these simulations. The analytical calculations are based on geometry only; and not on MD.

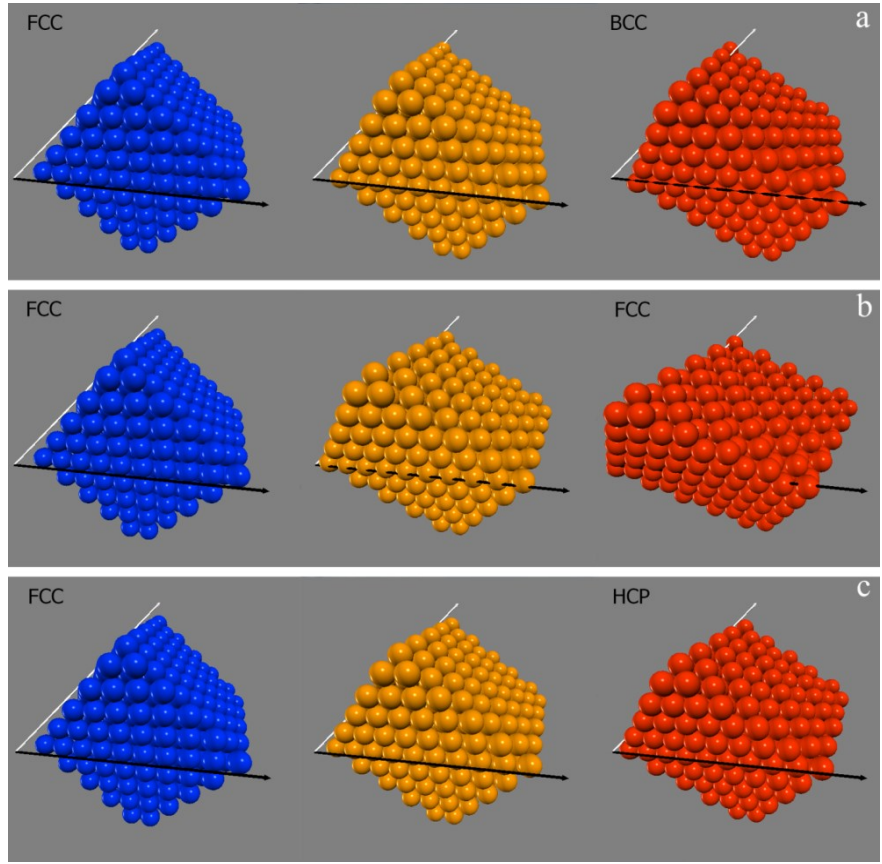

Figure 27. Snapshots of 3D simulation movies of (a) fcc→ bcc martensitic transformation, (b) fcc→ fcc mechanical twinning, and (c) fcc→hcp martensitic transformation, with, in blue, the initial parent fcc cube with its $\{100\}_\gamma$ facets, in red, the resulting transformed daughter crystals, and in yellow, the intermediate states stopped at midway (half of the maximum distortion angle). The black arrow represents the invariant line **PO**, and the white arrow the direction **PK** (also invariant for the fcc→ hcp and fcc→ fcc transformations). From Ref. [123].

The habit planes are calculated according to the "untilted plane" criterion. For bcc-fcc transformations and for bcc-hcp transformations, they are $\{(7 + 2\sqrt{6},\ 2 + 2\sqrt{6},\ 5)\}_\alpha$ and $\{(\sqrt{6}, \sqrt{6},\ 2)\}_\alpha$, respectively. They are less than 15° away from the habit planes reported in literature. This agreement is thus quite good if one considers that there no free parameter at all in the theory. Of course, the interface plane of the martensite in these alloys is more complex that a plane simply untilted, and the dislocations, twins, and the twin-related variants that structure the martensite interface are not taken into account by the model for the moment, except for the (225) martensite (Figure 24).

In the same spirit, one-step crystallographic models of martensitic transformation using the hard-sphere assumption were proposed recently by Sowa [131]; the main difference with our approach is that they are based on group-subgroup chains of intermediate structures.

## 9 Angular distortive model for deformation twinning in hcp metals

### 9.1 *Application to extension twinning*

The cases of deformation twinning in bcc and hcp were not investigated in Ref. [123] because they are more numerous and complex than for fcc-fcc twinning. The investigations on deformation twinning in magnesium were started two years ago. Among hcp metals, magnesium was chosen because of the c/a ≈ 1.625 ratio close to the ideal hard-sphere packing value $\gamma = c/a = \sqrt{8/3} \approx 1.633$. For extension twinning, the displacements of the Mg atoms were quite easy to determine [132]. A brief explanation is given here. The twin-parent misorientation is a rotation of 86° around the common **a**-axis. Instead of working directly with this misorientation, it is simpler working on a "prototype" stretch twin with a 90° misorientation, and do the first calculations in the orthogonal basis *xyz* shown in Figure 28a. The lattice stretch results from an exchange between the basal and prismatic planes, as shown in Figure 28b. This exchange can be visualized by considering the XYZ supercell containing 4 atoms: it is a consequence of a simultaneous rotation of angle η of the three atoms inside the cell. The lattice stretch induces a slight rotation of the $(0\bar{1}12)$ plane, called obliquity. The obliquity angle ξ shown in Figure 28c can now be calculated and continuously compensated in order to leave the $(0\bar{1}12)$ plane untitled during the twinning distortion such that the final twin-parent misorientation is (86°, **a**). The analytical equations of trajectories of all the magnesium atoms are established as functions of a unique angular parameter (as for martensitic transformations). It was shown that the $(0\bar{1}12)$ habit plane is not fully invariant during the twinning process; it is just untilted and restored when the distortion is complete. Indeed, the distance OV in this plane is not constant (Figure 28d). In addition, there is volume change of 3% during the distortion (Figure 28e), as expected from Kepler-Hales' theorem for a transition between two hcp states. This volume change cannot be totally accommodated elastically; which means that plasticity is required to accommodate the twinning distortion in polycrystalline alloys.

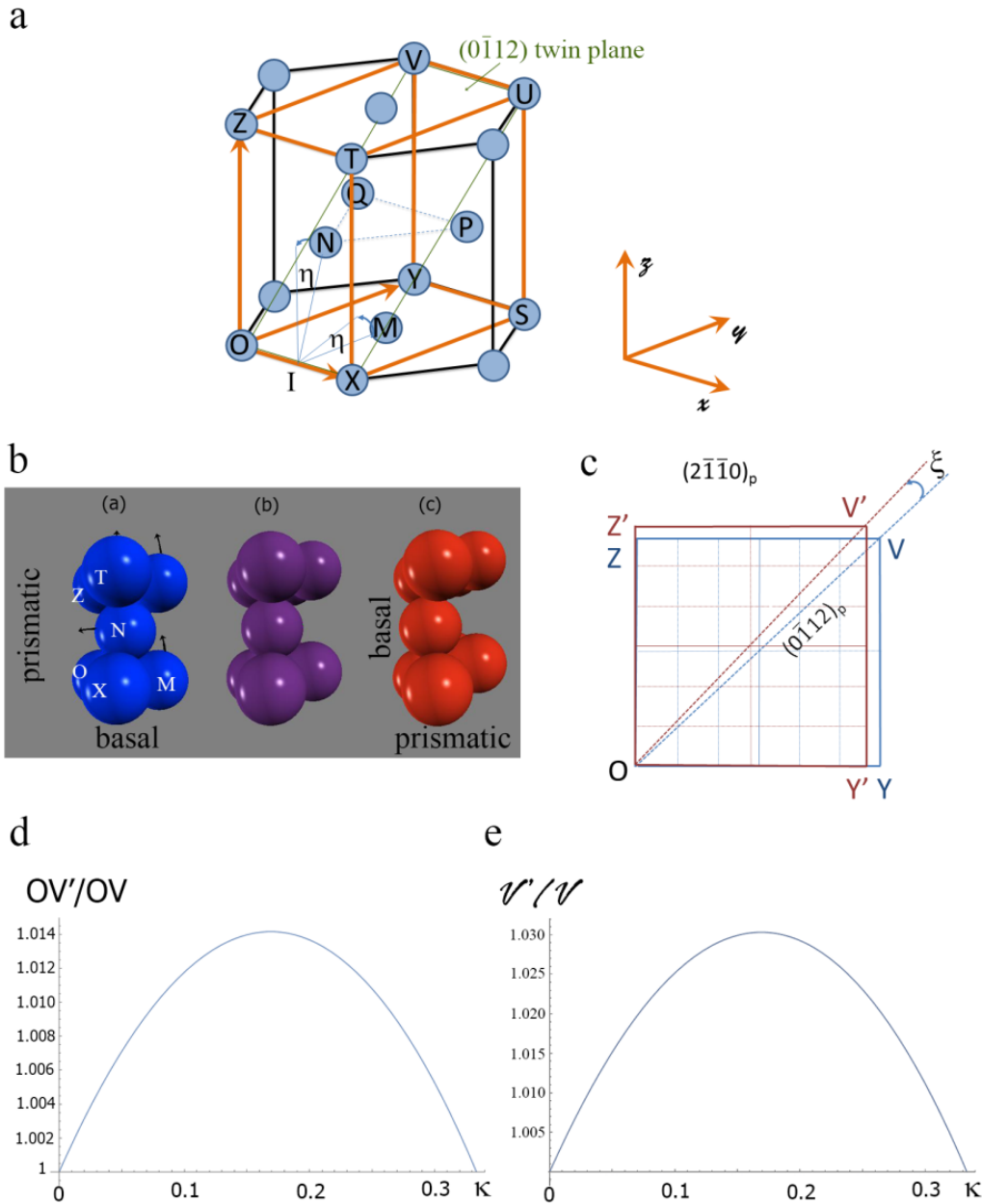

Figure 28. Hard-sphere displacive model of extension twinning in magnesium. (a) Unit cell with labels O,X,M,N,Z,T given to the magnesium atoms. (b) Schematic view of the atomic displacements associated with a stretch distortion, (c) obliquity ξ of the {0$\bar{1}$12} plane that should be compensated, (d) evolution of the distance OV belonging to the {0$\bar{1}$12} plane, and (e) change of the unit volume during the distortion process, as functions of the angular parameter κ that is the cosine of the angular parameter η. Adapted from Ref. [132].

This work conciliates inside a unique model the (90°, **a**) twins observed in nano-pillars [133] and the (86°, **a**) usual twins observed in bulk samples. Both result from the same atomic displacements and lattice distortion and differ only by their obliquity. At this step, the model does not imply to imagine twinning dislocations, disconnections, or any complex mechanism involving dislocations. Some elastic strains are induced some and dislocations are emitted by the distortion, and that could be interesting to determine how they can be predicted from the distortion matrix and from the usual slip systems in magnesium, but that is beyond the scope of the present model.

As the distortion is not a simple shear, a new criterion had to be introduced to substitute the Schmid's law. It is often reported that twins can form in hcp metals despite low or even negative Schmid factor [134]-[137]. This "non-Schmid" behaviour is usually explained by invoking that the local stress fields in the grains differ from the applied stresses, but another explanation is worth being considered. The intermediate state at the maximum volume change implies the largest strains perpendicularly to the $\{10\bar{1}2\}$ habit plane; so, it is conceivable that this state constitutes an energetic gap for the twin formation. Twinning occurs when the work performed by the external stresses accompanying the twin formation is higher than this energetic gap, or at least positive. The general formula of the work W done during a lattice distortion is $\boldsymbol{\mathcal{E}} = \mathbf{D} - \mathbf{I}$, where **D** is the distortion matrix and **I** the identity matrix occurring in a fixed stress field **Γ** is

$$W = \mathbf{\Gamma}_{ij} . \boldsymbol{\mathcal{E}}_{ij} \tag{17}$$

Whereas the usual shear model with Schmid's law predicts the formation of extension twins during uniaxial tensile tests with parent crystals tilted in the range $[-43°, 47°]$ as shown in Figure 29a, the hard-sphere displacive model associated with the work criterion predicts a range of $[-59°, 59°]$, as shown in Figure 29b. Experimental works would be needed to confirm this prediction. It should be noted that for fcc-fcc deformation twinning, the difference between Schmid's law and the work criterion at maximum volume change is hardly perceptible, as shown by Figure 29d-f (unpublished result); which could explain why non-Schmid behaviour could not be observed in mechanically twinned fcc metals.

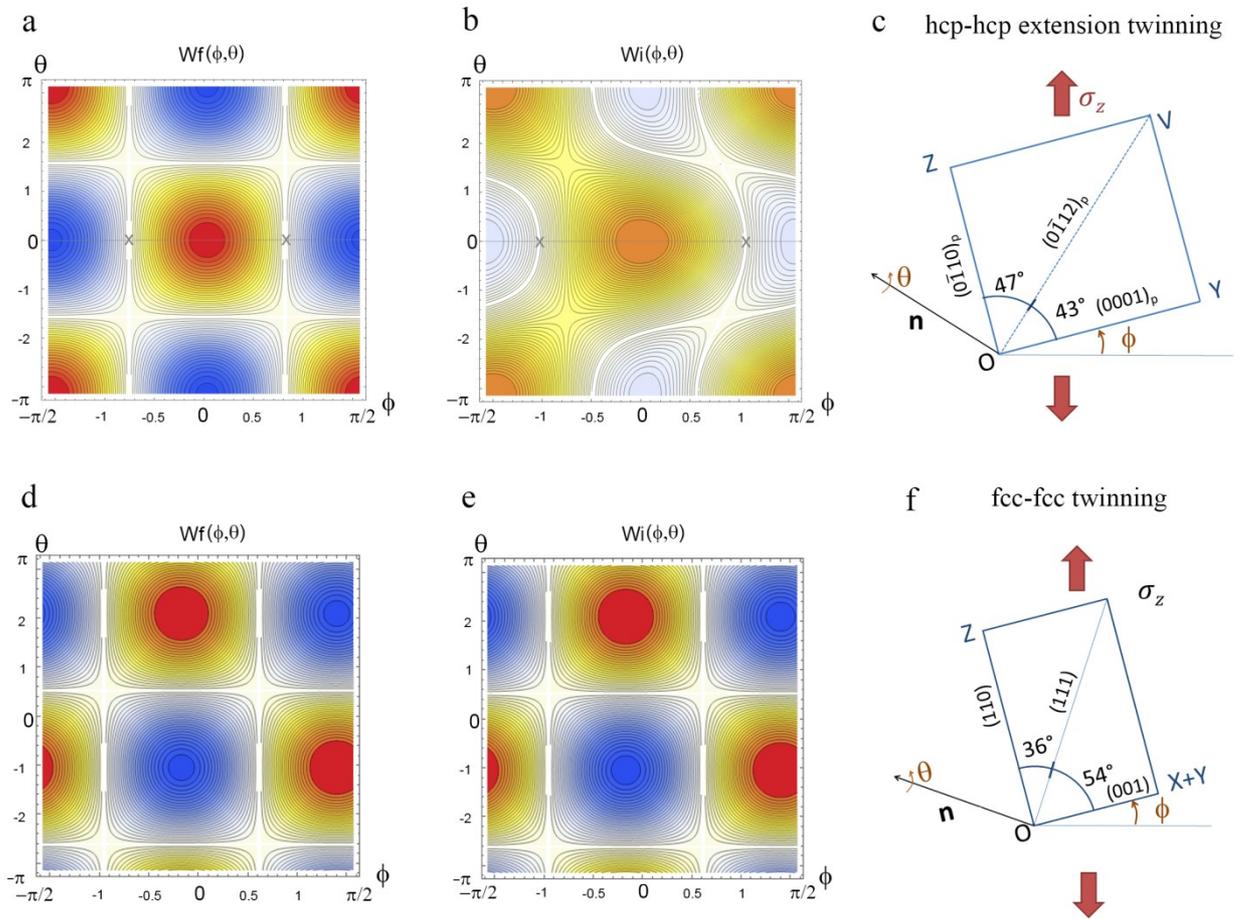

Figure 29. Interaction work during a tensile stress along the **z**-axis of a parent crystal tilted by an angle ϕ around the **x**-axis and rotated by an angle θ around the **n**-axis (normal to the twinning plane), with (a,b,c) for extension twinning in hcp metals, the **x**-axis is the [100] axis, and (d,e,f) for fcc-fcc deformation twinning, the **x**-axis is the [$\bar{1}$10] axis. (a,d) Interaction work $W_f$ calculated with the complete distortion (simple shear) matrix. $W_f$ is proportional to the usual Schmid factor. (b,e) Interaction work $W_i$ calculated with the intermediate distortion matrix corresponding to the maximum volume change. (c,f) Schematic view of the orientation of the parent crystal. The distortion matrix of extension twinning in hcp metal is given in Ref. [132]. The distortion matrix for fcc-fcc deformation twinning is given in Ref. [123].

*9.2   Comparison with the pure-shuffle model*

Another model of extension twinning without twinning dislocations exists. In 2009, Li and Ma "observed" the results of MD simulations and proposed an "*atomic shuffling dominated mechanism*" [138]. In 2013, Wang *et al.* also noticed in the MD simulations a nucleation step obeying a "*pure shuffle mechanism*", the growth being accomplished "*via the conventional glide-shuffle mechanism*" [139]. These authors also used the expressions "*unit cell reconstruction*" [140], and "*twinning with zero twinning shear*" [141], which can make think that the mechanism is diffusive on short distances. Li and colleagues define shuffling by "*inhomogeneous displacements of an ensemble of atoms in the layers immediately adjacent to the TB[twin boundary]*" [138], and twinning growth is imagined as a boundary migration where "*the TBs can migrate freely and fully consume the parent grains*" [141]. To avoid confusion, it is worth recalling what a "shuffle" is, because the term is not used with the same meaning everywhere. A shuffling is historically invocated when some atoms in the unit cell do not follow the same trajectories as those located at the nodes of the lattice. As written by Christian, Olson and Cohen, "*a shuffle only rearranges the atom positions within a unit cell*" [142]. For example, as the hcp unit cell contains 2 atoms, the martensitic transformations between fcc and hcp or between bcc and hcp require an additional shuffling of one-half of the atoms, as shown previously in Figure 26. The shuffling equations are analytically determined for these transformations in Ref.[123]. It is not surprising that shuffling and distortion are correlated, as experimentally found by Wang et al. for the bcc-hcp transformation in titanium [143] because the lattice distortion and the displacements of the atoms in the lattice are intrinsically correlated. Similarly, deformation twinning in hcp imagined as an hcp-hcp transformation involves shuffling. In the example of extension twinning in magnesium, a quarter of the atoms (those at the nodes of the XYZ lattice) follow the trajectory given by the distortion matrix, and three quarters of the atoms shuffle inside their local XYZ cell. Consequently, traditionally shuffling cannot exist by itself; if it exists, it always accompanies a lattice distortion. With Christian, Olson and Cohen's words, "*shuffles can only be avoided if both structures have primitive unit cells containing only one atom*" [142]. This meaning has digressed over time with the introduction of the concept of "shuffle transformations". Christian, Olson and Cohen define them as follows: "*A pure shuffle transformation requires that some unit cell of one lattice is almost identical with a cell of the other lattice*". Delanay [144] uses a more precise definition: *"A shuffle is a coordinated movement of atoms that produces, in itself, no lattice distortive deformations but alters only the symmetry or structure of the crystal; a sphere before the transformation remains the same sphere after the transformation."* He adds later: *"Shuffle transformations are not necessarily pure; small distortive deformations may additionally occur. They therefore also include those transformations involving dilatational displacements, in addition to the pure shuffle displacements, provided that they are small enough not to alter significantly the kinetics and morphology of the transformation".* Therefore, a pure shuffle transformation is a transformation that should not change significantly the lattice, which is not the case for extension twinning, and that is why the idea of "pure shuffle" is unclear. It is indeed difficult to imagine that the atoms can move independently of the lattice in which they are contained. This would be possible with a lattice containing many and largely spaced atoms of small size, but is completely impossible with a hard-sphere model in which the atoms are in contact, as proved by the calculations of Ref. [132]: when the atoms move the lattice is distorted, and reciprocally. There is no possibility to accommodate elastically the atomic displacements in the unit cell while maintaining the unit cell dimension. Liu *and*

*al.* are more cautious than Li and Zhang in the use of vocabulary; they clearly specify that *"UCR [unit cell reconstruction] produces tetragonal deformation instead of simple shear"* [140]. However, for all these researchers, the lattice distortion, if it exists, is seen only as a consequence of the shuffling mechanism, without clear link with the external stresses, without relation with the classical "shear" models of twinning. It would be usefully if the authors could specify their opinion on the distortion associated (or not) to the "pure-shuffle" mechanism. Even if the hard-sphere model comes to the same conclusion that *"the {10-12} twinning plane cannot remain invariant during twinning"* and that a mechanism *"without the need of twinning dislocations"* is possible [141], it is difficult to agree with the conclusion that the *"[the extension twinning] mechanism distinctively differs from other twinning modes"*. We have shown that deformation twinning can be modelled by a distortive mechanism in which all the atoms can move collectively at the speed of sound, or at lower speeds if time is needed to reorganize the dislocations emitted by the distortion and relax the back-stresses in the surrounding matrix. Therefore, to our point of view, the mechanism of extension twinning in magnesium does not differ from the other twinning modes. All the twinning modes should imply an angular distortion, and most of them (but not all of them) become a simple shear when the distortion is complete. Despite this slight divergence of interpretation, it is important to note that the hard-sphere displacive model and the shuffle model developed by Wang, Liu, and colleagues are not so dissimilar. The trajectories of the atoms shown in Figure 3 of Ref. [139] and reported in Figure 30a seem to be the same as those shown in Figure 28a. The trajectories reported in Figure 4 of Ref. [133] shown in Figure 30b are different, but the rotation and mirror symmetries between the figures makes an exact comparison difficult.

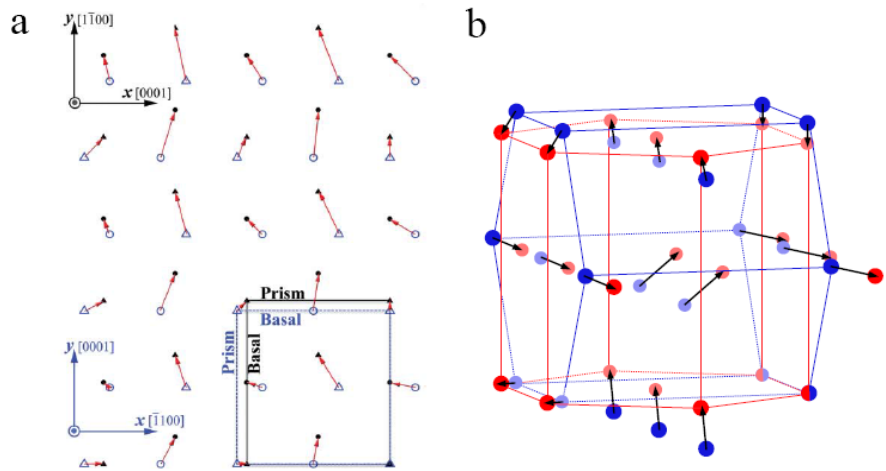

Figure 30. Shuffle models of extension twinning in hcp metals proposed by (a) Wang *et al.*, from Ref. [139], and by (b) Liu *et al.*, from Ref. [133].

The advantages of the distortive model of deformation twinning is that it can be combined with mechanical calculations, such as that of the work W of equation (17) performed during twinning, in order to explain why twins are formed and for which orientations they are formed. Contrarily to the shuffle model, the distortive model makes macroscopic predictions. On another hand, it is true that it does not yet explain the microscopic characteristics at the twin interface, such as the fact that the interface plane is often not straight and that basal-prismatic segments observed in TEM. Partisans of "pure-shuffle" model insist on these features as if they were a specific property of extension twinning; however, the existence of a segmented interface is a quite general property of displacive transformations. The formation of terraces and ledges are observed for the deformation twins in the TWP steels, and for many diffusive and displacive transformations, as proved by the HRTEM observations made by Ogawa and Kajiwara in martensitic iron alloys [145]. At mesoscale, the lenticular shape of extensions in magnesium is often observed for other deformation twin modes and in other structures (Figure 31). Thus, neither the lenticular shape nor the basal-prismatic segments are a proof of a pure-shuffle mechanism; they can be formed as well displacively.

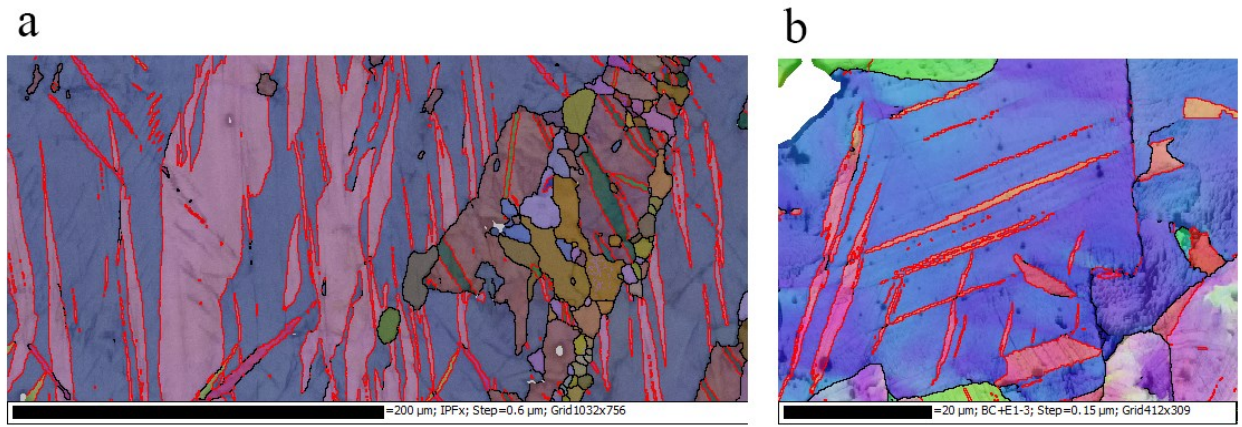

Figure 31. EBSD maps of lenticular deformation twins, (a) formed in recrystallized pure (hcp) magnesium, and (b) in a (fcc) TWIP steel (kindly given by K. Zhu, ArcelorMittal). The twin boundaries are marked in red; they are identified by the misorientations (87° ± 5°, $[001]_{hex}$) and (60°± 5°, $<111>_{fcc}$) in (a) and (b) respectively.

The atoms can move displacively, collectively, rapidly, such that at macroscopic scale the $\{10\bar{1}2\}$ twinning plane remains untilted, and at the microscopic scale, the interface between the twin and its surrounding matrix is made of ledges that are here segments of XYZ cells.

In brief, it seems that the "pure-shuffle" and the "distortive" visions could be mixed because they are not antagonist. Both approaches have strong arguments [146]; they agree to say that the dislocation twinning and the complexity of the topological model are actually unnecessary to explain the twinning mechanism. In addition, both discard the classical assumption that extension twinning is a simple shear mechanism. However, Li and Ma are probably too cautious when they say that the $\{10\bar{1}2\}$ twinning mechanism *"distinctively differs from other twinning modes"*, and that *"this should not*

*be deemed as the failure of the classical theory"*. There is nothing special in the mechanism of extension twinning in magnesium and it is conceivable that all the deformation twinning modes and all the martensitic transformations obey the same crystallographic rules. The angular distortive paradigm proposes a way to get this unification.

### *9.3 Application of the distortive model to other twinning modes*

There is another important difference between the pure-shuffle and the distortive approaches. The former relies on MD computer simulations to explore new twinning modes and extract some hints and rules to build crystallographic models, whereas the latter just requites a sheet of paper and a pen; the computer is just used to help the analytical calculations that, at least in theory, could be done by hand. The displacive model is purely geometric; this simplicity is a force to understand and explain the various twinning modes in metals. For example, the hard-sphere displacive model was also recently applied to $\{10\bar{1}1\}$ contraction twinning in hcp metals [147], as shown in Figure 32; leading to similar conclusions, i.e. the volume of the unit cell is not constant and the habit plane is not invariant during the process; it is just untilted and restored, and only then, the distortion matrix of complete twinning becomes a simple shear matrix. The result is a shear of amplitude s = 0.358 on the plane $(0\bar{1}1)$ along the axis $[18,\bar{5},\bar{5}]$, which was not predicted by Bevis and Crocker's theory. It was recently realized by writing the shear direction with four indices, i.e. $-[\bar{41}, 28, 13, 5]$, that this shear mode is exactly that given by Serra, Pond and Bacon in their list of three possible $\{10\bar{1}1\}$ twinning systems given in Table 1 of Ref.[148] calculated from the translation vectors between the parent and twin crystals (i.e. vectors of the dichromatic structure) [149]. This confirms the validity of the calculations [147] and also proves that an approach different from a disconnection model is possible.

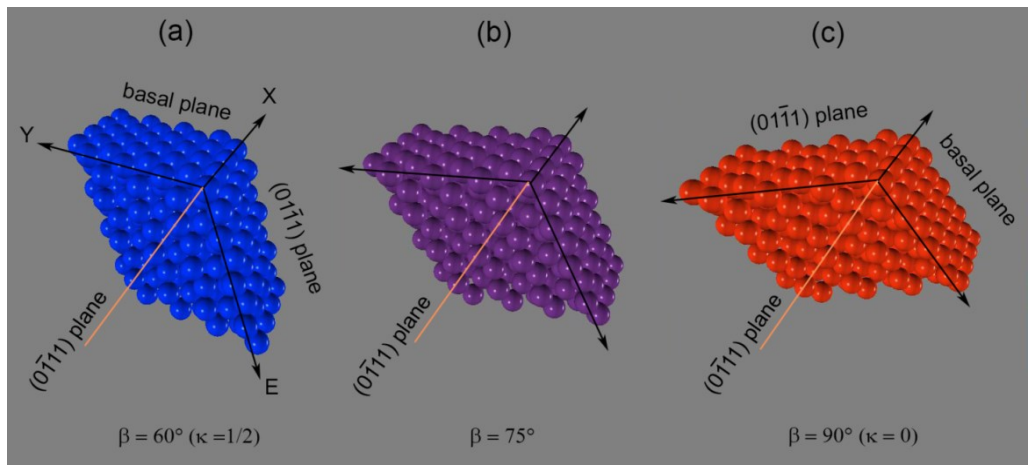

Figure 32. Snapshots of three states during the formation of a (56°, **a**) contraction twin on the $(0\bar{1}11)$ plane in magnesium simulated with the displacive hard-sphere model. (a) Initial hcp crystal, (b) intermediate state at midway, (c) newly formed hcp crystal. This figure is made to help the understanding of the crystallographic exchange between the basal plane and the $(01\bar{1}1)$ plane. From [147].

There is no technical issue to apply the same approach to other twinning modes in magnesium or to bcc-bcc twinning modes; and a general theory of deformation twinning compatible with a hard-sphere assumption can be developed in the near future.

It is important to note that the habit plane is not an invariant plane, but just a plane restored from the intermediate state when the transformation is complete. The possibility that the plane is transformed into a new plane different from the initial one is offered by the new paradigm. We recently found by a EBSD study of a magnesium single crystal a new and unconventional twinning mode that is incompatible with simple shear but corresponds exactly to this case [150].

### 9.4  *Comparison with earlier atomistic models of hcp twinning*

Dubertret and Le Lann proposed in 1980 models of twinning in hcp metals [151][152] that were cited by Wang et al. [139] as references for the concept of shuffling. These models were nearly forgotten despite or because of the huge literature on deformation twinning. As those detailed in the previous sections, they are based on hard-spheres, but a direct comparison is difficult because the atomic displacements are expressed from tetrahedrons with an odd case where $\frac{c}{a} = \sqrt{3}$. The main initial idea is the same as that used in the pure-shuffle model, and the odd ratio was probably chosen to let the atoms move and exchange their positions in the lattice without changing the lattice dimensions in order to be in a pure-shuffle case. Despite this this odd choice and the absence of correlation between the atomic displacements and the lattice distortion, Dubertret and Le Lann clearly understood that $\{10\bar{1}2\}$ extension twinning and $\{10\bar{1}1\}$ contraction twinning are obtained by atomic displacements that produce a crystallographic exchange (correspondence) between the basal and a prismatic planes in the former case, and between the basal plane and a $\{1\bar{1}01\}$ plane in the latter case. They called these exchanged planes, the *"corrugated planes"*. Going deeper into historical considerations, it is interesting to note the initial 1968 Kronberg's model [153] on which Dubertret and Le Lann's works are based also uses a hard-sphere representation of the atoms and creates a correlation between the atomic trajectories and the lattice distortion, as shown in Figure 33.

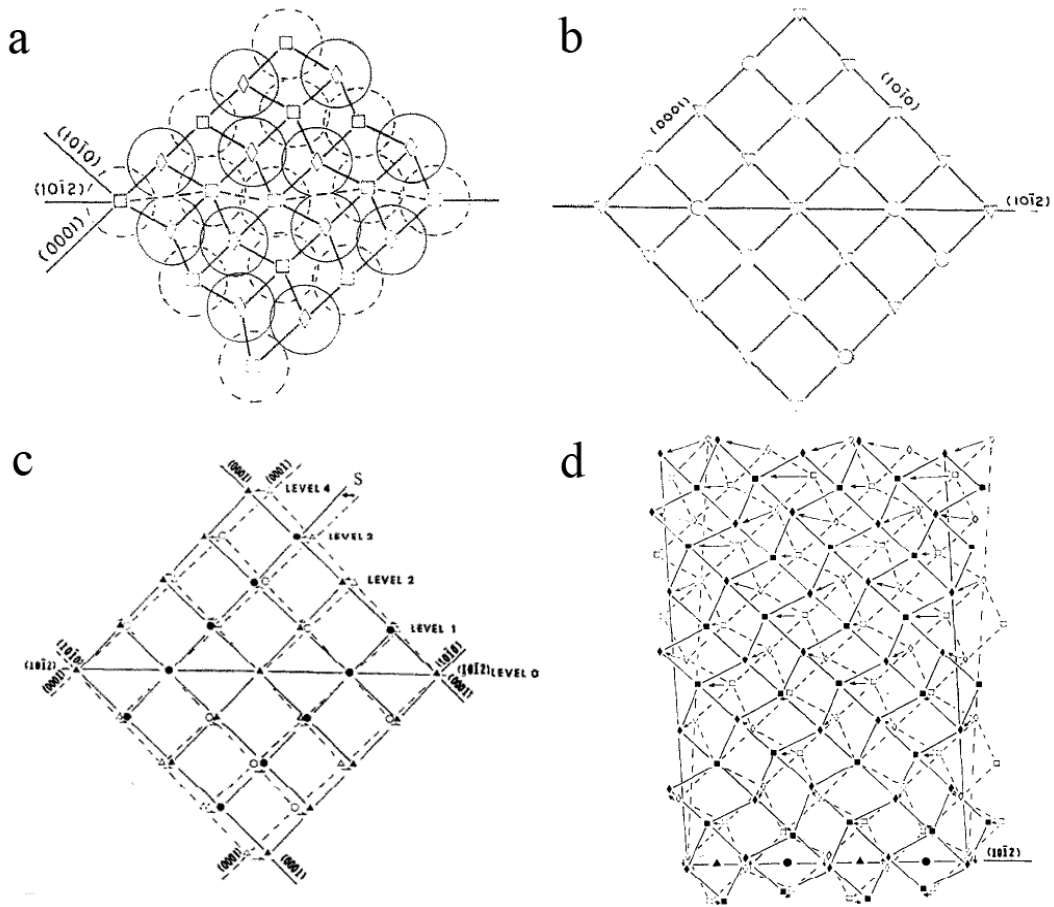

Figure 33. Kronberg's model of extension twinning in hcp metals. (a) Initial crystal with the "hard-sphere" atoms projected along the $[010]_{hex}$ direction, and (b) corresponding lattice. (c) Lattice distortion during twinning, and (d) lattice distortion and atomic displacements during twinning. From Ref. [153].

The model of Ref. [132] is a kind of "rediscovery" of Kronberg's work [153]; but it now includes the calculation of the distortion, correspondence and orientation matrices, and analytical equations of the atomic trajectories.

# 10 Future works and perspectives

## 10.1 *The distortion angle as a natural order parameter*

Thermodynamic models of phase transitions are based on a parameter called "order parameter". In phase field models (see for example Ref. [154][155]), the order parameter is simply the proportion of phases, i.e. a real number between 0 (parent phase) and 1 (daughter phase), with the interface encoded by a real value between 0 and 1. It is usual to distinguish second-order and first-order phase transitions (note that the meaning of term "order" is different from that of "order parameter"). Some models try to find a physical parameter that controls the order parameter, but the choice depends on the type of transition.

Second-order transitions are generally modelled by the Landau theory [156]. The order parameter is arbitrarily chosen depending on the physical property that is judged of interest; it can be the variation of the local density for liquid-crystal transitions, as in Landau's paper [156], the atomic displacement or polarization for ferroelectrics, the variation of the probability of site occupancy for order-disorder transitions, etc. The free energy is expressed by a Taylor decomposition into a polynomial form of the order parameter, and the polynomial coefficients depend on the temperature and pressure. The order parameter is solution of the null derivative of this polynomial; it is zero at high temperature and non-null below the critical transition temperature $T_c$, and it continuously varies as a function of the temperature (Figure 34ab).

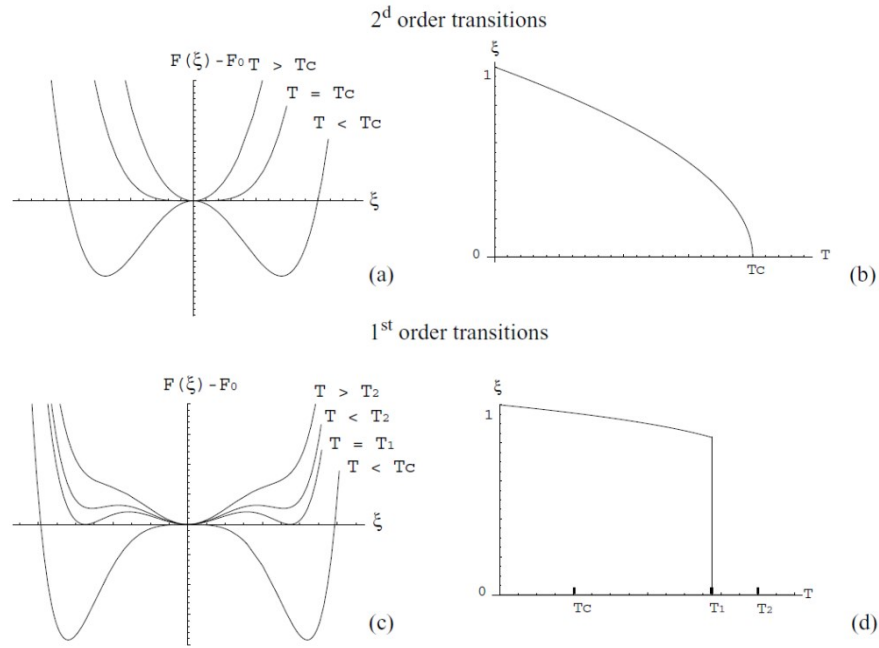

Figure 34. The free energy F function of the order parameter ξ and temperature T, and the order parameter function of the temperature T, in the case of (a,b) second order transitions, or (c,d) first order transitions.

The Landau theory [156] is phenomenological because of the ad-hoc choices of the polynomial exponents and coefficients, but it gives a first important mathematical framework of the symmetries. Landau stated that the order parameter is a multidimensional parameter that has many degrees of freedom as the number of dimensions of the irreducible space chosen for the irreducible representation of the group of the parent phase. This concept is now didactically explained in many modern textbooks. In the example of cubic-tetragonal ferroelectric transitions, one irreducible space is the 1-dimensional space **z** (the direction that will becomes the **c**-axis of the tetragonal phase), and the other irreducible space is the 2-dimensional space (**x**, **y**). Landau's model establishes a link between crystallography and thermodynamics. The use of the group representations theory in physics was quite natural as is was already introduced few years earlier in fundamental physics, mainly by Wigner [157] and Weyl [158]. Group representation was and still is considered as a major part of group theory;

it has acquired important success in materials science, mainly in spectroscopy with the calculations of the Raman frequencies.

The martensitic transformations and deformation twinning are first-order structural transitions. It is agreed that the order parameter is a discontinuous function of temperature (Figure 34cd). What is the order parameter for these transformations? Defining it is an old and difficult problem. Clapp for example wrote in Ref. [159]: "*Certainly one of the stumbling blocks in defining a martensitic transformation is that there is no obvious "order parameter" associated with it, such as one has with ferromagnetic transformations (magnetic moment), order-disorder (long range order parameter), etc. Of course, one always has the possibility of using the volume fraction transformed as such an order parameter, analogous to the use of the fractional density of superconducting electrons for superconducting transformations, but this is a much more erratic quantity (markedly dependent on sample history) in the case of interest here*." Different order parameters have been proposed beside the "volume fraction" mentioned by Clapp. For example, Flack [96], and more recently Clayton and Knap [160], proposed to use the shear strain(s). Beside the problem of defining the order parameter, Landau's theory cannot be applied directly to first-order transitions because it is based on the idea that the transition is a loss of symmetry; which is generally not the case for first-order transformations[3], for example fcc-bcc transformations. Some symmetry elements are lost, but others are created, i.e. there is not anymore a group-subgroup relationship between the parent and daughter phases. These transitions are also called in physics "reconstructive" (the meaning here is different from that used in metallurgy where it is synonymous of "diffusive"). An interesting and powerful generalization of Landau' theory to reconstructive transitions was proposed by Tolédano and Dmitriev [162]. The effective order parameter is a variation of a density function that is a sum of wave functions. The periodic form of this parameter permits to decompose the free energy as a truncated Fourier series of the order parameter (and not anymore a Taylor series). Many examples of applications, from crystal to quasicrystals, are studied and described in Ref. [162]. The density-wave description is an important element of generalization that establishes a first link with the wave propagation mode envisioned in section 5.1. However, the systematic use of a latent lattice, relevant for order-disorder transitions, would require more physical justifications for displacive transformations. Besides, Tolédano and Dmitriev's book [162] aims at physicists, but it does not use the classical vocabulary and concepts of metallurgy, and it does not respond to some basic questions raised by metallurgists, for example about orientation relationships and the habit planes.

---

[3] The distinction between second and first order transitions is quite arbitrary. All the structural transitions imply an accommodation process in the surrounding matrix and thus a dissipation of energy. Of course, if the distortion is small as in the "displacive" transitions (with the physical meaning, i.e. small atomic displacements), the accommodation is mainly elastic, the transition is reversible and the hysteresis is small; whereas if the transition is "reconstructive" (with the physical meaning, i.e. breaking of the atomic bounds and formation of new ones), the distortion is large and imply plasticity. It is often assumed that the former are second-order and the latter are first-order. but the distinction is more quantitative than qualitative. It is also usual to assume that when a group-subgroup

Some of the concepts detailed in the paper could constitute a connection between physics and metallurgy, but important work of mutual understanding remains to be done to build a real bridge between the two communities. The angular-distortive models

---

relation exists, the transition is second-order, and when this relation does not exist, the transition is first order; however, one can imagine a cubic-tetragonal distortion with very high distortion amplitudes that would be irreversible (and thus would be of first order). Mnyukh [161] came to the extreme conclusion that the distinction between the first and second order transitions is artificial, and that all the transitions are first-order and only differ by the amplitude of the discontinuity.

presented in the previous sections show that a one-dimensional order parameter (simply the angle of distortion) can be chosen for many martensitic transformations. This parameter is physical; it controls both the atomic trajectories and the lattice distortion. It is however not yet clear to us how this parameter could enter into the framework of Landau's theory or into its advanced reconstructive version. Indeed, there is a fundamental difference in the way the symmetries are mathematically treated. Let us explain it with the simple questions: how many variants (domains) and how many types of domain boundaries are created by a phase transition? We got this problem when we had to find a method to reconstruct the prior parent grains from EBSD data (Figure 18). Instead of using the theory of group representation (based on action of the group by conjugation) as in Landau's and Tolédano and Dmitriev's theories, it was shown that the action by left multiplication is more appropriate to respond to these questions. The distinct orientations of the domains are called orientational variants. For some special parent/daughter orientation relationships, the parent and daughter phases have some symmetry operations in common; these symmetries form a subgroup of the point group of the parent phase called intersection group. The orientational variants are the cosets based on this subgroup. More explicitly, we call $\mathbb{G}^\gamma$ and $\mathbb{G}^\alpha$ the point groups of the parent and daughter phases, respectively; there is no need here to imagine them as abstract groups that should be decomposed into irreducible representations; a point group here is simply the set of 3x3 matrices of symmetries tabulated in the International Tables for Crystallography. Let us also call $\mathbf{T}_0^{\gamma \to \alpha}$ the coordinate transformation matrix deduced from the orientation relationship. The symmetries of the daughter crystal expressed in the parent basis are given by the set of matrices $\mathbf{T}_0^{\gamma \to \alpha} \mathbb{G}^\alpha (\mathbf{T}_0^{\gamma \to \alpha})^{-1}$. The intersection group is thus $\mathbb{H}^\gamma = \mathbb{G}^\gamma \cap \mathbf{T}_0^{\gamma \to \alpha} \mathbb{G}^\alpha (\mathbf{T}_0^{\gamma \to \alpha})^{-1}$. The set $\mathbb{H}^\gamma$ is a subgroup of $\mathbb{G}^\gamma$, i.e. $\mathbf{H}^\gamma \leq \mathbf{G}^\gamma$. The distinct orientational variants $\alpha_i$ are defined by the cosets $\alpha_i = g_i^\gamma \mathbf{H}^\gamma$ and their orientations relatively to the parent basis are $\mathbf{T}_0^{\gamma \to \alpha_i} = [\mathbf{B}_0^\gamma \to \mathbf{B}_0^{\alpha_i}] = g_i^\gamma \mathbf{T}_0^{\gamma \to \alpha}$ with $g_i^\gamma \in \alpha_i$. One must understand "$g_i^\gamma \in \alpha_i$", as a matrix $g_i^\gamma$ arbitrarily chosen in the coset of matrices $\alpha_i$. By convention, $g_1^\gamma$ is the identity matrix. The number of orientational variants $N^\alpha$ is the number of cosets on $\mathbf{H}^\gamma$; i.e. the cardinal of $\mathbf{G}^\gamma / \mathbf{H}^\gamma$; it is given by the Lagrange formula: $N^\alpha = |\mathbf{G}^\gamma| / |\mathbf{H}^\gamma|$. More details are given in Ref. [106].

Now, we consider the different distortion matrices at the origin of the orientational variants. A distortional variant is defined by its effect on the initial shape of a crystal. An initial parent crystal of shape whose symmetries form a group $\mathbf{G}^\gamma$ becomes after complete distortion $\mathbf{D}_0^{\gamma \to \alpha}$ a crystal (of daughter phase) whose shape has symmetries given by the matrices $\mathbf{D}_0^{\gamma \to \alpha} \mathbf{G}^\gamma (\mathbf{D}_0^{\gamma \to \alpha})^{-1}$. The "distortional" intersection group is formed by the symmetries that are common to the crystal before and after distortion, i.e. $\mathbf{K}^\gamma = \mathbf{G}^\gamma \cap \mathbf{D}_0^{\gamma \to \alpha} \mathbf{G}^\gamma (\mathbf{D}_0^{\gamma \to \alpha})^{-1}$ with $\mathbf{K}^\gamma \leq \mathbf{G}^\gamma$. The distortional variants $d_i$ are defined by the cosets $d_i = g_i^\gamma \mathbf{K}^\gamma$. The number of distortional variants $M^\alpha$ is the number of cosets on $\mathbf{K}^\gamma$; it is the cardinal of $\mathbf{G}^\gamma / \mathbf{K}^\gamma$ and is given by the Lagrange formula: $M^\alpha = |\mathbf{G}^\gamma| / |\mathbf{K}^\gamma|$. Generally, $\mathbf{K}^\gamma \leq \mathbf{H}^\gamma$, which means that the number of distortional variants is higher than the number of orientational variants: $M^\alpha \geq N^\alpha$. More details are given in Ref. [123].

Let us come back to the orientational variants. The special misorientation between them, simply called *operators*, are defined by the double-cosets $\mathbf{H}^\gamma \backslash \mathbf{G}^\gamma / \mathbf{H}^\gamma$. The number of operators is given Burnside's formula. In addition, the algebraic structure of the variants and their operators is a groupoid, and the groupoid composition table (imagined for the purpose) can be used as a crystallographic signature of the phase transition

[106]. These ideas are not completely new; important results about coset and double-coset decompositions were already discovered by Janovec [163][164] in ferroectrics, and the intersection group was found by Kalonji and Cahn for transitions without group-subgroup relation in their theoretical study of grain boundaries [165]. The mathematical notion of groupoid unifies these notions and allows their generalization. A groupoid structure for the distortional variants, similar to that of the orientational variants, seems conceivable. Since the distortion matrices simply depend on a 1-dimensional order parameter (the angle of distortion), these matrices or their associated angular order parameter could be used to build a $M^{\alpha}$-dimension space for the polynomial form of the free energy. In a broad way, one would replace the group representation theory used in Landau's theory by a groupoid theory based on cosets and double-cosets. One could hope finding a polynomial form of the free energy whose solutions would constitute a groupoid. We recall that groupoids were initially introduced by Brandt in 1926 [166][167] for quadratic forms. The main idea here would be to find a groupoid on the set of solutions of the polynomial form of the free energy that would be isomorph to the groupoid of distortional variants; and one way to show the isomorphism could be to compare the composition tables of both groupoid structures. The idea is still vague and such researches would require the help of mathematicians.

### *10.2 Dynamics of phase transformation and accommodation phenomena*

During a martensitic transformation all the atoms move collectively but, as previously discussed, it does not mean that all the atoms move together *at the same time*. Imagining that the transformation propagates in the material as a wave permits to escape to the Frenkel and Cottrell's issue about the unrealistic critical stress required for an instantaneous and homogenous shear. It is indeed possible to use the continuous analytical form of the angular distortion matrix of a structural transformation **D**(*β*), where $\beta$ is the distortion angle, to generate heterogeneously in space and time the martensite product. One way could be to find an analytical equation for the order parameter $\beta$ that depends both on space and time. For example, let us introduce (i) the position of the parent interface $\mathbf{L}(t)$ that moves at a wave velocity $c$, and (ii) a length of accommodation (LA), without detailing for the moment if the accommodation is elastic or plastic. An interesting form of the order parameter $\beta(\mathbf{r}, t)$ is given by a linear ramp function

$$\beta(\mathbf{r},t) = \begin{cases} \text{If } r \leq L(t): & \beta = \beta_f \\[6pt] \text{If } L(t) < r < L(t) + LA: & \beta = \dfrac{(r - L(t))}{LA} \beta_s + \dfrac{(L(t) + LA - r)}{LA} \beta_f \\[6pt] \text{If } r > L(t): & \beta = \beta_s \end{cases}$$

Where $\beta_f$ is the final distortion angle, i.e. $\mathbf{D}(\beta_f)$ is the matrix of complete distortion, and $\beta_s$ is the initial distortion angle, i.e. $\mathbf{D}(\beta_s)$ = Identity, and where $L(t) = c.t$. The regular letters r and L mean the norm of the vectors $\mathbf{r}$ and $\mathbf{L}$. This ramp function is shown in Figure 35.

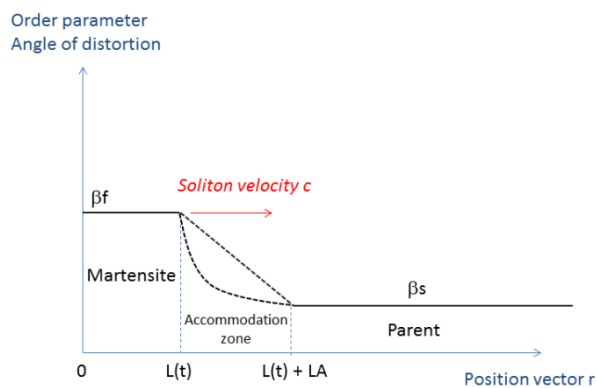

Figure 35. Ramp function for the angular parameter β of the lattice distortion transformation $\mathbf{D}(\beta)$.

In addition, one can imagine that **L**(t) is a function of the position vector **r**. For example the length of **L**(t) can be inversely proportional to the amplitude of strain in the position vector **r**, such that. $\boldsymbol{L}(\boldsymbol{r},t) = \frac{c.t}{\|(\mathbf{D}(\beta_f)-\mathbf{I}).\boldsymbol{n}\|+\varepsilon} \boldsymbol{n}$ , where $\boldsymbol{n} = \frac{\boldsymbol{r}}{r}$ , and ε is an additional constant added to avoid infinity issues. The model, applied to the 2D hexagon-square distortion presented in Figure 12 allows the simulation of simple dynamic-crystallographic-atomistic movies; some snapshots are given in Figure 36. The movies show a hexagonal lattice of hard-disks constituting a round crystal that is continuously and heterogeneously distorted to be transformed into a square lattice of hard-disks forming an elliptic crystal at completion. The length of accommodation (LA) was chosen to be equal to the size of the crystal in order to minimize the change of distances $d$ between the neighbouring atoms. Actually, LA could be arbitrarily chosen such that $d$ does not deviate from the initial interatomic distance (atom diameter) by more than a fixed and low value that agrees with elasticity. In addition, it seems possible to replace the linear ramp by a more complex form (for example the one indicated by the dashed curve in Figure 35) that could minimize the mean or maximum interatomic distance ($d_i$-$d$) on the set of the atoms $i$. This approach could be also investigated to check whether or not the existence of parallel dense directions and planes would favour the wave propagation, which could then be used to determine a new criterion to predict the "natural" parent-daughter OR (see section 7.3). This way also starts establishing a link with the Ginzburg-Landau theory [168] in which the gradient of the order parameter is taken into consideration in the analytical form of the free energy, with here the advantage of including the atoms, the lattices and many crystallographic tools that are of interest for martensitic transformations (distortion matrix, correspondence matrix etc.). This vague idea will investigated once the author will get a better understanding of Landau and Ginzburg-Landau theories.

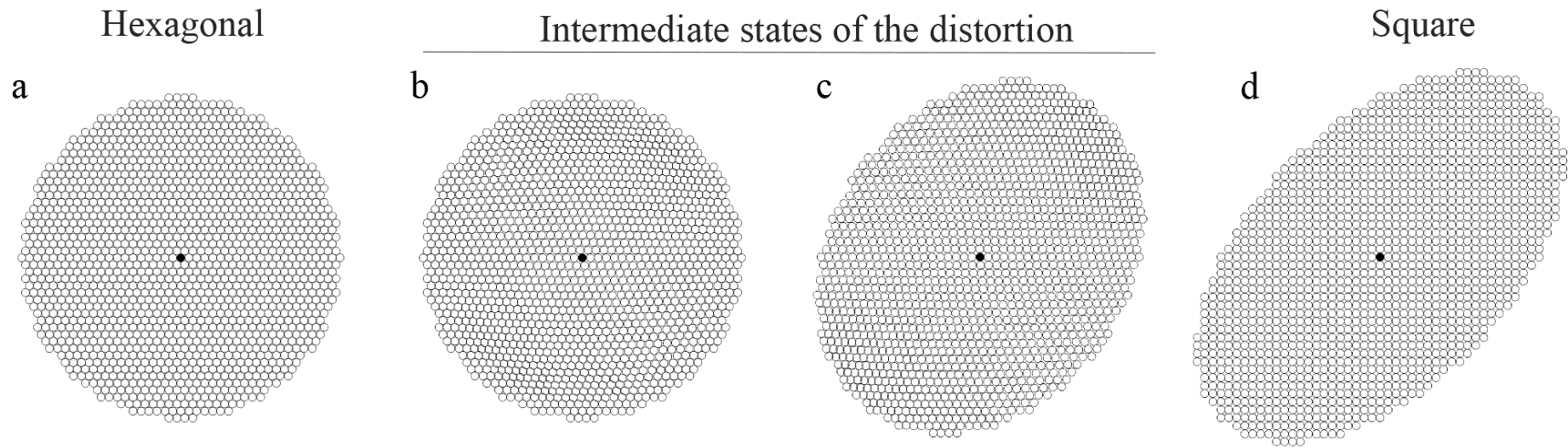

Figure 36.  Quasi-elastic lattice distortion during the hexagonal-square transformation of a free crystal. (a) Initial parent hexagonal phase. (b,c) Intermediate states showing the formation of the daughter square lattice in the centre of the parent hexagonal lattice (the remaining parent hexagonal phase can be identified close to the surface). (d) Daughter square phase after completely transformation. The initial circular shape of the crystal became elliptic after transformation. There is no "twinning dislocation" involved in the mechanism.

# 11 Conclusions

The concept of simple shear has been the corner stone of the classical theories of the crystallography of martensitic transformations and deformation twinning for 150 years. The early models of the fcc-bcc transformations proposed by Young, Kurdjumov and Sachs, and Nishiyama, those of the bcc-hcp transformations made by Burgers, the phenomenological theory of martensitic transformations (PTMC) initiated by Greninger and Troiano and developed by Weschler, Read and Liebermann, and Bowles and Mackenzie, the theory of deformation twinning developed by Bilby, Bevis, and Crockers among others, all are based on compositions of simple shears, or imply an invariant plane strain (IPS), which is a generalized form of shear that accounts for the volume change induced by the transformation. The mathematical treatment is irreproachable, although complex due to the initial premises. The predictive character of these theories is more limited than what is often claimed; which may explain why they have evolved in time, by the addition of new shears (double or triple-shear versions of PTMC, double-shear versions of twinning), i.e. by adding new layers of complexity, as the "epicycles" were added to the Ptolemean model of our solar system in order to get better precisions. An important problem of the shear paradigm was raised by Frenkel's calculations (corrected later by Orowan and Cottrell); it shows that an instantaneous simple shear is impossible under realistic stresses or without the help of dislocations. In order to tackle this issue, Cottrell and Bilby, followed by Sleeswyk, Venables, and Christian among others, came to imagine additional mechanisms in which the deformation twins are *created by* a sequential and coordinated movements of partial dislocations, called "*twinning dislocations*". Sleeswyk also proposed that the twinning dislocations dissociate into "*emissary dislocations*" before moving away from the interface to reduce the stresses. Often, TEM images of partial Shockley dislocations in front of twins or martensite are shown to prove the existence of these "twinning dislocations". The concept of dislocation-mediated mechanism of twinning was generalized to displacive transformations by Hirth, Pond and co-workers; the twinning dislocations were replaced by "*disconnections*" specifically designed to model the ledges at the parent/martensite interfaces. In this framework, the mechanism of the martensitic transformation is envisioned as the result of a displacement of the interface, and the martensite growth condition depends on the glissile character of the disconnections. Important questions are not addressed by these models. The pole mechanism and its derivatives do not convincingly explain how the dislocations/disconnections are created and how they propagate in a sequential and coordinated way at speeds close to the acoustic velocity. Besides, the fundamental question "how do the atoms move?" remains unsolved.

In order to respond this simple question in the case of fcc, bcc, hcp metals, the assumption was made that the atoms move as hard-spheres. In order to respect this assumption, the concept of simple shear had to be replaced by that of angular distortion because the volume change of the intermediate states is higher than allowed by simple elasticity. The sole input data is the final orientation relationship between the parent and the twin/martensite daughter phases. There is no other free parameter. A unique 1-dimensional angular parameter specific to each transition allows the analytical calculation of the trajectories of all the atoms. The lattice distortion and shuffling (when required) appear as both sides of the coin. Obviously, the distortion can be decomposed into infinitesimal combinations of stretches and rotations, or LIS and IPS, as in PTMC, but this decomposition would artificial and of low interest.

In the framework of this new paradigm, it is not required that all the atoms move collectively at the same time; martensite may actually be formed as a "lattice distortion wave" that propagates at high velocity and around which an accommodation zone can be spread on large distances to maintain the interatomic distances below the elastic limit. It is conceivable that the accommodation is elastic in the first stages of nucleation and growth, and becomes plastic when the martensite size becomes comparable to the grain size. In this way, contrarily to the usual dislocation/disconnection–mediated models, the dislocations are not the *cause* but appear as a *consequence* of the lattice distortion. The partial dislocations shown by TEM in front of the mechanical twins may be created and emitted by the twin itself and not by hypothetical pole mechanisms.

When the lattice transformation is achieved, the angular distortive paradigm allow finding the same results as in shear-based theories in classical cases, as for examples for the {225} martensite in high-carbon steels, for the deformation twins in fcc metals and for the extension twinning in hcp metals. As the angular distortive paradigm is however more general than the shear one and opens new possibilities beyond the calculations of the atomic trajectories. The habit planes can be determined with a criterion that is less restrictive than with the shear paradigm; it is assumed that they are planes untilted by the distortion but not necessarily fully invariant. Different accommodation mechanisms are possible: variant pairing, special dislocation arrays that close the rotational gaps (disclinations), etc. These mechanisms can be studied in a second stage; there is no need to imply them directly in the mechanism of transformation; there is no need to impose gliding conditions on the interface in order to understand the propagation of a martensitic transformation. Undercooling or external stresses are important to get sufficient chemical or mechanical energy to create the elastic or plastic accommodation zone required by the lattice distortion. In athermal martensite, the stress field accumulated around the product phase blocks the transformation; an additional undercooling is thus required to continue to form martensite in these deformed zones. In isothermal martensite, dislocations climbing and recovery phenomena may occur; which reduce the blocking strength in the accommodation zone and allows the transformation to continue with time.

Even if quite new, the angular-distortive paradigm already permitted to obtain significant results in various transformations that are classical in metallurgy. It allows the calculations of the irrational {225} and {557} habit planes in martensitic steels without any adjustment; it naturally introduces a unique "order parameter" (the distortion angle); it establishes quantitative correlations between deformation twinning and martensitic transformations, and it explains the unconventional twinning modes recently discovered in magnesium. It also predicts a volume change during the twinning process, it and proposes a solution to the issue raised by the formation of twins in crystals oriented in stress fields with negative-Schmid factors. Dedicated experiments are conceivable to test these predictions. Interesting links between thermodynamics and crystallography are now offered, even if mathematical tools different from those currently used in the physics of phase transitions are required.

As the angular-distortive paradigm is only purely geometric, it can be blamed for its over-simplicity, and that is true that the few initial assumptions can be criticized. a) The hard-sphere hypothesis ignores the electronic structure and real interatomic potential, b) the "arbitrary" choice of the orientation relationship is based on the assumption of the existence of a "natural" distortion without clear absolute criterion to define it, c) the formation of martensite or twins conceived as "phase transformation waves" is not yet physically and mathematically detailed, and c) the accommodation mechanisms are not yet correlated to the plasticity modes of the parent and daughter phases. Consequently, the approach will evolve in the future by replacing the hard-spheres by elastic spheres, by incorporating more fundamental physics with DFT calculations, and more mechanics with crystalline plasticity simulations. Investigations will be continued to establish a mathematical formalism for the wave propagation and to try to deduce a criterion that defines the "natural" orientation relationship. The features observed in the EBSD pole figures are qualitatively explained as the plastic traces of the lattice distortion (as if the transformation waves were "frozen" by plasticity), and we hope to quantitatively simulate them uniquely with the distortion matrices; this would bring a response to the initial question that triggered these researches ten years ago.

## Acknowledgments

I would like to show my gratitude to Prof. Roland Logé, director of LMTM, and to our Ph.D student Annick Baur for our discussions and common work on martensite. PX group is also acknowledged for the laboratory funding.

## References

[1] Cahn RW. *Twinned crystals*. Advanced in Physics 1954;3:363-444.

[2] Christian JW, Mahajan S. *Deformation Twinning*. Prog. Mater. Sci.1995;39:1-157.

[3] Mahajan S, Williams DF. *Deformation Twinning in Metals and Alloys*. Int. Metal. Rev. 1973;18:43-61.

[4] Hardouin Duparc OBM. *A review of some elements for the history of mechanical twinning centred on its German origins until Otto Mügge's K1 and K2 invariant plane notation*. J. Mater. Sci. DOI: 10.1007/s10853-016-0513-4.

[5] Thomson W, Tait PG. *Treatise on Natural Philosophy*, Cambridge, 1867; vol. I (1) 170-171, pp. 105-106 (pp.123-124 in the second edition 1879, §§149-150 in the Elements of Natural Philosophy, a simplified version of the Treatise).

[6] Mügge O. *Ueber homogene Deformationen (einfache Schiebungen) an den triklinen Doppelsalzen BaCdCl4.4aq.*,Neues Jahrbuch für Mineralogie, Geologie und Palaeontologie Beilage-Band. 1889;6:274–304 (+ Taf. IX).

[7] Haüy RJ, *Traité de Minéralogie*; 1801. 5 volumes, freely available at http://catalogue.bnf.fr/ark:/12148/cb373669857.public. The drawings are in the last volume.

[8] https://en.wikipedia.org/wiki/Atom

[9] Barlow W. *Probable Nature of the Internal Symmetry of Crystals*. Nature. 1883;29:186-188.

[10] Bragg L, Nye JF. *A Dynamical Model of a Crystal Structure*. Proc. R. Soc. Lond. A. 1947;190:474–481.

[11] Hall EO. *Twinning and diffusionless transformations in metals*. Butterworths Scientific Publications, London; 1954.

[12] Kihô H. *The Crystallographic Aspect of the Mechanical Twinning in Metals*. J. Phys. Soc. Japan. 1954;9:739-747.

[13] Jaswon MA, Dove DB. *The Crystallography of Deformation Twinning*. Acta Cryst. 1960;13:232-240.

[14] Bilby BA, Crocker AG. *The Theory of the Crystallography of Deformation Twinning*. Proc. R. Soc. Lond. A. 1965;288:240-255.

[15] Bevis M, Crocker AG. *Twinning Shears in Lattices*. Proc. Roy. Soc. Lond. A. 1968;304:123-134.

[16] Crocker AG, Bevis M. *The Science Technology and Application of Titanium*. ed. R. Jaffee and N. Promisel, Pergamon Press, Oxford; 1970. p. 453-458.

[17] Niewczas M. *Lattice correspondence during twinning in hexagonal close-packed crystals*. Acta Mater. 2010;58:5848-5857.

[18] Schmid E, Boas W. *Kristallplastizität: Mit Besonderer Berücksichtigung der Metalle.* in German (1st ed.). Springer. ISBN 978-3662342619; 1935.

[19] Barrett CD, El Kadiri H, Tschopp MA. *Breakdown of the Schmid law in homogeneous and heterogeneous nucleation events of slip and twinning in magnesium*. J. Mech. Phys. Solids. 2012; 60:2084-2099.

[20] Kumar MA, Beyerlein IJ, Tomé CN. *Effect of local stress fields on twin characteristics in HCP metals*. Acta Mater. 2016;116:143-154.

[21] Taylor GI. *The deformation of crystals of beta-brass*. Proc. Roy. Soc. A. 1928;118:1-24.

[22] Christian JW. *Some Surprising Features of the Plastic Deformation of Body-Centered Cubic Metals and Alloys*. Metall. Trans. A. 1983;14:1237-1256.

[23] Gröger R, Bailey AG, Vitek V. *Multiscale modeling of plastic deformation of molybdenum and tungsten: I. Atomistic studies of the core structure and glide of ½<111> screw dislocations at 0K*. Acta Mater. 2008;56:5401-5411.

[24] Wayman CM. *Shear Transformations and Microstructures*. Metallography 1975;8:105-130, republished in Mater. Charact. 1997;39:235-260.

[25] Bain EC. *The nature of martensite*. Trans. Amer. Inst. Min. Metall. Eng. 1924;70:25-35.

[26] Young J. *The Crystal Structure of Meteoric Iron as determined by X-Ray Analysis*. Roy. Soc. Proc. A 1926;112:630-642.

[27] Kurdjumov G, Sachs G. *Über den Mechanismus der Stahlhärtung*. Z. Phys. 1930;64:325-343.

[28] Wassermann G. Archiv Eisenhüttenwesen. 1933;6:347-351.

[29] Nishiyama Z. *X-Ray Investigation of the Mechanism of the Transformation from Face-Centred Cubic Lattice to Body-Centred Cubic*. Sci. Rep. Tohoku Univ. 1934;23:637-668.

[30] Burgers WG. *On the process of transition of the cubic-body-centered modification into the hexagonal-close-packed modification of zirconium*. Physica 1934;1:561-586.

[31] Kelly A, Groves GW. *Crystallography and Crystal Defects*. 1st ed., Longman, London; 1970.

[32] Greninger AB, Troiano AR. *The Mechanism of Martensite Formation*. Metals Trans. 1949;185:590-598.

[33] Jaswon MA, Wheeler JA. *Atomic displacements in austenite-martensite transformation*. Acta Cryst. 1948;1:216–224.

[34] Weschler MS, Liebermann DS, Read TA. *On the theory of the formation of martensite*. Trans. AIME 1953;197:1503–1515.

[35] Bowles JS, Mackenzie JK. *The crystallography of martensitic transformations I*. Acta Metall. 1954;2:129-137.

[36] Bowles JS, Mackenzie JK. *The crystallography of martensitic transformations II*. Acta Metall. 1954;2:138-147.

[37] Christian JW. *The Theory of Transformations in Metals and Alloys*. Pergamon Press, Oxford; 1965, last version; 2002. ISBN: 9780080440194.

[38] Nishiyama Z. *Martensite Transformation*, English edition, Academic Press, New York; 1978.

[39] Bhadeshia HKDH. *Worked examples in the geometry of crystals.* 2d ed. Brookfield, The Institute of Metals; 1987.

[40] Bhadeshia HKDH, Honeycombe R. *Steels: Microstructure and Properties.* Butterworth–Heinemann. ISBN 978-0750680844; 2006.

[41] Nishiyama Z, Shimizu K. *On the sub-bands in a martensite plate*. J. Electron Microscopy 1956;4:51.

[42] Shimizu K. *Japanese great pioneer and leader, Zenji Nishiyama, on studies of martensitic transformations*. J. Phys. IV. 2003;112:11-16.

[43] Machlin ES, Cohen M. *Habit Phenomenon in the Martensite Transformation*. Trans. AIME. 1951;1019-1029.

[44] Bowles JS, Dunne DP. The role of plastic accommodation in the (225) martensite transformation. Acta Metall. 1969;17:677-685.

[45] Dunne DP, Wayman CM. The assessment of the double shear theory as applied to ferrous martensitic transformations. Acta Metall. 1971;19:425-438.

[46] Wayman CM. *The growth of martensite since E.C. Bain (1924)-Some Milestones*. Mater. Sci. Forum. 1990;56-58:1-32.

[47] Dunne DP. *An historical account of the development of the Bowles-Mackenzie theory of the crystallography of martensitic transformation*. ICOMAT-08, edited by G.B. Olson, D.S. Lieberman, and A. Saxena, TMS; 2009, p. 47-53.

[48] Zhang MX, Kelly PM. *Crystallographic features of phase transformations in solids*. Prog. Mater. Sci. 2009;54:1101-1170.

[49] Bowles JS, Barrett CS. Crystallography of Transformations. Progress in Metal Physics 3, Editor B. Chalmers, London, Pergamon Press Ltd; 1952, p. 1-41.

[50] Kelly PM. *Crystallography of Lath Martensite in Steels*. Mater. Trans. JIM. 1992;33:235-242.

[51] Qi L, Khachaturyan AG, Morris Jr JW. *The microstructure of dislocated martensitic steel: Theory*. Acta Mater. 2014;76:23-39.

[52] Mackenzie JK, Bowles JS. *The crystallography of martensitic transformation IV Body-centred cubic to orthorhombic transformations*. Acta Metall. 1957;5:137-149.

[53] Williams AJ, Cahn RW, Barrett CS. *The crystallography of the $\beta \to \alpha$ transformation in titanium*. Acta Metall. 1954;2:117-128.

[54] Bogers AJ, Burgers WG. *Partial dislocations on the {110} planes in the bcc lattice and the transition of the fcc into the bcc lattice*. Acta Metall. 1964;12:255-261.

[55] Cottrell AH. *Dislocations and Plastic Flow in Crystals*. Oxford Clarendon Press, 1st edition; 1953.

[56] Zener C. *Elasticity and Anelasticity of Metals*. Chicago, University Press; 1948.

[57] Olson GB, Cohen MA. *Mechanism for the strain-induced nucleation of martensitic transformations*. J. Less. Common Met. 1972;28:107–118.

[58] Olson GB, Cohen MA. *General mechanism of martensitic nucleation: Part II. FCC → BCC and other martensitic transformations*. Metall. Mater. Trans. A. 1976;7:1905–1914.

[59] Le Lann A, Dubertret A. *A structural Description of the F.C.C.→B.C.C. Martensitic Transformation in Terms of {011} <01$\bar{1}$>$_{f.c.c}$ Simulation Double Shear*. phys. Stat.sol. (a) 1983;78:309-321.

[60] Frenkel J. *Zur theorie der elastizit. atsgrenze und der festigkeit kristallinischer kurper*. Z. Phys. 1926;37:572-609.

[61] Smallman RE, Bishop RJ. *Modern Physical Metallurgy & Materials Engineering*. 6th edition, Butterworth-Heinemann; 1999, chapter 4 "Defects in Solids".

[62] Kittel C. *Introduction to solid state physics*. 8th edition; 2005, chapter 21 "Dislocations".

[63] Peierls R. *The Size of a Dislocation*. Proc. Phys. Soc. 1940;52:34-37.

[64] Nabarro FRN. *Dislocations in a simple cubic lattice*. Proc. Phys. Soc. 1947;59:256-272.

[65] Bragg L, Lomer WM. *A dynamical model of a crystal structure II*. Proc. Roy. Soc. A. 1949. DOI: 10.1098/rspa.1949.0022.

[66] Hirth JP. *A Brief History of Dislocation Theory*. Met. Trans. A. 1985;16:2085-2090.

[67] Seitz F, Read WT. *Theory of plasticity of solids III.* J. Appl. Phys. 1941;12:470-486.

[68] Vladimirskii KV. *Twinning in calcite* (in Russian). Zh. Eksp. Teor. Fiz. 1947;17:530.

[69] Frank FC, van der Merwe JH. *One dimensional dislocations I*. Statistic theory. Proc. Roy. Soc. A. 1949;198:205-216.

[70] Sleeswyk AW. *1/2⟨1 1 1⟩ screw dislocations and the nucleation of {1 1 2}⟨1 1 1⟩ twins in the b.c.c. lattice*. Phil. Mag. 1963;8:1467-1486.

[71] Christian JW. *Twinning and martensitic transformation*. J. Phys. 1974;C7-35:65-76.

[72] Cottrell AH, Bilby BA. *A mechanism for the growth of deformation twins in crystals*. Phil. Mag. 1951;42:573-581.

[73] Sleeswyk AW. *Perfect dislocation pole models for twinning in f.c.c and b.c.c lattices*. Phil. Mag. 1974;29:407-421.

[74] Venables JA. *On dislocation pole models for twinning*. Phil. Mag. 1974;30:1165-1169.

[75] Mahato B, Sahu T, Shee SK, Sahu P, Sawaguchi T, Kömi J, Karjalainen L-P. *Simultaneous twinning nucleation mechanisms in an Fe-Mn-Si-Al twinning induced plasticity steel*. Acta Mater. 2017;132:264-275.

[76] Thomas G, Whelan MJ. *Helical dislocations in quenched aluminium-4% copper alloys*. Phil. Mag. 1959;4:511-527.

[77] Venables JA. *The electron microscopy of deformation twinning*. J. Phys. Chem. Solids 1964;25:685-692.

[78] Mahajan S. *Nucleation and growth of deformation twins in Mo-35 at.% Re alloy*. Phil. Mag. 1972;26:161-171.

[79] Mahajan S, Chin GY. Formation of deformation twins in F.C.C. crystals. Acta Metal. 1973;21:1353-1363.

[80] Zárubová N, Ge Y, Heczko O, Hannula SP. *In situ TEM study of deformation twinning in Ni-Mn-Ga non-modulated martensite*. Acta Mater. 2013;61;5290-5299.

[81] Bergeon N, Guenin G, Esnouf C. *Microstructural analysis of the stress-induced $\varepsilon$ martensite in a Fe-Mn-Si-Cr-Ni shape memory alloy: Part I – calculated description of the microstructure*. Mater. Sci. Engng A. 1998;242:77-86.

[82] Weitz T, Karnthaler HP. *Transformation strains in martensitic phase transitions of Co alloys*. Phase Transitions. 1999;67:695-705.

[83] Zhao H, Song M, Ni S, Shao S, Wang J, Liao X. *Atomic-scale understanding of stress-induced phase transformation in cold-rolled Hf*. Acta Mater. 2017;131:271-279.

[84] Zhao H, Hu X, Song M, Ni S. *Mechanisms for deformation induced hexagonal close-packed structure to face-centered cubic structure transformation in zirconium*. Scripta Mater. 2017;132:63-67.

[85] Hirth JP, Pond RC. *Steps, Dislocations and disconnections as interface defects relating to structure and phase transformations*. Acta Mater. 1996;44:4749–4763.

[86] Pond RC, Ma X, Hirth JP. *Geometrical and physical models of martensitic transformations in ferrous alloys*. J. Mater. Sci. 2008;43:3881-3888.

[87] Howe JM, Pond RC, Hirth JP. *The role of disconnections in phase transformations*. Prog. Mater. Sci. 2009;54:792-938.

[88] Pond RC, Hirth JP, Serra A, Bacon DJ. *Atomic displacements accompanying deformation twinning: shear and shuffles*. Mater. Res. Lett. 2016;DOI: 10.1080/21663831.2016.1165298.

[89] Hirth JP, Wang J, Tomé CN. *Disconnections and other defects associated with twin interfaces*. Prog. Mater. Sci. 216;83:417-471.

[90] Capolungo L, Beyerlein IJ. *Nucleation and stability of twins in hcp metals*. Phys. Rev. B. 2008;78:024117.

[91] Wang J, Zeng Z, Weinberger CR, Zhang Z, Zhu T,. Mao SX. *In situ atomic-scale observation of twinning dominated deformation in nanoscale body-centred cubic tungsten*. Nature Mater. 2015;14:594-600.

[92] Wilson KG. *Problems in Physics with Many Scales of Length*. Scientific American. 1979:158-179.

[93] Meyers M.A. *On the growth of lenticular martensite*, Acta Metall. 1980;28:757-770.

[94] Barsch GR, Krumhansl JA. *Twin Boundaries in Ferroelastic Media with Interface Dislocations*. Phys. Rev. Lett. 1984;53:1069-1072.

[95] Barsch GR, Horovitz B, Krumhansl JA. *Dynamics of Twin Boundaries in Martensites*. Phys. Rev. Lett. 1987;59:1251-1254.

[96] Flack F. *Ginzburg-Landau Theory and Solitary Waves in Shape-Memory Alloys*. Z. Phys. B – Condensed Matter. 1984;54:159-167.

[97] Kashchenko MP, Chashchina VG. *Key Role of Transformation Twins in Comparison of Results of Crystal Geometric and Dynamic Analysis for Thin Plate Martensite*. Phys. Metals Metallography, 2013;114:821–825.

[98] Volterra V. *Sur l'équilibre des corps élastiques multiplement connexes*. Ann. Ecole Normale Sup. Paris. 1907;24:401-518.

[99] Romanov AE. *Mechanics and physics of dislinations in solids*. Eur. J. Mech. A/Solids. 2003;22:727-741.

[100] Kleman M, Friedel J. *Disclinations, dislocations, and continuous defects: A reappraisal*. Rev. Mod. Phys. 2008;82:61-115.

[101] Klimanek P, Klemm V, Romanov AE, Seefeldt M. *Disclinations in Plastically Deformed Metallic Materials*. Adv. Eng. Mater. 2001;3:877-884.

[102] Müllner P, Romanov AE. *Internal Twinning in deformation twinning*. Acta Mater. 2000;48:2323-2337.

[103] Müllner P, King A.H. *Deformation of hierarchically twinned martensite*. Acta Mater. 2010;58:5242-5261.

[104] Kelly PM. *Martensite Crystallography – The Apparent Controversy between the Infinitesimal Deformation Approach and the Phenomenological Theory of Martensitic Transformations*. Metall. Mater. Trans. A. 2003;34:1783-1786.

[105] Gey N, Humbert M. *Specific analysis of EBSD data to study the texture inheritance due to the $\beta \rightarrow \alpha$ phase transformation*. J. Mater. Sci. 2003;38:1289-1294.

[106] Cayron C. *Groupoid of orientational variants*. Acta Cryst. 2006;62:21-40.

[107] Cayron C, Artaud B, Briottet L. *Reconstruction of parent grains from EBSD data*. Mater. Charact. 2006;57:386-401.

[108] Cayron C. *ARPGE: A computer program to automatically reconstruct the parent grains from electron backscatter diffraction data*. J. Appl. Cryst. 2007;40:1183-1188.

[109] Bunge HJ, Weiss W, Klein H, Wcislak L, Garbe U, Schneider JR. *Orientation relationship of Widmannstätten plates in an iron meteorite measured with high-energy synchrotron radiation*. J. Appl. Cryst. 2003;36:137-140.

[110] Nolze G, Geist V, Saliwan Neumann R, Buchheim M. *Investigation of orientation relationships by EBSD and EDS on the example of the Watson iron meteorite*. Cryst. Res. Technol. 2005;40:791-804.

[111] Cayron C, Barcelo F, de Carlan Y. *The mechanism of the fcc-bcc martensitic transformation revealed by pole figures*. Acta Mater. 2010;58:1395-1402.

[112] Cayron C, Barcelo F, de Carlan Y. *Reply to "Comments on 'The mechanism of the fcc-bcc martensitic transformation revealed by pole figures'"*. Scripta Mater. 2011;64:103-106.

[113] Thiaudière D, Hennet L, King A, de Carlan Y, Béchade JL, Cayron C. *Soleil and ESRF ultra-fast in situ X-ray diffraction experiments to track the hcp phases in martensitic steels*. 2012, unpublished results.

[114] Pitsch W. *The Martensite Transformation in Thin Foils of Iron-Nitrogen Alloys*. Phil. Mag. 1959;4:577-584.

[115] Cayron C. *One-step model of the face-centred-cubic to body-centred-cubic martensitic transformation*. Acta Cryst. 2013;69:498-509.

[116] Cayron C. *EBSD imaging of orientation relationships and variants groupings in different martensitic alloys and Widmanstätten iron meteorites*. Mater. Charac. 2014;94:93-110.

[117] Cayron C. *Continuous atomic displacements and lattice distortion during fcc–bcc martensitic transformation*. Acta Mater. 2015;96:189-202.

[118] Kong LT. *Phonon dispersion measured directly from molecular dynamics simulations*. Comp. Phys. Com. 2011;182:2201-2207.

[119] Sinclair CW, Hoagland RG. *A molecular dynamics study of the fcc→bcc transformation at fault intersections*. Acta Mater. 2008;56:4160-4171.

[120] Sandoval L, Hurbassek HM. *The Bain versus Nishiyama-Wassermann path in the martensitic transformation of Fe*. New J. Phys. 2009;11:103027.

[121] Wang B, Sak-Saracino E, Gunkelmann N, Urbassek HM. *Molecular-dynamics study of the $\alpha \leftrightarrow \gamma$ phase transition in Fe-C*. Comp. Mater. Sci. 2014;82:399-404.

[122] Meiser J, Urbassek HM. *Martensitic transformation of pure iron at a grain boundary: Atomistic evidence for a two-step Kurdjumov-Sachs−Pitsch pathway*. AIP Adv. 2013;6:085017.

[123] Cayron C. *Angular distortive matrices of phase transitions in the fcc-bcc-hcp system*. Acta Mater. 2016;111:417-441.

[124] Baur AP, Cayron C, Logé RE. *{225}$\gamma$ habit planes in martensitic steels: from the PTMC to a continuous model*. Sci. Rep. 2017;7:40938.

[125] Shimizu K, Oka M, Wayman CM. *The association of martensite platelets with austenite stacking faults in an Fe-8Cr-1C alloy*. Acta Metall. 1970;18:1005-1011.

[126] Shibata A, Murakami T, Morito S, Furuhara T, Maki T. *The Origin of Midrib in Lenticular Martensite*. Mater. Trans. 49;2008:1242-1248.

[127] Cayron C, Baur AP, Logé RE. *A crystallographic model of the {557} habit planes in low-carbon martensitic steels*. https://arxiv.org/abs/1606.04257

[128] Peet MJ, Bhadeshia HKDH. *Surface Relief Due to Bainite Transformation at 473K (200°C)*. Metal. Mater. Trans. A. 2011;42A:3344-3348.

[129] Burgers WG. *On the process of transition of the cubic-body-centered modification into the hexagonal-close-packed modification of zirconium*. Physica 1934;1:561-586.

[130] Nishiyama Z. *Martensitic Transformation*. Ed. by M.E. Fine, M. Meshii, C.M. Waymann, Materials Science Series, Academic Press, New York; 1978.

[131] Sowa H. *Sphere packing as tool for a description of martensitic transformations*. Acta Cryst. A. 2017;73:1-7.

[132] Cayron C. *Hard-sphere displacive model of extension twinning in magnesium.* Mater. Design. 2017;119:361-375.

[133] Liu B-Y, Wang J, Li B, Lu L, Zhang X-Y, Shan Z-W, Li J, Jia C-L, Sun J, Ma E. *Twinning-like lattice reorientation without a crystallographic plane*. Nat. Com. 2014;5:3297.

[134] Barnett MR, Keshavarz Z, Beer AG, Ma X. *Non-Schmid behavior during secondary twinning in a polycrystalline magnesium alloy*. Acta Mater. 2008;56:5-15.

[135] Luo JR, Godfrey A, Liu W, Liu Q. *Twinning behavior of a strongly basal textured AZ31 Mg alloy during warm rolling*. Acta Mater. 2012;60:1986-1998.

[136] Jonas JJ, Mu S, Al-Samman T, Gottstein G, Jiang L, Martin E. *The role of strain accommodation during the variant selection of primary twins in magnesium*. Acta Mater. 2011;59:2046-2056.

[137] Beyerlein IJ, Capolungo L, Marshall PE, McCabe RJ, Tomé CN. *Statistical analyses of deformation twinning in magnesium*. Phil.Mag. 2010;90:2161-2190.

[138] Li B, Ma E. *Atomic Shuffling Dominated Mechanism for Deformation Twinning in Magnesium*. Phys. Rev. Lett. 2009;103:035503.

[139] Wang J, Yadav SK, Hirth JP, Tomé CN, Beyerlein IJ. *Pure-Shuffle Nucleation of Deformation Twins in Hexagonal-Close-Packed Metals*. Mater. Res. Lett. 2013;1:126-132.

[140] Liu B-Y, Wan L, Wang J, Ma E, Shan Z-W. *Terrace-like morphology of the boundary created through basal-prismatic transformation in magnesium*. Scripta Mater. 2015;100:86-89.

[141] B. Li, X.Y. Zhang, *Twinning with zero shear*, Scripta Mater. 125 (2016) 73-79.

[142] Christian JW, Olson GB, Cohen M. *Classification of Displacive Transformations: What is a Martensitic Transformation?* J. Phys. IV 1995;C8:3-10.

[143] Wang HL, Hao YL, He SY, Du K, Li T, Obbard EG, Hudspeth J, Wang JG, Wang YD, Wang Y, Prima F, Lu N, Kim MJ, Cairney JM, Li SJ, Yang R. *Tracing the coupled atomic shear and shuffle for a cubic to a hexagonal crystal transition*. Scripta Mater. 2017;133:70-74.

[144] Delaey L. *Diffusionless Transformations*, in Phase Transformations in Materials. ed G. Kostorz, Wiley-VCH Verlag GmbH & Co. KGaA, Weinheim, FRG. doi: 10.1002/352760264X.ch9; 2001.

[145] Ogawa K, Kajiwara S. *High-resolution electron microscopy study of ledge structures and transition lattices at the austenite-martensite interface in Fe-based alloys*. Phil. Mag. 2004;84:2919-2947.

[146] Tu J, Zhang S. *On the $\{10\bar{1}2\}$ twinning growth mechanism in hexagonal close-packed metals*. Mater. Design. 2016;96:143-149.

[147] Cayron C. *Hard-sphere displacive model of deformation twinning in hexagonal close-packed metals. Revisiting the case of the (56°, a) contraction twins in magnesium*. Acta Cryst. A. 2017;73:346-356.

[148] Serra A, Pond RC, Bacon DJ. *Computer Simulation of the Structure and Mobility of Twinning Dislocations in H.C.P. Metals*. Acta Metall. Mater. 1991;39:1469-1480.

[149] Serra A, Pond RC, Bacon DJ. *The Crystallography and Core Structure of Twinning Dislocations in H.C.P. Metals*, Acta Metall. Mater. 1988;36:3183-3203.

[150] Cayron C, Logé RE. *Unconventional twinning modes in magnesium*. https://arxiv.org/abs/1707.00490. Under review.

[151] Le Lann A, Dubertret A. *A Development of Kronberg's Model for $\{10\bar{1}2\}$ Twins in H.C.P. Metals*. Phys. Stat. Sol. A. 1979;51:497-507.

[152] Dubertret A, Le Lann A. *Development of a New Model for Atom Movement in Twinning*. Phys. Stat. Sol. A. 1980;60:145-151.

[153] Kronberg ML. *A structural mechanism for the twinning process on {10$\bar{1}$2} in hexagonal close packed metals.* Acta Metall. 1968;16:29-34.

[154] Qin RS, Bhadeshia HK. *Phase field method*. Mater. Sci. Techn. 2010; 26:803-811.

[155] Levitas VI. *Thermodynamically consistent phase field approach to phase transformations with interface stresses*. Acta Mater. 2013;61:4305-4319.

[156] Landau L. *On the theory of phase transitions*. Zh. Eksp. Teor. Fiz. 1937;7:19-32, translated and reprinted in Ukr. J. Phys. 2008;53:25-35.

[157] Wigner E. *Gruppentheorie und ihre Anwendung auf die Quantenmechanik der Atomspektren*. Springer Fachmedien Wiesbaden GmbH; 1931.

[158] Weyl H. *Gruppentheorie Und Quantenmechanik*. S. Hirzel, Leipzig; 1931.

[159] Clapp PC. *How Would we Recognize a Martensitic Transformation if it Bumper us on a Dark and Austy Night?* J. Phys. IV. 1995;C8-5:11-19.

[160] Clayton JD, Knap J. *A phase field model of deformation twinning: Nonlinear theory and numerical simulations*. Physica D. 2011;240:841-858.

[161] Mnyukh Y. *Second-Order Phase Transitions, L. Landau and His Successors*. Am. J. Cond. Mat. Phys. 2013;3:25-30.

[162] Tolédano JC, Dmitriev V. *Reconstructive Phase Transitions in Crystals and Quasicrystals*. World Scientific, Singapore; 1996.

[163] Janovec V. *Group analysis of domains and domain pairs*. Czech. J. Phys. B. 1972;22:975-994.

[164] Janovec V, Dvorakova E. *The coset and double coset decomposition of the 32 crystallographic point groups*. Acta Cryst. A. 1989;45:801-802.

[165] Kalonji G, Cahn JW. *Symmetry constraints on the orientation dependence of interfacial properties: the group of the Wulff plot*. J. Phys. 1982;C6-43:25-31.

[166] Brandt W. *Über eine Verallgemeinerung des Gruppenbegriffes*. Math. Ann. 1926;96:360-366.

[167] Brown R. *From groups to groupoids: a brief survey*. Bull. London Math. Soc. 19; 1987:113-134.

[168] Ginzburg VL, Landau LD. *Zh. Eksp. Teor. Fiz.* 1950;20:1064. English translation in Landau LD. Collected papers, Oxford: Pergamon Press; 1965, p. 546.